\documentclass[useAMS,usenatbib,times]{mn2e}
\usepackage{amsmath}
\usepackage{graphicx}
\usepackage{amssymb}
\usepackage{subfigure}
\def\pageoffset#1#2{\hoffset=#1\relax\voffset=#2\relax} 
\pageoffset{0pc}{2.5pc} 
 at 10truept
\def\and  {\it {et al.} \rm}

\def\spose#1{\hbox to 0pt{#1\hss}}
\def\simlt{\mathrel{\spose{\lower 3pt\hbox{$\mathchar"218$}}
     \raise 2.0pt\hbox{$\mathchar"13C$}}}
\def\simgt{\mathrel{\spose{\lower 3pt\hbox{$\mathchar"218$}}
     \raise 2.0pt\hbox{$\mathchar"13E$}}}
\def\be{\begin{equation}}
\def\ee{\end{equation}}
\def\bce{\begin{center}}
\def\ece{\end{center}}
\def\bea{\begin{eqnarray}}
\def\eea{\end{eqnarray}}
\def\ben{\begin{enumerate}}
\def\een{\end{enumerate}}

\def\brr{\begin{array}}
\def\err{\end{array}}

\def\btau{\overline{\tau}}

\def\nh1{n_{\rm HI}}

\def \p1dk {P_{\rm 1D}(k)}
\def \simlt {\mathrel{\spose{\lower 3pt\hbox{$\mathchar"218$}}
     \raise 2.0pt\hbox{$\mathchar"13C$}}}
\def \simgt {\mathrel{\spose{\lower 3pt\hbox{$\mathchar"218$}}
     \raise 2.0pt\hbox{$\mathchar"13E$}}}

\def \P {{\cal P}}

\def \M {{\bf M}}

\def \be {\begin{equation}}
\def \en {\end{equation}}
\def \bea {\begin{eqnarray}}
\def \ena {\end{eqnarray}}

\def \bi {\begin{itemize}}
\def \ei {\end{itemize}}

\usepackage[usenames]{color}
\definecolor{Blue}{rgb}{0,0.08,0.65}
\definecolor{Red}{rgb}{0.65,0.08,0.05}
\definecolor{Green}{rgb}{0.15,0.45,0.25}

\newcommand{\noun}[1]{\textsc{#1}}
\providecommand{\tabularnewline}{\\}
\global\long\def\V#1{\boldsymbol{#1}}
\global\long\def\M#1{\mathbf{#1}}
\global\long\def\TR{^{\mathrm{T}}}

\global\long\def\RM#1{\mathrm{#1}}
\global\long\def\abs#1{|#1|}
\global\long\def\Abs#1{\left|#1\right|}
\global\long\def\avg#1{\langle#1\rangle}
\global\long\def\Avg#1{\left\langle #1\right\rangle }
\global\long\def\norm#1{\Vert#1\Vert}
\global\long\def\Norm#1{\left\Vert #1\right\Vert }

\global\long\def\Spin#1{{}_{p}\!#1}
\global\long\def\Qmap{\mathcal{Q}}
\global\long\def\Qdata{\mathcal{L}}
\global\long\def\Qprior{\mathcal{R}}

\begin{document}

%\title[{{ASKI}:All-sky  convergence maps...}]{{ASKI}:All-sky  convergence maps from pixelized  partial  shear data}
% towards a full-sky lensing map making pipeline}]{{ASKI}: towards a full-sky lensing map making pipeline}
\title[{{ASKI}:  full-sky lensing map making algorithms}]{{ASKI}:  full-sky lensing map making algorithms}\author[C. Pichon, E. Thi\'ebaut, S. Prunet, K. Benabed, S. Colombi, T. Sousbie, R. Teyssier]{
C. Pichon{$^{1,2,3}$}, E. Thi\'ebaut{$^{2,1}$}, S. Prunet{$^{1}$}, \newauthor
K. Benabed{$^{1}$}, S. Colombi{$^{1}$}, T. Sousbie{$^{1,4}$}, R. Teyssier{$^{3,1,5}$}
\\
\\
$^{1}$Institut d'astrophysique
de Paris \& UPMC (UMR 7095), 98, bis boulevard Arago , 75 014, Paris,
France.\\
 $^{2}$Observatoire de Lyon (UMR 5574), 9 avenue Charles
Andr\'e, F-69561 Saint Genis Laval, France.\\
 $^{3}$CEA/IRFU/SAP, l'Orme des Merisiers,
91170, Gif sur Yvette, France\\
$^{4}$ Tokyo University, Physics Dept 7-3-1 Hongo Bunkyo-ku,JP Tokyo 113-0033 Japan \\
$^{5}$ Institute f\"ur Theoretische Physik,  Universit\"at Z\"urich, Winterthurerstrasse 190  CH-8057 ZŸrich 
\\
}
\maketitle

\begin{abstract}
Within the context of upcoming full-sky lensing surveys, the edge-preserving non-linear 
 algorithm
\noun{Aski} (\emph{All Sky $\kappa$ Inversion}) is presented. Using
the framework of Maximum A Posteriori inversion, it aims at recovering the {\sl optimal} full-sky 
convergence map from noisy surveys with masks.
\noun{Aski} {contributes} 
two steps: (i) CCD images of possibly crowded galactic fields are
deblurred using {\sl automated} edge-preserving deconvolution; (ii) once
the reduced shear is estimated using standard techniques, the partially masked convergence map is also inverted
via an edge-preserving method.
% A critical feature of both components of the
%\noun{Aski} algorithm is that the penalty can be applied in model
%space (i.e.  \emph{resp}. the Fourier and the harmonic coefficients of the
%corresponding sky brightness distribution and $\kappa$ maps), while
%the optimization iterates back and forth between data space (i.e.  \emph{resp}. the pixels
%of the image and of the ellipticity map) and model space. This freedom  is essential in order to deal simultaneously with masks and edge preserving penalties.
% \noun{Aski} uses
%the efficient variable metric limited memory algorithm \noun{OptimPack}, which
%allows both optimizations to scale to high resolutions. The deblurring is implemented on
%Cartesian maps as large as  $16\,384^{2}$ pixels while the inversion is
%carried on the sphere for HEALPix resolutions as large as $n_{{\rm side}}=4096$. 

The efficiency of the deblurring of the image is quantified  by the relative  gain in the quality factor of the reduced
shear,  as estimated by \noun{Sextractor}. Cross validation
as a function of the number of stars removed yields an automatic estimate
of the optimal level of regularization for the deconvolution of the
galaxies. It is found that when the observed field is crowded, this
gain can be quite significant for realistic ground-based eight-metre
class surveys. The most significant improvement occurs when both positivity
and edge-preserving {\normalsize $\ell_{1}-\ell_{2}$} penalties are imposed during the iterative deconvolution. 
 
The quality of the convergence inversion  is
investigated on noisy maps derived from the \noun{horizon}-4$\pi$ N-body simulation with SNR within
the range $\ell_{{\rm cut}}=500-2500$, with and without Galactic
cuts, and quantified using one-point statistics ($S_{3}$ and $S_{4}$),
power spectra, cluster counts, peak patches and the skeleton. 
It is
found that (i) the reconstruction is able to interpolate and extrapolate
within the Galactic cuts/non-uniform noise; (ii) its sharpness-preserving
penalization avoids strong biasing near the clusters of the map (iii)
it reconstructs well the shape of the PDF as traced by its skewness
and kurtosis (iv) the geometry and topology of the reconstructed map
is close to the initial map as traced by the peak patch distribution
and the skeleton's differential length (v) the two-points statistics
of the recovered map is consistent with the corresponding smoothed
version of the initial map (vi) the distribution of point sources
is also consistent with the corresponding smoothing, with a significant
improvement when $\ell_{1}-\ell_{2}$ prior is applied. The contamination
of B-modes when realistic Galactic cuts are present is also investigated.
Leakage mainly occurs on large scales. The non-linearities implemented
in the model are significant on small scales near the peaks in the
field. 
\end{abstract}


\date{10/10/08}

}

\section{Introduction}\label{sec:Introduction}
%

% \citep{2007MNRAS.376}
% 
% \bibliographystyle{mn2e}
%\bibliography{aski}
%\end{document}

In recent years, weak shear measurements have become a major source
of cosmological information. By measuring the bending of the rays
of light emerging from distant galaxies, one can gain some knowledge
of the distribution of matter between the emitter and ourselves, and
thus probe the properties and evolution history of dark matter \citep{2001PhR...340..291B}.
This technique has led to significant results in a broad spectrum
of topics, from measurements of the projected dark matter power spectrum
(for the latest results see \citet{Fu:2007}), 3D estimation of the
dark matter spectrum \citep{Kitching:2006p542}, studies of the higher
order moments of the dark matter distribution, selection of source
candidates for subsequent follow-ups \citep{schirmer:2007p920}, and
reconstruction of the mass distribution from small \citep{Jee:2007p996}
to large scales \citep{Massey:2007p176}. 
In view of these successes, numerous surveys have been planned specifically
to use this probe either from ground-based facilities (eg VST-KIDS%
\footnote{\texttt{http://www.astro-wise.org/projects/KIDS/}%
}, DES%
\footnote{\texttt{https://www.darkenergysurvey.org/}%
} Pan-STARRS%
\footnote{\texttt{http://pan-starrs.ifa.hawaii.edu/}%
}, LSST%
\footnote{\texttt{http://www.lsst.org/}%
}) or space-based observatories (EUCLID%
\footnote{\texttt{http://www.dune-mission.net/}%
}, SNAP%
\footnote{\texttt{http://snap.lbl.gov/}%
} and JDEM%
\footnote{\texttt{http://universe.nasa.gov/program/probes/jdem.html}%
}). More generally, it is clear that weak lensing will be a major player
in the future, as it has been identified by different European and
US working groups as one of the most efficient way of studying the
properties of dark energy%
\footnote{see, on the European side \texttt{http://www.stecf.org/coordination/
}and on the US side\texttt{,} \texttt{http://www.nsf.gov/mps/ast/aaac.jsp}
and \texttt{http://www.nsf.gov/mps/ast/detf.asp}%
}.
Data processing is an important issue in the exploitation of weak
lensing of distant galaxies. The signal comes from the excess alignment
of the ellipticities of the observed galaxies. Assuming one can ignore
or deal with spurious alignments due to intrinsic effects \citep{Hirata:2004p1549,aubert04,PB},
or due to spurious lensing effects \citep{Bridle:2007p1602}, the
weak lensing signal will thus come from a small statistically coherent
ellipticity on top of the random one of each object. Any result obtained
with weak lensing on distant galaxies is thus conditioned by the quality
with which shape parameters of the galaxies are recovered. This issue
has of course been raised by the weak lensing community and tackled
by the SHear Testing Program working group \citep{2007MNRAS.376,2006MNRAS.368.1323H}
whose effort have allowed for a fair comparison of the existing techniques.
Schematically, the measurement of the shape parameters of the galaxies
can be seen as a two-step process. First, one must correct for the
non-idealities of the images due to atmospherical seeing (for ground-based telescopes), and telescope and camera aberrations. Indeed, these
effects translate into an asymmetrical beam, which is varying between
two images, and even possibly in the field of one image. Typically,
the asymmetry induced by the instrumental response is much larger
than the ellipticity to be measured. After this preprocessing step,
a shape determination algorithm can be applied, and some estimation
of the ellipticity of the object recovered. Stars, defects in the
images, and objects too close to each other after deconvolution have
to be removed from the final catalogue so as to avoid contamination
from erroneous shape measurements. 

After these operations, one obtains a catalogue of position and shape
parameters. Many techniques exist for recovering the weak shear signal
from this catalogue. For example, a lot of efforts have been devoted
to the measurement of the shear two-point functions. The most used
method is the two-point functions; however  measurement of the so called \emph{Mass Aperture} averaged 
two-point function, which is the result of the convolution of the
shear two-point functions by a compensated filter \citep{2002A&A...396....1S} is becoming  the preferred method \citep{Fu:2007}.
This scheme includes the separation between the curl-free convergence-field two-point function, and the residual curl mode that can arise
from incomplete PSF correction or intrinsic galaxy alignment \citep{Crittenden:2002p2070}.
For three-point functions, different resummation schemes have been
proposed, either using direct measurement of the shear \citep{Bernardeau:2002p2223,Benabed:2005p2447}
or using the \emph{Mass Aperture }filter \citep{Takada:2003p2287,Kilbinger:2005p2383}. 

Other applications (source detection and fit, some tomography algorithms)
call for an estimation of the map of the convergence field. A convergence
map can also be used to measure the two- and three-point functions
as well, even if, as we will see later this is not optimal. 
For these reasons an important amount of work  has already been
devoted to the reconstruction of the convergence map \citep{ka11,ka12,k14}. The problem
in this reconstruction lies in the inversion of the non-local equations
linking the convergence field $\kappa$, and the ellipticities of
the galaxies, while controlling the noise and avoiding pollution from
the spurious curl modes. Moreover, even assuming that the ellipticity
catalogue was a noise free estimation of a curl-free underlying shear,
the inversion could only be exact up to a global translation given 
the functional form of the equation. Thus Bayesian techniques that use \emph{a
priori} properties on the solution to regularize the inversion problem
are well suited to the reconstruction of $\kappa$. Previous works
on the topic have explored different sets of \emph{a priori }and regularization
techniques \citep{Marshall:2002p2821,Starck:2005p2878, Seitz:1998p2768,Bridle:1998p2820}.
The primary goal of those investigations being the measurement of the mass
distribution in clusters, most of them are dealing only with finite 
regions of the sky.  For the same reason those papers have been extended
to include strong-lensing effects that can be observed around the
cluster whose mass is being reconstructed using their lensing effect
\citep{Cacciato:2006p3147,Bradac:2005p3138,Halkola:2006p3208,Jee:2007p996}. 

In this paper, we will focus on the {optimal} reconstruction of the $\kappa$
{\sl field} from very large, and possibly full-sky maps, of the sky. We
will thus only be interested in the weak lensing regime including
the onset of the quasi-linear regime, where the non-linearities of
the relation linking the ellipticities of the galaxies to the shear
cannot be safely neglected. We will propose a self calibrated regularization technique,
that can be compared to multi resolution methods  or wavelet approach
\citep{Starck:2005p2878,starckforever}, and use a $\ell_{1}-\ell_{2}$ regularization
scheme to perform a sharp feature preserving inversion. One of the
biggest issues we will have to cope with is the incomplete coverage
of the sky. We will show how our technique can deal with irregular
coverage and masked portions of the sky. 

Specifically, Section \ref{sec:deblurring}
shows how self calibrated non-parametric $\ell_{1}-\ell_{2}$ deblurring can {\sl improve}
the construction of reduced shear, hence convergence maps. Section
\ref{sec:A-Model-for} describes the model for the reduced shear,
the corresponding inverse problem, and the optimization procedure. Section
\ref{sec:Validation} investigates the quality of the global reconstruction;
in particular, it probes the asymmetry/kurtosis of the recovered maps,
its topology (total length and differential length of the skeleton), the
recovered power spectra, the point source catalogue with and without
galactic star cut. The leaking of B-modes induced by the Galactic
cut is also investigated. Finally, Section \ref{sec:Discussion} discusses
implications for upcoming full-sky surveys and wraps up.\\
Appendix~\ref{sec:Efficient-Star-Removal} describes the 
star removal algorithm (implemented for the cross validation estimation of  the optimal level of smoothing
required to deconvolve the crowded images), Appendix~\ref{sec:Detailled-model} details the $\kappa$ inverse problem on the sphere
while Appendix~\ref{sec:From-the-sphere} derives the local plane corresponding approximation. Appendix~\ref{sec:simu-2-map} 
describes the construction of realistic $\kappa$ maps from large N-body simulations.
\section{Deblurring of crowded fields}\label{sec:deblurring}
The first step involved in reconstructing a full-sky map of the convergence
on the sky requires estimating ellipticity and orientation maps from
wide angle CCD images of large patches of the sky. Whether the experiment
is ground-based, or space-born, it is advisable to correct for the
effect of the instrumental response, in particular when mapping more
crowded regions closer to the galactic plane. Indeed, the PSF-induced
partial overlapping of galaxies within the field of view will bias
the estimation of the reduced shear.
What we will describe here would correspond to a method belonging to the ``orange" quadrant of the classification proposed in table 3 of \cite{2007MNRAS.376}.
{Current methods have been designed for deblurring of isolated objects and are consequently less efficient in deblurrinng blended objects.
}
 As a first step towards building
a full-sky map maker, let us therefore address the issue of deblurring
crowded fields via regularized non parametric model fitting, and assess
its efficiency in the weak lensing context. 

In particular we will show that
cross validation
as a function of the number of stars removed yields an automatic estimate
of the optimal level of regularization for the deconvolution of the
galaxies. When the observed field is crowded, this
gain can be quite significant for realistic ground-based eight-metre
class surveys. The most significant improvement occurs when both positivity
and edge-preserving {\normalsize $\ell_{1}-\ell_{2}$} penalties are imposed during th
e iterative deconvolution.

\subsection{Deblurring as an inverse problem}\label{sec:inverse-deblurring-problem}
\subsubsection{Regularized solution}
Since observed objects are incoherent sources, the observed image
depends linearly on the sky brightness distribution:\[
y(\boldsymbol{\omega})=\int h(\boldsymbol{\omega},\boldsymbol{\omega}')\, x(\boldsymbol{\omega}')\,\mathrm{d}\boldsymbol{\omega}'+e(\boldsymbol{\omega})\,,\]
where $y(\boldsymbol{\omega})$ is the observed distribution in the
direction $\boldsymbol{\omega}$, $h(\boldsymbol{\omega},\boldsymbol{\omega}')$
is the atmospheric and instrumental point spread function (PSF) which
is the distribution of observed light in the direction $\boldsymbol{\omega}$
due to light coming from direction $\boldsymbol{\omega}'$, $x(\boldsymbol{\omega}')$
is the true sky brightness distribution and $e(\boldsymbol{\omega})$
is the noise. After discretization:
\begin{equation}
\V y=\M H\cdot\V x+\V e\,,\label{eq:model2}\end{equation}
 where $\V y$ is the vector of pixel intensities in the observed
image (the data), $\M H$ is the matrix which accounts for the atmospheric
and instrumental blurring, $\V x$ is the (discretized or projected
onto a basis of functions) object brightness distribution and $\V e$
accounts for the errors (pixel-wise noise and modelisation approximations).
Deblurring requires estimating the best sky brightness distribution
given the data. Since the atmospheric and instrumental PSF results
in a smoother distribution than the true one, it is well known that
de-blurring is an ill-conditioned problem (\citep{1972JOSA...62...55R,1979MNRAS.187..145S,1982RvGSP..20..219T,1998MNRAS.301..419P,2001MNRAS.326..597P}).
In other words, straightforward deblurring by applying $\M H^{-1}$
to the data $\V y$ would result in uncontrolled amplification of
noise: a small change in the input data would yield unacceptably large
artifacts in the solution. Regularization must be used to overcome
ill-conditioning of this inverse problem. This is achieved by using
additional prior constraints such as requiring that the solution be
as smooth as possible, while being still in statistical agreement
with the data and while imposing that the brightness distribution
is positive. Following this prescription, the Maximum A Posteriori
(MAP) solution $\V x_{\mu}$ is the one which minimizes an objective
function $\Qmap(\V x)$:\begin{equation}
\V x_{\mu}=\arg\min_{\V x\ge0}\Qmap(\V x)\,,\quad\mbox{with:\ }\Qmap(\V x)=\Qdata(\V x)+\mu\,\mathcal{R}(\V x)\,,\label{eq:optim-problem}\end{equation}
where $\Qdata(\V x)$ is a likelihood penalty which enforces agreement
of the model with the data, $\mathcal{R}(\V x)$ is a \emph{regularization}
penalty which enforces prior constraints set on the model, and $\mu>0$
is a so-called \emph{hyper-parameter} which allow the tuning of the
relative weight of the prior with respect to the data. Hence the MAP
solution is a compromise between what can be inferred from the data
alone and prior knowledge about the parameters of interest. Assuming
Gaussian statistics for the errors $\V e$ in equation~(\ref{eq:model2}),
the likelihood penalty writes:\begin{equation}
\Qdata(\V x)=(\M H\cdot\V x-\V y)^{\TR}\cdot\M W\cdot(\M H\cdot\V x-\V y)\,,\label{eq:Qdata}\end{equation}
where the weighting matrix $\M W$ is equal to the inverse of the
covariance matrix of the errors: $\M W\equiv\RM{Cov}(\V e)^{-1}$.

The most effective regularization for ill-conditioned problems such
as deconvolution of blurred images consists in imposing a smoothness
constraint \citep{2005opas.conf..397T}. Then the regularization penalty
writes:\begin{equation}
\Qprior(\V x)=\sum_{j}\phi(\Delta x_{j})\,,\label{eq:general-smoothness-penalty}\end{equation}
where $\Delta x_{j}$ is the local gradient of $\V x$ and $\phi$
is some cost function. The local gradient of $\V x$ can be approximated
by finite differences: $\Delta\V x=\M D\cdot\V x$ where $\M D$ is
a linear finite difference operator. For instance, in 1-D: $\Delta x_{j}=(\M D\cdot\V x)_{j}=x_{j+1}-x_{j}$.
To enforce smoothness, the cost function $\phi$ must be an increasing
function of the magnitude of its argument. Very common choices for
$\phi$ are: the $\ell_{2}$ norm, the $\ell_{1}$ norm, or an $\ell_{1}-\ell_{2}$
norm. For our deblurring problem, we have considered different priors
(quadratic or $\ell_{1}-\ell_{2}$ smoothness) possibly with an additional
positivity constraint. We have used generalized cross validation (GCV,\citep{1990smod.conf.....W})
applied to the circulant approximation of the quadratic problem to
estimate the optimal regularization level $\mu$. These different
possibilities and their effects on the recovered images are discussed
in details in what follows.

\textcolor{black}{Finally, to solve for the constrained optimization
problem (\ref{eq:optim-problem}), we used the }\textcolor{black}{\noun{vmlmb} }\textcolor{black}{
algorithm from }\textcolor{black}{\noun{OptimPack \citep{Thiebaut:spie2002:bdec}}}\textcolor{black}{.
}\textcolor{black}{\noun{Vmlmb} }\textcolor{black}{ (for }\textcolor{black}{\emph{Variable
Metric, Limited Memory, Bounded}}\textcolor{black}{) makes use of
a BFGS \citep{Nocedal_Wright-2006-numerical_optimization} update
of the approximation of the Hessian (matrix of second partial derivatives)
of $\Qmap(\V x)$ to derive a step to improve the parameters at every
iteration. This strategy only requires computing the objective function,
$\Qmap(\V x)$, and its gradient (partial derivatives) $\nabla_{\V x}\Qmap(\V x)$
with respect to the parameters $\V x$. The BFGS update is limited
to a few last steps so that the memory requirements remains modest,
that is $ $a few times the number of sought parameters, and the algorithm
can be applied to solve very large problems (in our case, there are
as many parameters as the number of pixels in the sought image). Finally,
}\textcolor{black}{\noun{Vmlmb} }\textcolor{black}{ accounts for bound
constraints by means of gradient projections \citep{Nocedal_Wright-2006-numerical_optimization}.
For a convex penalty $\Qmap(\V x)$, }\textcolor{black}{\noun{Vmlmb}}\textcolor{black}{
is guaranteed to converge to the unique feasible minimum of $\Qmap(\V x)$
which satisfies the bound constraints; for a non-convex penalty, }\textcolor{black}{\noun{Vmlmb}}\textcolor{black}{
being based on a descent strategy, it will find a local minimum depending
on the initial set of parameters.}
\begin{figure}
\includegraphics[width=1\columnwidth]{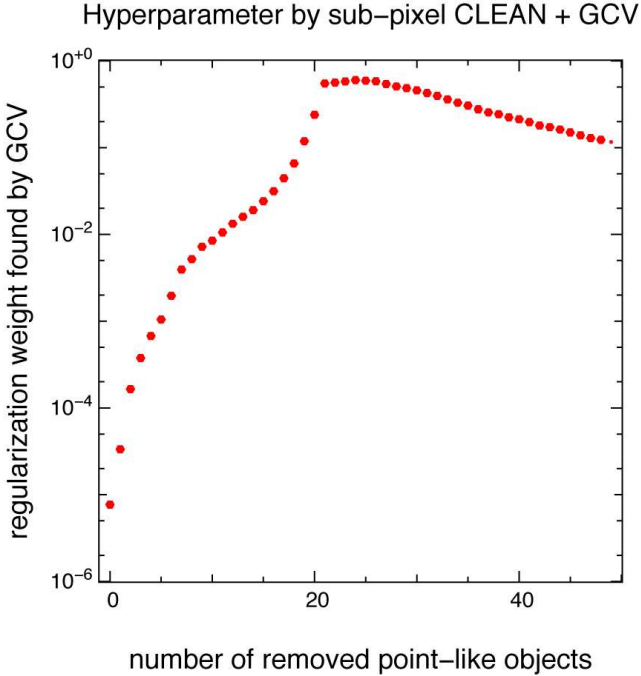}
\caption{Hyper-parameter chosen by GCV as a function of the number of stars
removed by our star removal algorithm (Appendix~\ref{sec:Efficient-Star-Removal}).
Note that this curve reaches a maximum corresponding to the moment
when all the stars have been removed. Indeed stars correspond to high
frequency correlated signal, while the wings of galaxies (for which
the core has been erroneously removed) also give rise to such signal.
In between, when all stars have been removed, while no galaxies has
yet been deprived of its core, the amount of correlated high frequency
signal reaches a minimum, or equivalently the GCV estimated value
of $\mu$ reaches a maximum.\label{fig:GCV}}
\end{figure}
\subsubsection{Quadratic regularization and Wiener proxy}
\begin{table}
\begin{tabular}{|r|r|r|r|r|r|}
\hline 
$n_{{\rm side}}$ & 128 & 256 & 512 & 1024 & 2048 \tabularnewline
\hline
time for one step (s) & 0.13 & 0.59 & 2.13 & 8.48 & 34.3 \tabularnewline
\hline
number of steps (s) & 13 & 12 & 9 & 13 & 24 \tabularnewline
\hline
total time (s)  & 2.6 & 10.4 & 33.4 & 171.1 & 1129.3 \tabularnewline
\hline
\end{tabular}
\caption{the performance of the optimization of the linearized inversion problem
$n_{{\rm side}}S_{{\rm FS}}^{\ell_{{\rm cut}}}$ as a function of
$n_{{\rm side}}$ for an octo \noun{Opteron} in\noun{ OpenMP}.\noun{\label{tab:the-performance-of}}}
\end{table}
Using the finite difference operator $\M D$ and an $\ell_{2}$ norm
for the regularization and ignoring for the moment the positivity
constraint, the MAP solution is the minimum of a quadratic penalty
which simply involves solving a (huge) linear problem:\begin{eqnarray}
\V x_{\mu} & = & \arg\min_{\V x}\left\{ (\M H\cdot\V x-\V y)^{\TR}\cdot\M W\cdot(\M H\cdot\V x-\V y)\right.\nonumber \\
 &  & \quad\quad\quad\quad\left.+\mu\,(\M D\cdot\V x)^{\TR}\cdot(\M D\cdot\V x)\right\} \nonumber \\
 & = & \left(\M H^{\mathrm{T}}\cdot\M W\cdot\M H+\mu\,\M D^{\mathrm{T}}\cdot\M D\right)^{-1}\cdot\M H^{\mathrm{T}}\cdot\M W\cdot\V y\,,\label{eq:quadratic-solution}\end{eqnarray}
providing the Hessian matrix $\M H^{\mathrm{T}}\cdot\M W\cdot\M H+\mu\,\M D^{\mathrm{T}}\cdot\M D$
is non-singular, which is generally the case for $\mu>0.$ Owing to
the large size of the matrices involved in this equation (there are
as many unknown as the number of pixels), the linear problem has to
be iteratively solved (by a limited memory algorithm such as \noun{vmlm})
unless it can be diagonalized as explained below. The solution, equation~(\ref{eq:quadratic-solution}),
involves at least one parameter, $\mu$, which needs to be set to
the correct level of regularization: too low would give a solution
plagued by lots of artifacts due to noise amplification, too high
would result in an oversmoothed solution with small details blurred.
The optimal level of smoothing can be computed by generalized cross
validation (GCV) by minimizing with respect to $\mu$ the function
\citep{Golub:Heath:Wahba:1979,1990smod.conf.....W}:\begin{equation}
\mathrm{GCV}(\mu)=\frac{\left(\M A_{\mu}\cdot\V y-\V y\right)^{\mathrm{T}}\cdot\M W\cdot\left(\M A_{\mu}\cdot\V y-\V y\right)}{\left[1-\mathrm{tr}(\M A_{\mu})/N\right]^{2}}\,,\label{eq:GCV}\end{equation}
where $N$ is the number of data (size of $\V y$) and $\M A_{\mu}=\nabla_{\V y}(\M H\cdot\V x_{\mu})$
is the so-called \emph{influence matrix}, in our case:\begin{equation}
\M A_{\mu}=\M H\cdot\left(\M H^{\mathrm{T}}\cdot\M W\cdot\M H+\mu\,\M D^{\mathrm{T}}\cdot\M D\right)^{-1}\cdot\M H^{\mathrm{T}}\cdot\M W\,.\label{eq:influence-matrix}\end{equation}
Computing the value of $\mathrm{GCV}(\mu)$ involves: (i) solving
the problem to find the regularized solution $\V x_{\mu}$ and compute
$\M A_{\mu}\cdot\V y=\M H\cdot\V x_{\mu}$; (ii) estimate the trace
of $\M A_{\mu}$ perhaps by using Monte Carlo methods \citep{Girard-1989-fast_monte_carlo_cross_validation}
since the influence matrix is very large. The computational cost of
stages (i) and (ii) is similar to that of a few solvings of the quadratic
problem. Since this has to be repeated for every different value of
the regularization level, finding the optimal value of $\mu$ by means
of GCV can be very time consuming unless the problem can be approximated
by a diagonal quadratic problem (for which matrix inversions are both
fast and trivial).

For this purpose, we introduce the proxy problem corresponding to
white noise and circulant approximations of the operators $\M H$
(convolution by the PSF) and $\M D$ (finite differences). Then the
weighting matrix becomes:\[
W_{i,j}=\delta_{i,j}/\sigma^{2}\,,\quad\mbox{where\ }\sigma^{2}=\mathrm{Var}(n_{i}).\]
where $\sigma^{2}=\mathrm{Var}(e_{i})$ is the variance of the noise.
In the special case where the PSF is \emph{shift-invariant}, $\M H$
is a convolution operator which can be approximated by a block Toeplitz
with Toeplitz block matrix that can be coputed very quickly by means
of FFT's:\begin{equation}
\M H\cdot\V x\simeq\M F^{-1}\cdot\mathrm{diag}(\M F\cdot\V h)\cdot(\M F\cdot\V x)\,,\label{eq:fast-convolution}\end{equation}
where $\V h$ is the PSF (the first row of $\M H$), $\M F$ is the
forward DFT operator, and $\mathrm{diag}(\V v)$ is the diagonal matrix
with its diagonal given by the vector $\V v$. This discrete convolution
equation assumes that $F_{u,j}=\exp(-2\,\mathrm{i}\,\pi\,\sum_{n}u_{n}\, j_{n}/N_{n})$
where $N_{n}$ is the length of the $n^{\mathrm{th}}$ dimension,
$j_{n}=0,\ldots,N_{n}-1$ and $u_{n}=0,\ldots,N_{n}-1$ are the indices
of the position and discrete Fourier frequency along this dimension.
In this case, the inverse DFT is simply $\M F^{-1}=\M F^{\mathrm{H}}/N_{\mathrm{tot}}$
with $N_{\mathrm{tot}}$ the total number of elements in $\V x$ and
the $\mathrm{H}$ exponent standing for the conjugate transpose. With
these approximations and definitions of the DFT, the likelihood term
writes:\begin{equation}
\Qdata(\V x)=\frac{1}{\sigma^{2}}\,\norm{\M H\cdot\V x-\V y}^{2}\simeq\frac{1}{N_{\mathrm{tot}}\,\sigma^{2}}\,\sum_{u}\abs{\hat{h}_{u}\,\hat{x}_{u}-\hat{y}_{u}}^{2}\,,\label{eq:diagonal-likelihood-approximation}\end{equation}
where $\hat{h}_{u}$ is the transfer function (the DFT of the point
spread function) and $\hat{y}_{u}$ and $\hat{x}_{u}$ respectively
the DFT of the data and of the sought image. Note that the exact normalization
factor, here $1/N_{\mathrm{tot}}$, depends on the particular definition
of the DFT. 

Similarily, ignoring edges effects, the finite difference operator
$\M D$ along $n^{\mathrm{th}}$ direction can be approximated by:\begin{equation}
\M D_{n}\cdot\V x\simeq\M F^{-1}\cdot\mathrm{diag}(\hat{\V d}_{n})\cdot(\M F\cdot\V x)\,,\label{eq:fast-finite-differences}\end{equation}
where $\hat{\V d}_{n}$ is the DFT of the first row of $\M D_{n}$;
then the quadratic regularization writes:\begin{equation}
\Qprior(\V x)=\norm{\M D\cdot\V x}^{2}=\sum_{n}\norm{\M D_{n}\cdot\V x}^{2}\simeq\frac{1}{N_{\mathrm{tot}}}\,\sum_{u}r_{u}\,\left|\hat{x}_{u}\right|^{2},\label{eq:diagonal-regularization-approximation}\end{equation}
with:\begin{equation}
r_{u}=\sum_{n}\abs{\hat{d}_{n,u}}^{2}=4\,\sum_{n}\sin^{2}\left(\frac{\pi\, u_{n}}{N_{n}}\right)\,,\label{eq:diagonal-regularization-weights}\end{equation}
for first order finite differences and our choice for the DFT. Note
that any $r_{u}\ge0$ being an increasing function of the length
$\abs u$ of the spatial frequency could be used instead and would
result in imposing a smoothness constraint although with a different
behaviour. Finally putting all these circulant approximations together,
the quadratic problem to solve is diagonalized in the DFT space and
trivially solved to gives the DFT of the MAP solution:\begin{equation}
\hat{x}_{\mu,u}=\frac{\hat{h}_{u}^{\star}\,\hat{y}_{u}}{\vert\hat{h}_{u}\vert^{2}+\mu\,\sigma^{2}\, r_{u}}\,,\label{eq:solwiener}\end{equation}
the asterisk exponent denoting the complex conjugate. Note that this
circulant approximation of the solution is very fast to compute as
it involves just a few FFT's. This expression of the MAP solution
is very similar to what would give Wiener filter which would be exactly
achieved by setting the term $\mu\, r_{u}$ equals to the reciprocal
of the expected image powerspectrum in equation (\ref{eq:solwiener}).
Since, in our case, the image powerspectrum is unknown a priori, we
have to choose the \emph{regularization shape} $r_{u}$ and derive
the optimal level of smoothing, for instance, by means of GCV. Thanks
to the circulant approximation made here, GCV criterion is now very
easy to compute as:\[
\M A_{\mu}\simeq\M F^{-1}\cdot\mathrm{diag}(\hat{\V a}_{\mu})\cdot\M F\,,\quad\text{with: }\hat{a}_{\mu,u}=\frac{\vert\hat{h}_{u}\vert^{2}}{\vert\hat{h}_{u}\vert^{2}+\mu\,\sigma^{2}\, r_{u}}\,,\]
and $\mathrm{tr}(\M A_{\mu})=\sum_{u}\hat{a}_{\mu,u}/N_{\mathrm{tot}}$,
hence:\begin{equation}
\mathrm{GCV}(\mu)=\frac{N_{\mathrm{tot}}\,\sum_{u}t_{\mu,u}^{2}\,|\hat{y}_{u}|^{2}}{\sigma^{2}\,[\sum_{u}t_{\mu,u}]^{2}}\,,\label{eq:GCV-diagonalized}\end{equation}
with\begin{equation}
t_{\mu,u}=1-\hat{a}_{\mu,u}=\frac{\mu\,\sigma^{2}\, r_{u}}{\abs{\hat{h}_{u}}^{2}+\mu\,\sigma^{2}\, r_{u}}\,.\end{equation}
In practice, for the optimization of equation~(\ref{eq:optim-problem}),
equation~(\ref{eq:solwiener}) is taken as a starting point together
with the choice of $\mu$ given by the minimum of equation (\ref{eq:GCV-diagonalized}).
Then the optimization of equation~(\ref{eq:optim-problem}) is carried
with possibly non stationary weights, while iterating back and forth
between model and data space.
\begin{figure*}
\includegraphics[width=2.\columnwidth]{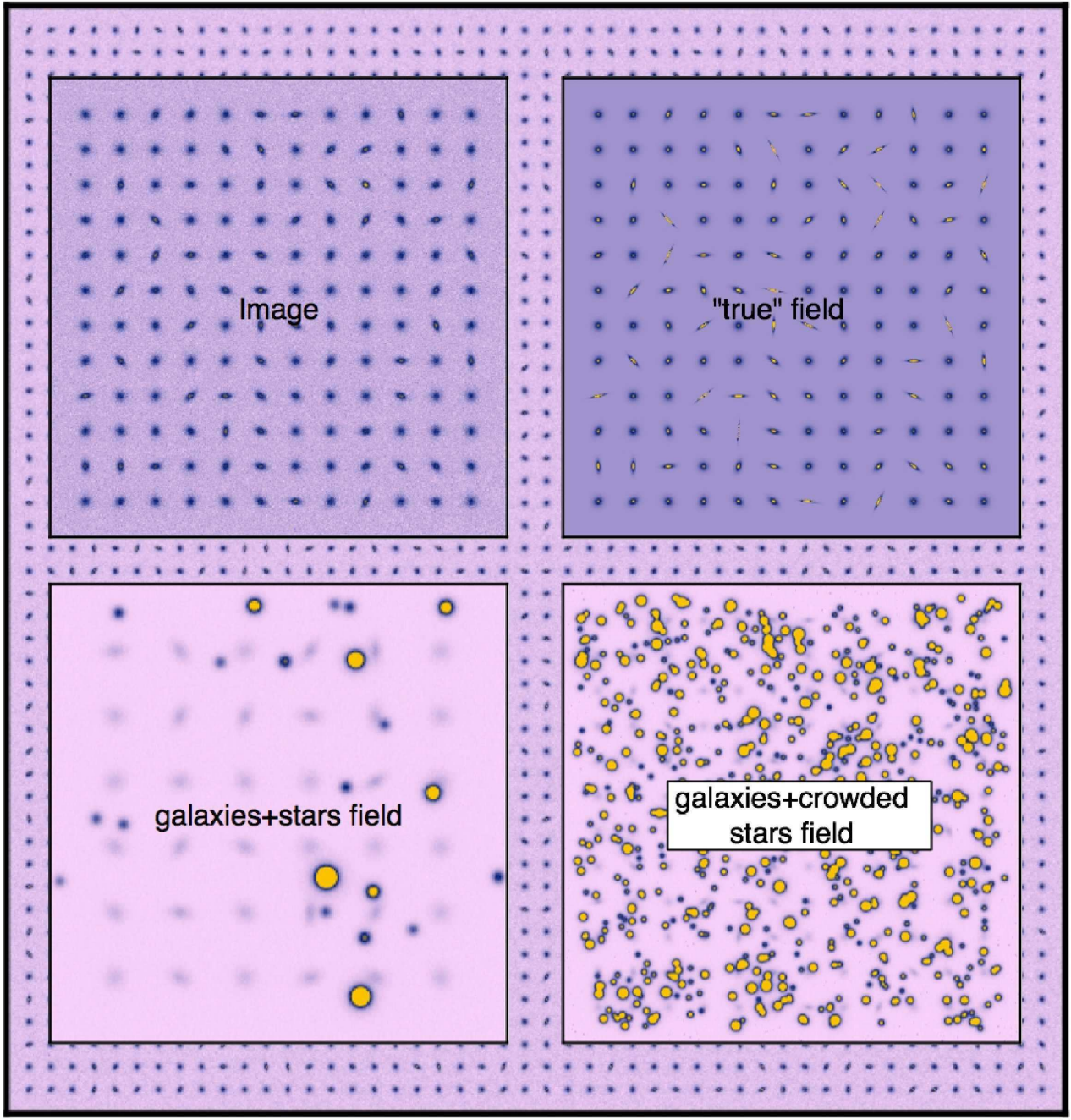}
\caption{An example of virtual fields generated with \noun{skymaker} to be
fed to \noun{sextractor} before and after deconvolution using the
different regularizations described in the text.\emph{ From top to
bottom and left to right,} a galaxy field image and the corresponding
{}``true'' field, a galaxies field with stars and a crowded galaxy
field with stars ($10^{6}$ stars/arcmin$^{2}$). The exposure time
is $10$ seconds and the seeing is $1"$ for the VLT with VIMOS. The
background field corresponds to the actual size of the corresponding observed 
images. \label{fig:field}}
\end{figure*}
\subsubsection{Crowded fields and star removal}\label{sub:Crowded-fields-and}
Even though the estimation of the ellipticities does not require  {\it per
se} the deconvolution of the galaxies, it is shown below that this
estimation is significantly improved by deconvolution when the fields
of view are crowded and polluted by foreground stars: indeed galaxies
and stars overlap less when deconvolved, which reduces the fraction
of erroneous measurements. Unfortunately, when these stars are present,
they significantly bias the estimation of the hyper parameter, $\mu$,
since stars correspond to high frequency correlated signal which leads
to an underestimation of the optimal level of smoothing (for the galaxies)
by cross validation. This is best seen in Figure \ref{fig:GCV} which
displays the evolution of the hyper-parameter which minimizes GCV
as a function of the number of stars removed by our star removal algorithm,
see Appendix \ref{sec:Efficient-Star-Removal}. Interestingly, it
suggests that GCV could be used as a classifier.
\subsubsection{$\ell_{1}-\ell_{2}$ penalty and positivity}\label{sec:L2-L1-penalty} 
The drawback of using a quadratic ($\ell_{2}$) norm in the regularization
is that it tends to over-smooth the regularized map especially around
sharp features as point-like sources (i.e. stars) and the core of
galaxies. This is because the regularization prevents large intensity
differences between neighboring pixels and result in damped oscillations
(Gibbs effect). Such ripples hide any faint details in the vicinity
of sharp structures. To avoid this, it would be better to use a regularization
which smoothes out small local fluctuations of the sought distribution
(here the deblurred image), presumably due to noise, but let larger
local fluctuations arise occasionally (see \cite{image} and reference therein). This can be achieved by using
a $\ell_{1}-\ell_{2}$ cost function $\phi$ in equation~(\ref{eq:general-smoothness-penalty}).
A possible $\ell_{1}-\ell_{2}$ sparse cost function is  \citep{Mugnier_Fusco_etal-2004-JOSAA-Mistral}:
\begin{equation}
\phi(r)\equiv2\,\varepsilon^{2}\,\left[\Abs{\frac{r}{\varepsilon}}-\log\left(1+\Abs{\frac{r}{\varepsilon}}\right)\right]\,.\label{eq:l2-l1-norm}\end{equation}
For a small, respectively large, pixel differences $r$, $\phi(r)$
has the following behavior\[
\phi(r)\sim\left\{ \begin{array}{ll}
r^{2} & \text{when }\abs r\ll\varepsilon\,,\\
2\,\abs{\varepsilon\, r} & \text{when }\abs r\gg\varepsilon\,,\end{array}\right.\]
which shows that, as required, the $\ell_{1}-\ell_{2}$ penalty behave
quadratically for small \emph{residuals} $r$'s (in magnitude and
w.r.t. $\varepsilon$) and only linearly for large $r$'s.
The derivative, needed for the optimization algorithm, of the $\ell_{1}-\ell_{2}$
penalty writes:\[
\phi'(r)=\frac{2\,\varepsilon\, r}{\varepsilon+\Abs r}\,.\]
An additional possibility to improve the restitution of faint details
with level close to that of the background is to apply a strict positivity
constraint. This is achieved by using \noun{vmlmb}, a modified limited
memory variable metric method \textcolor{black}{\noun{ \citep{Thiebaut:spie2002:bdec}}},
which imposes simple bound constraints by means of gradient projection.
This yields a reduction of aliasing by bounding the allowed region
of parameter space which can be explored during the optimization.
\subsection{Numerical experiments}
The public package \noun{SkyMaker \citep{skymaker}} was used to generate
galactic and stellar fields from ellipticity and magnitude catalogues.
Table \ref{tab:skymaker-parameters-used} summarizes the main parameter
corresponding to the VLT with a VIMOS instrument, a worse case situation
compared to upcoming space missions.

A regular grid of $12\times12$ galaxies of magnitude 20 with random
orientation is produced twice (with the same random seed), one corresponding
to a fixed seeing and a given exposure time, while the other assumes
zero noise and zero seeing for a set of $512\times512$ pixels images,
see Figure \ref{fig:field}.

\begin{figure}
\includegraphics[width=1\columnwidth]{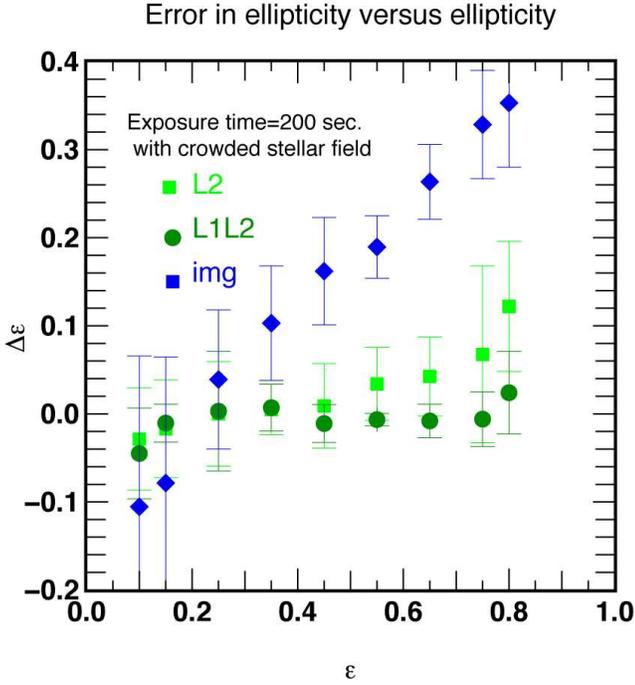}
\caption{the error in ellipticity as a function of the ellipticity (measured
by \noun{sextractor}) for a set of 50 images (such as those shown
on Figure \ref{fig:field}) either directly on the image (\emph{medium
diamond}s), deconvolved with $\ell_{2}$ gradient penalty function
with enforced positivity (\emph{light square}s) and $\ell_{1}-\ell_{2}$
gradient penalty function with positivity (\emph{dark circle}s). For
each set, the ellipticity is also measured directly on the raw image.
Note that, as expected, the error on the bias is largest for circular
galaxies, since deconvolution will tend to over amplify departure
from circular symmetry. \label{fig:recov-ellip}}
\end{figure}

The background level and the amplitude of the background noise is
first estimated automatically from the histogram of the pixel values
and fed to \noun{Sextractor \citep{sex}} which then estimates the
position, the flux, the orientation and the ellipticity for all the
galaxies in the field. Here the ellipticity is defined as $1-b/a$, where $a$ and $b$ are the long and short axis.
This procedure is reproduced 50 times with
different realizations. The measured and the recovered ellipticity
are compared, together with flux and orientation for all the galaxies
in the field. In this set of simulations the prior knowledge of the
position of the galaxy is used to minimize errors which might arise
while using \noun{sextractor}: the recovered galaxy is chosen to be
that which is closest to the known input position. The median and interquartile
of the error (difference between the {}``true'' and recovered) in
ellipticity versus the ellipticity is computed for a range of exposure
time; this procedure is iterated for the three deconvolution techniques
used in this paper (Wiener, $\ell_{2}$ with positivity, $\ell_{1}-\ell_{2}$
with positivity). An example of such a plot is shown in Figure \ref{fig:recov-ellip}.
Clearly the bias in the recovered ellipticity increases with the ellipticity
and the amount of noise in the image (via poorer seeing or shorter
exposure time). As expected, the Wiener deconvolution is the least
efficient of the three methods, since the linear penalty does not
avoid some level of Gibbs ringing. In contrast the $\ell_{2}$ penalty
with positivity avoids partially such ringing, while the $\ell_{1}-\ell_{2}$
penalty works best at recovering the input eccentricity with a consistent
level of bias below 10\,\% for an ellipticity in the range $[0.1,0.8[$.
Note that this bias is relative, not absolute. If an alternative shear estimator that
doesnÕt consider deconvolution
 is accurate 
to a level of, say 1\%, the expected bias after deconvolution will be below 0.1 \%.

Interestingly, there is also a residual bias (even for longer exposure
times) for small ellipticity galaxies, which arises because noise
induced departure from sphericity is amplified by the deconvolution. 
\begin{figure*}
\includegraphics[width=1\columnwidth]{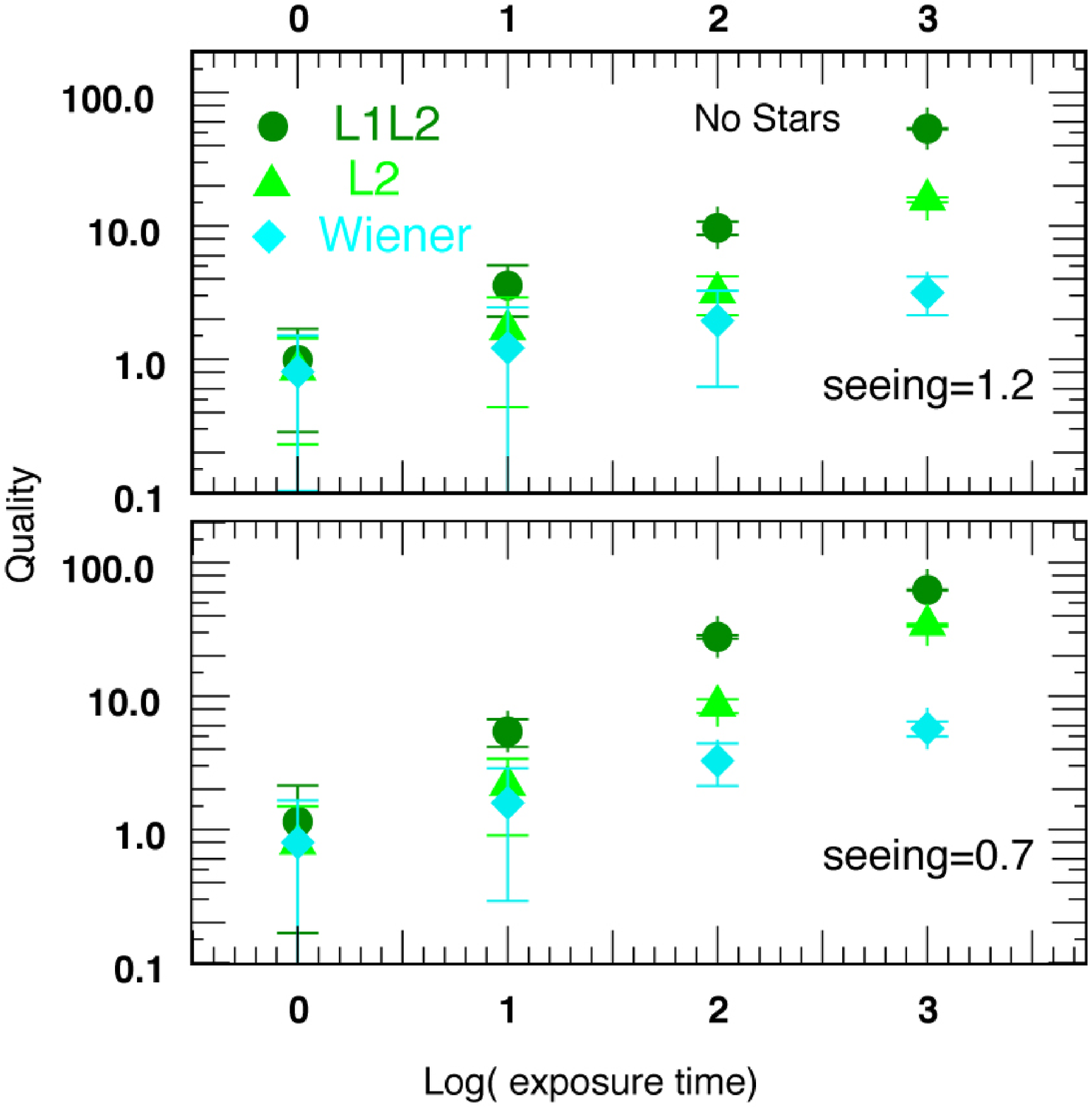}\includegraphics[width=1\columnwidth]{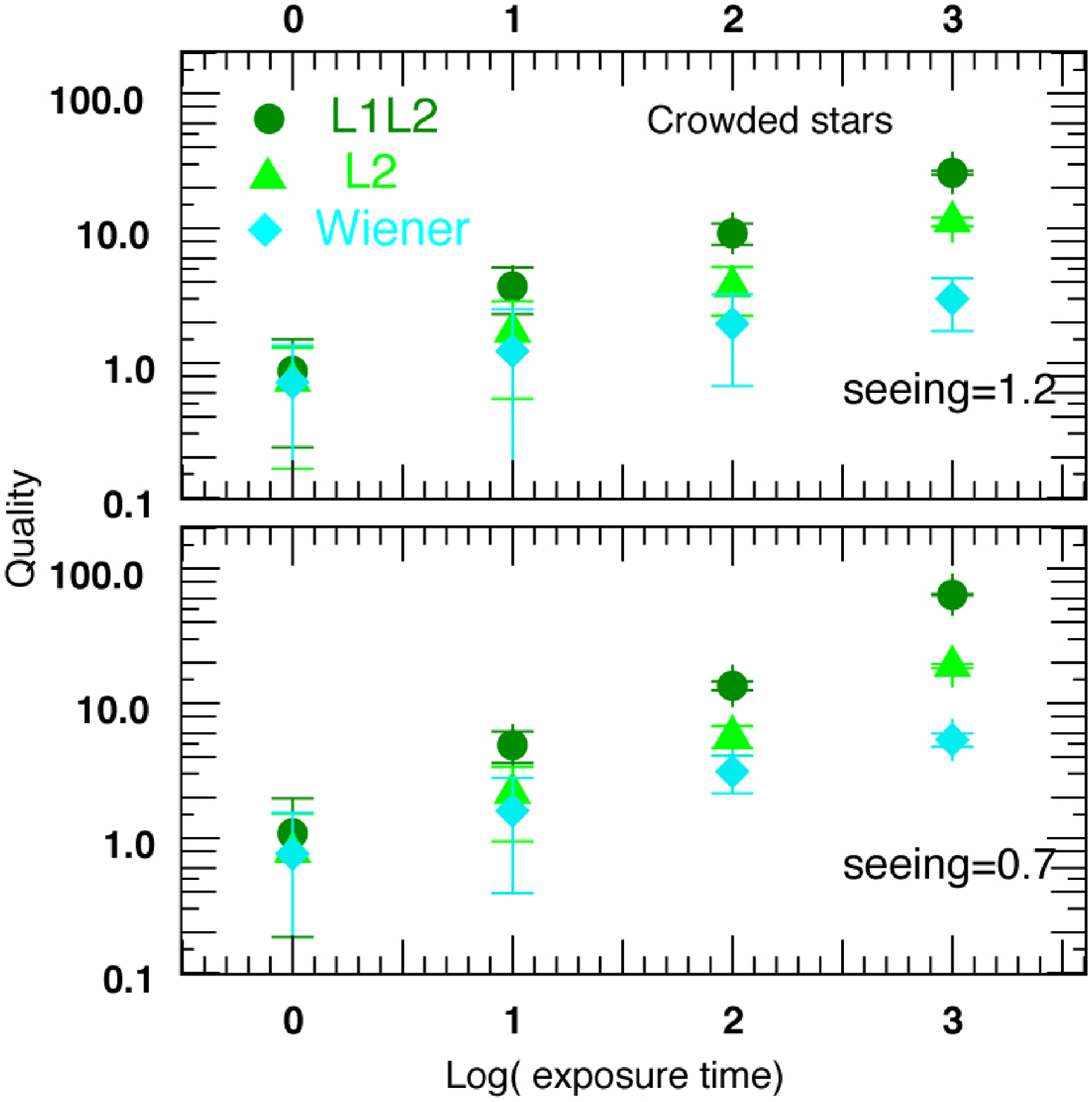}
\caption{\emph{Left panel}: the relative error quality factor (see main text)
as a function of the log exposure time for the three methods, respectively
Wiener filtering (\emph{diamonds}), $\ell_{2}$ gradient penalty function
with enforced positivity (\emph{triangles}) and $\ell_{1}-\ell_{2}$
gradient penalty function with positivity (\emph{circles}). Two seeing
conditions are investigated, corresponding to a good ($0.7"$) and
a fair ($1.2"$) seeing condition. These simulations assume that no
star are present in the field, and correspond to a set of non overlapping
galactic disks with random orientation and magnitude $20$ in V (see
Fig.~\ref{fig:field}). The telescope setting correspond to the VIMOS
instrument on an 8 meter VLT.\emph{ Right panel}: the quality factor
as a function of the log exposure time, but this time while allowing
for stars in the field. The star count is $10^{5}$ stars per arcmin$^{2}$.
As discussed in the text, the penalty weight is estimated via generalized
cross validation (GCV) on a temporary image where all stars are automatically
removed via blind cleaning as described\textcolor{black}{ in Appendix
A. }Here the removal of stars is essential since the GCV hyper-parameter
(which sets the level of smoothing in the deconvolved image) varies
by orders of magnitudes in the process (see Fig \ref{fig:GCV} for
a discussion) and would be otherwise underestimated.\label{fig:Qfig}}
\end{figure*}
Note that the Wiener deconvolution is significantly faster than the
iterative deconvolution with positivity (with $\ell_{2}$ or $\ell_{1}-\ell_{2}$
penalties). Positivity improves significantly the deconvolution, but
will depend critically on the ability to estimate the background.
In the present simulations, the level of background is automatically
estimated while looking at the histogram of the pixels. Finally the
$\ell_{1}-\ell_{2}$ regularization significantly improves the restoration
of fields of stars and galaxies, because the stars and the cores of
galaxies are very sharp. These non-linear iterative methods are slower
than the Wiener filtering, but can account at no extra cost for non
uniform noise, or saturation and masking. Their convergence can be
considerably boosted when they are initiated by the Wiener solution.

For any such plot, two numbers are defined which summarize the trend.
The mean error (averaged over the various ellipticities) $\bar \epsilon$, and the mean
of the interquartile, $\Delta \bar \epsilon$ were measured. The quality factor, $QF$ is defined
to be the ratio of the sum of this mean error and the mean interquartile
for the image without deblurring, divided by the sum of the mean error
and the mean interquartile for the deconvolved image for the three
techniques (Wiener, $\ell_{2}$ and $\ell_{1}-\ell_{2}$). This reads
\[
QF_{{\rm method}}=\frac{\bar{\epsilon}_{\mathrm{image}}+\Delta\bar{\epsilon}_{\mathrm{image}}}{\bar{\epsilon}_{\mathrm{method}}+\Delta\bar{\epsilon}_{\mathrm{method}}}\,.\]
The evolution the quality of the ellipticity measurement is traced
versus seeing conditions and signal to noise (exposure time) in two
regimes: a galaxy-only field, and a galactic field with a crowded
star content where the number of stars per square degree reaches $10^{5}$
stars/arcmin$^{2}$. These two regime represent high and low Galactic
region respectively. Figure \ref{fig:Qfig} displays the evolution
of $QF_{\mathrm{Wiener}}$ (\emph{diamond}s), $QF_{\ell_{2}}$ (\emph{triangle}s),
and $QF_{\ell_{1}-\ell_{2}}$ (\emph{circle}s), as a function the
exposure time of $1$, $10$, $100$ and $1000$ seconds respectively,
and two seeing conditions of $1.2"$ and $0.7"$. No stars are present
in the field on the left panel of Figure \ref{fig:Qfig}, whereas
its right panel displays the three $QF$ estimators for a field with
a realistic $10^{5}$ stars per square degree. 
\noun{Aski} achieves efficient debluring in this regime.
 It remains to be shown that regularized deconvolution obtained through (sparse) parametric local decomposition of both PSF and objects (as done e.g. with shapelet-based methods) can properly deblur blended objects.

\begin{table}
\begin{tabular}{|l|l|}
\hline 
\noun{Object} & \noun{value}\tabularnewline
\hline
\hline 
Gain (e-/ADU) & 30.11\tabularnewline
\hline 
Full well capacity in e- & 300000\tabularnewline
\hline 
Saturation level (ADU) & 60000\tabularnewline
\hline 
Read-out noise (e-) & 1.3\tabularnewline
\hline 
Magnitude zero-point (ADU per second) & 21.254\tabularnewline
\hline 
Pixel size in arcsec. & 0.2\tabularnewline
\hline 
Number of microscanning steps  & 1\tabularnewline
\hline 
SB (mag/arcsec$^{2}$) at 1' from a 0-mag star & 16.0\tabularnewline
\hline 
Diameter of the primary mirror (in meters) & 8.0\tabularnewline
\hline 
Obstruction diam. from 2nd mirror in m. & 2.385\tabularnewline
\hline 
Number of spider arms (0 = none) & 4\tabularnewline
\hline 
Thickness of the spider arms (in mm) & 5.0\tabularnewline
\hline 
Pos. angle of the spider pattern & 45.0\tabularnewline
\hline 
Average wavelength analyzed (microns) & 0.80\tabularnewline
\hline 
Back. surface brightness (mag/arcsec$^{2}$) & 21.5\tabularnewline
\hline 
Nb of stars /$^{\square}$ brighter than MAG\_LIMITS & 1e5\tabularnewline
\hline 
Slope of differential star counts (dexp/mag) & 0.3\tabularnewline
\hline 
Stellar magnitude range allowed & 12.0,19.0\tabularnewline
\hline
\end{tabular}
\caption{\noun{SkyMaker} parameters used to generate the VIMOS/ VLT images
\label{tab:skymaker-parameters-used}}
\end{table}

Now that we have shown that state of the art automated positive edge-preserving deconvolution
of deep sky images is mandatory to get good quality shear estimates {(most importantly in the context of crowded fields)},
let us conclude this section by a leap forward, and assume from now
on that we have access not only to discrete measurements of ellipticities
over a significant fraction of the sky, but also that this point like
process has been re-sampled. Indeed, since it is beyond the scope
of this paper to carry out a full-sky deconvolution and reconstruction
at the resolution of $0.7"$ (This would amount to about $10^{12}$
pixels!), it is assumed from now on that a full-sky catalogue of vector
reduced shear exists and that the interpolation/re-sampling of the
corresponding map on a uniform grid over the sphere has been done,
together with an estimate of the corresponding shot noise. {In other words, we 
skip the critical step of optimal shear estimation, which has already been addressed by the STEP \citep{2006MNRAS.368.1323H,2007MNRAS.376} working group.} 
In this
paper, we extract the virtual catalogue from a state of the art simulation
(see below) we make use of the \noun{Healpix} Pixelisation \citep{1999elss.conf...37G},
a hierarchical equi-surface and iso-latitude pixelisation of the sphere,
which was developped to analyze polarized CMB type data. 
\section{A full-sky Map Maker}\label{sec:A-Model-for}
\begin{figure}
\includegraphics[width=1\columnwidth]{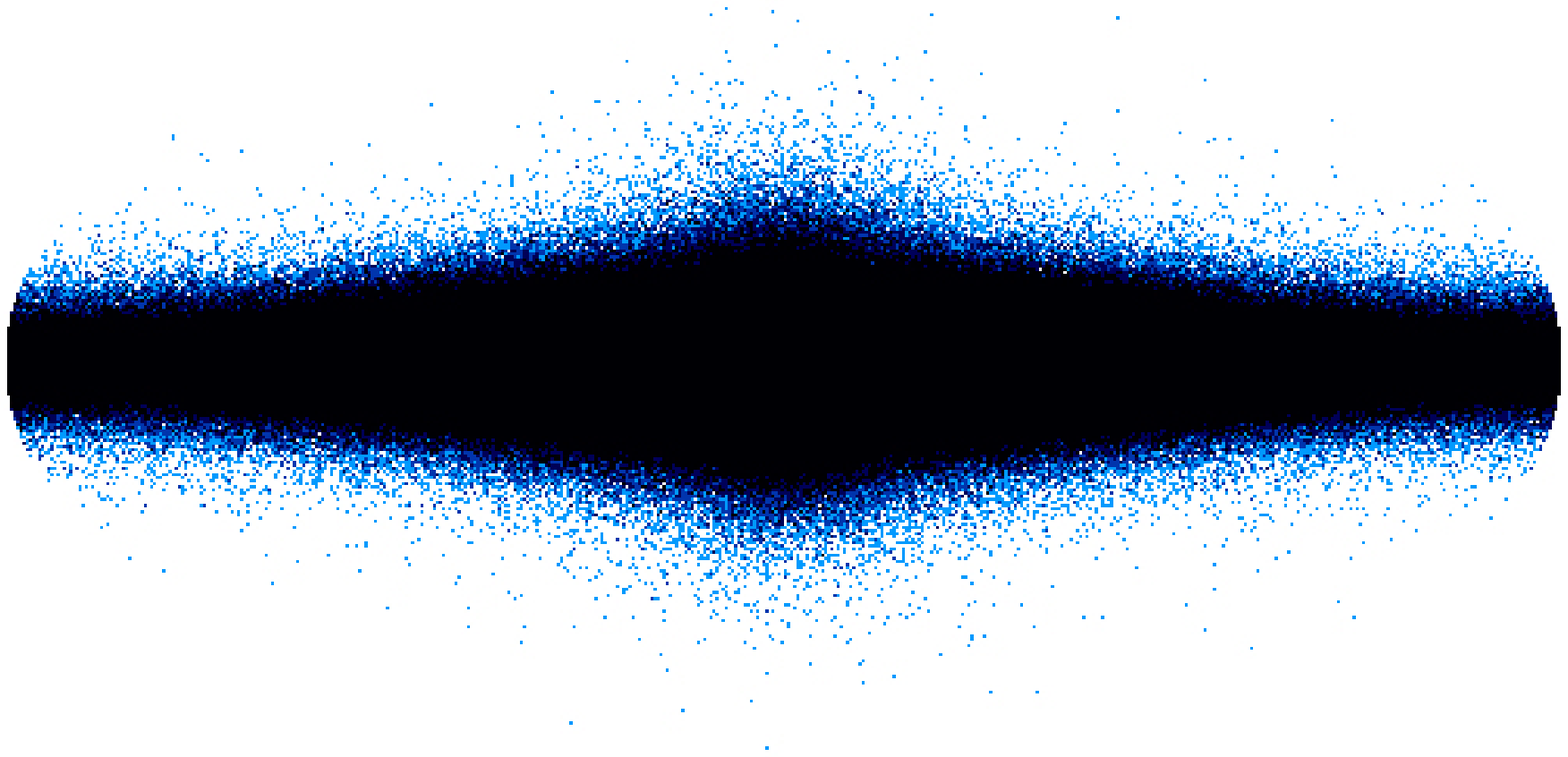}
\centering{\includegraphics[width=0.8\columnwidth]{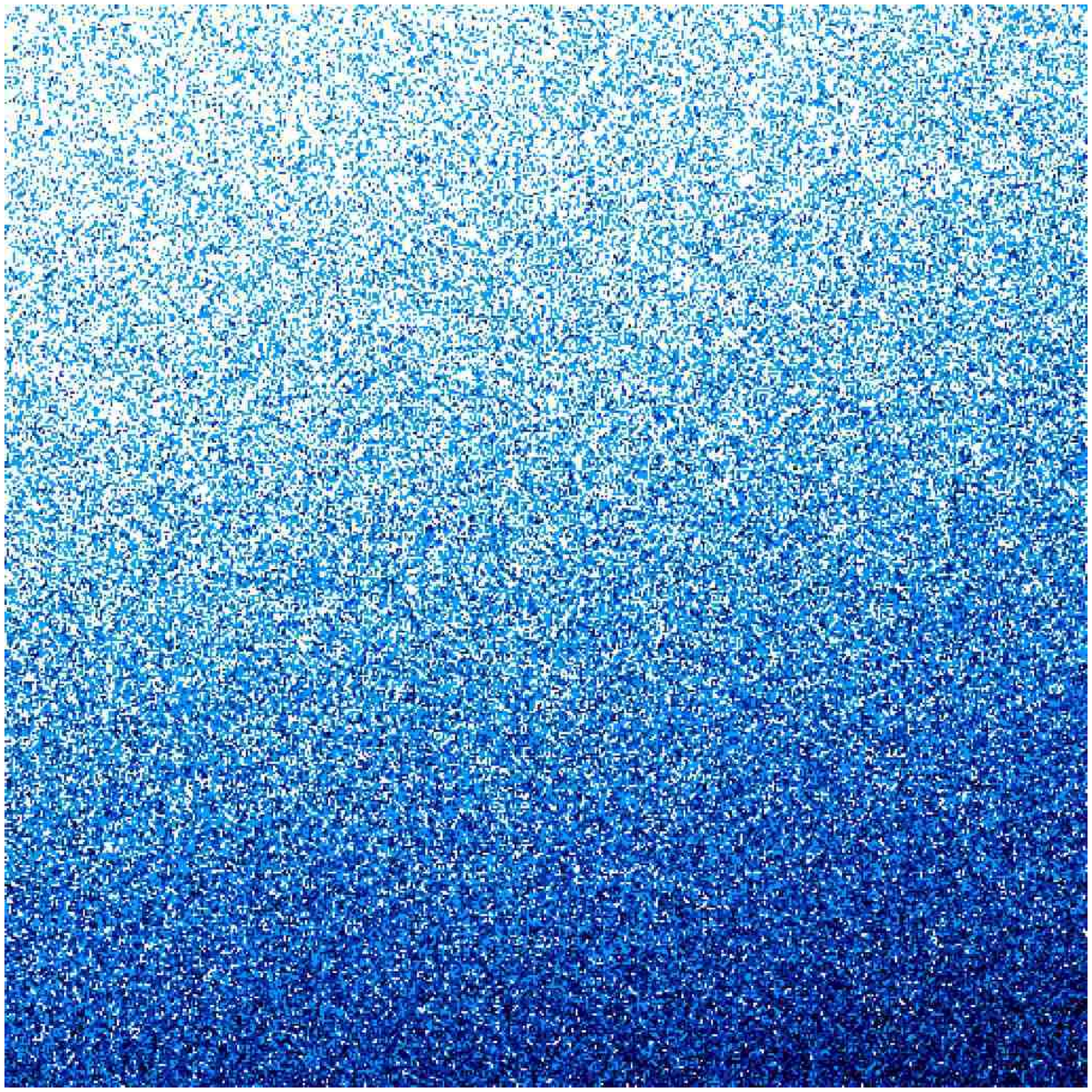}}
\caption{\emph{Top panel}: full-sky view of the mask;\emph{ Bottom panel:}
a zoom at coordinate $(l,b)=(30^{\circ},30^{\circ})$ showing the
distribution of stellar cuts. This cut corresponds to the inner central
region of the reconstruction shown in Figure \ref{fig:gaps}. \label{fig:mask}}
\end{figure}
\subsection{The inverse problem}\label{sec:Inverse-problem}
Our purpose is now to solve for the non-linear inverse problem of
recovering the $\kappa(\hat{\V n})$ map corresponding to a noisy
incomplete measurement of the 2-D field $\left(g_{1}(\hat{\V n}),g_{2}(\hat{\V n})\right)\TR$
of the ellipticity and orientation on the sphere (in the local tangent
plane):
\begin{equation}
g_{k}(\hat{\V n})=\frac{\gamma_{k}(\hat{\V n})}{1-\kappa(\hat{\V n})}+e_{k}(\hat{\V n})\,,\quad\mbox{for }k=1\mbox{ or }2\,,\label{eq:model}\end{equation}
where $\hat{\V n}$ is the sky direction, $\gamma$ and $\kappa$
are respectively the shear and the convergence, while $\mathbf{e}$
is a tensor field of the \emph{errors} which accounts for the measurement
noise (including the shot noise induced by the finite number of galaxies
within that pixel) and model approximations. 
\subsubsection{Spherical formulation}
On the sphere, the scalar field $\V{\kappa}$ and the tensor field
$\V{\gamma}$ are linear functions of the \emph{unknown} complex field
$\V a=\M Y\cdot\V{\kappa}$ whose coefficient are the spherical harmonic
coefficients of $\V{\kappa}$. After discretization and using matrix
notation, $\V{\kappa}$ and $\V{\gamma}$ write\begin{equation}
\kappa\equiv\mathbf{K}\cdot\mathbf{\V a}\quad\mbox{and}\quad\gamma\equiv\mathbf{G}\cdot\mathbf{\V a}\,,\label{eq:defk}\end{equation}
where $\M K=\mathbf{Y}$ and $\M G=\Spin{\mathbf{Y}}\cdot\M J$, denoting
$\M Y$ the scalar spherical harmonics and $\Spin{\mathbf{Y}}=(_{E}\mathbf{Y},_{B}\mathbf{Y})$
the parity eigenstates based on spin 2 spherical harmonics. These
eigenstates are defined in such a way that \[
\gamma_{1}\pm i\gamma_{2}=-\sum_{\ell m}(a_{\ell,m,E}\pm ia_{\ell,m,B})_{\pm2\,}\mathbf{Y}_{\ell m}\,,\]
so that we have \[
\left(\begin{array}{c}
\!\gamma_{1}\!\\
\!\gamma_{2}\!\end{array}\right)=\sum_{\ell,m}\left(\begin{array}{c}
\!\!-\mathbf{W}_{\ell,m}^{+}\!\!\\
\!\!+i\,\mathbf{W}_{\ell,m}^{-}\!\!\end{array}\right)\, a_{\ell,m,E}+\sum_{\ell,m}\left(\begin{array}{c}\!\!
-i\,\mathbf{W}_{\ell,m}^{-}\!\!\\
-\mathbf{W}_{\ell m}^{+}\!\!\end{array}\right)\, a_{\ell,m,B}\]
with $\mathbf{W}_{\ell,m}^{\pm}=({}_{2}\mathbf{Y}_{\ell,m}\pm{}_{-2}\mathbf{Y}_{\ell,m})/2$.
Here $\mathbf{J}$ operates on $\mathbf{\V a}$ as \begin{eqnarray}
\left(\M J\cdot\V a\right)_{\ell,m,E} & = & \sqrt{\frac{(\ell+2)(\ell-1)}{(\ell+1)\ell}}\, a_{\ell,m}\,,\label{eq:defF1}\\
\left(\M J\cdot\V a\right)_{\ell,m,B} & = & 0\,.\label{eq:defF2}\end{eqnarray}
Appendix~\ref{sec:Detailled-model} gives more explicit formulations
of the operators $\M K$ and $\M G$, using index notation on the
sphere.
\subsubsection{Flat sky formulation}\label{sec:flat-sky}
The flat sky limits (corresponding to large $\ell$'s) of equations.
(\ref{eq:defk})-(\ref{eq:defF1}) are (see Appendix \ref{sec:From-the-sphere}):
\begin{equation}
\mathbf{J}\approx(\mathbf{1},\mathbf{0})\,,\quad{\rm and}\quad\mathbf{Y}\approx\exp(i\V{\ell}\cdot\hat{\V n})\,,\label{eq:Flocal}\end{equation}
while the parity eigenstates read locally, in the fixed copolar basis
$e_x, e_y$:
\begin{eqnarray}
\mathbf{W}^{+}\!\!\!\!\!\!\! & \approx & \!\!\!\!\!\! -\cos(2\,\phi_{\V{\ell}})\exp(i\,\V{\ell}\cdot\hat{\V n})=-\frac{l_{x}^{2}-l_{y}^{2}}{l_{x}^{2}+l_{y}^{2}}\exp(i\,\V{\ell}\cdot\hat{\V n})\,,\nonumber \\
\mathbf{W}^{-}  \!\!\!\!\!\!\! & \approx & \!\!\!\!\!\!\!  -i\,\sin(2\,\phi_{\V{\ell}})\exp(i\,\V{\ell}\cdot\hat{\V n})=-i\,\frac{2\, l_{x}\, l_{y}}{l_{x}^{2}+l_{y}^{2}}\exp(i\,\V{\ell}\cdot\hat{\V n})\,.
\label{eq:yeblocal}\end{eqnarray}
In this limit, the \emph{unknowns}, $\V a$, represent the Fourier
coefficients of the convergence field, $\kappa$. Note that our definition
of $\gamma$ and $\kappa$ warrants that they are consistent with
the lens equation on the tangent plane --- solving for $\kappa$ in
equation~(\ref{eq:defk}) and plugging the solution into equations
(\ref{eq:yeblocal}) --- which reads locally in real space:
\begin{equation}
\nabla^{2}\kappa(\hat{\V n})=\left(\partial_{x}^{2}-\partial_{y}^{2}\right)\gamma_{1}(\hat{\V n})+2\,\partial_{x}\partial_{y}\gamma_{2}(\hat{\V n})\,,\label{eq:local-shear}\end{equation}
where $\gamma_{1}(x,y)$ and $\gamma_{2}(x,y)$ are the two components
of the E and B modes of the shear field. Also note that thanks to
equation~(\ref{eq:defF2}) the recovered map will \emph{not} have
B modes by construction. It can nevertheless be checked that the amplitude
of the B modes in the residuals is small compared to the amplitude
of the signal in the E modes, see Section \ref{sub:Residual-B-modes}.
\begin{figure*}
\includegraphics[width=0.75\textwidth]{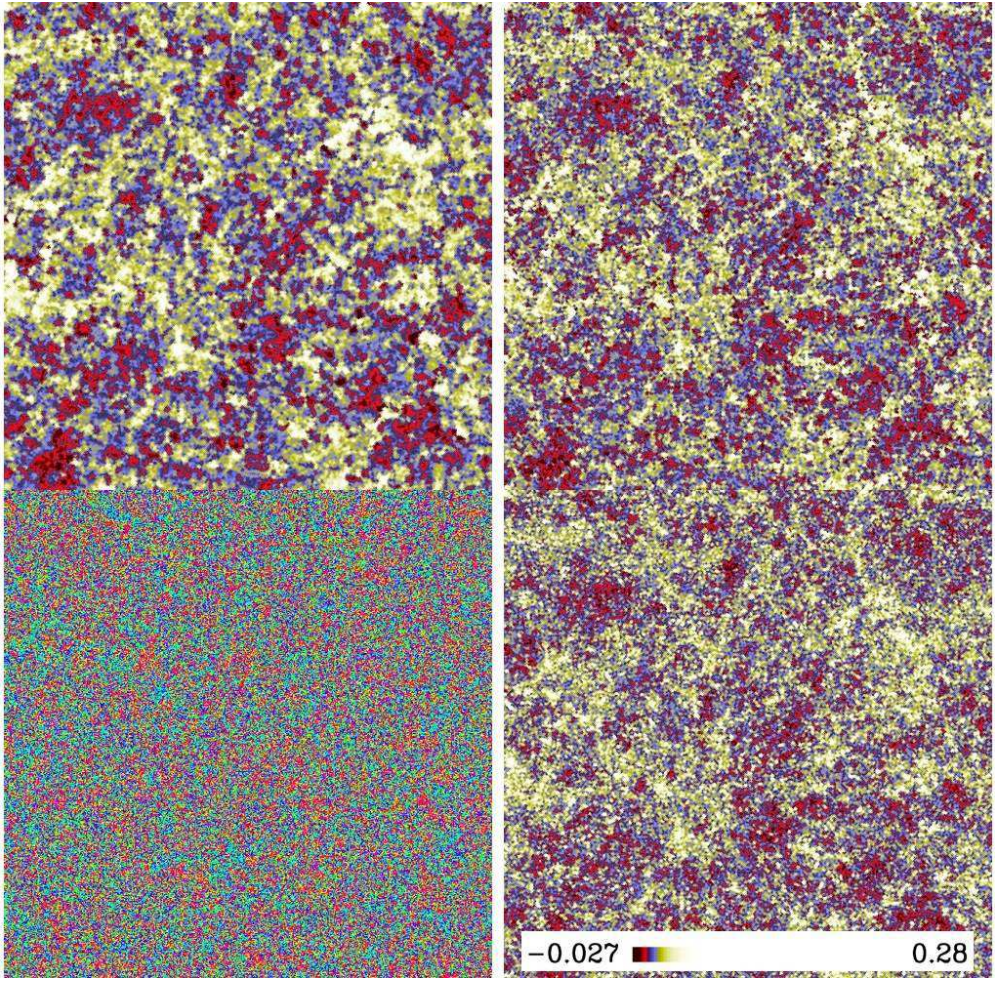}\caption{a zoom of the full-sky recovered $\kappa$ maps of a simulation $_{2048}S_{{\rm FS}}^{\ell_{{\rm cut}}}$
with $\ell_{{\rm cut}}=722,$\emph{ (top left panel) and }$1569$\emph{,
(top right panel)} (resp. 24, and 78 ngal/$\square$ arcmin)  at coordinates $(\phi,\theta)=(0,0)$ (the
color table corresponds to a histogram equalization);\emph{ bottom
left panel:} the corresponding data (the \emph{hue} color table codes
the shear orientation); \emph{bottom right panel}: the corresponding
underlying $\kappa$ map. \label{fig:L2-full-sky}}
\end{figure*}
\subsubsection{Cost function}
The considered problem can be stated as recovering $\V a$ given the
data $\V g$ according to the model in equation~(\ref{eq:model}).
In the same way as what has been done for deblurring the images (section
\ref{sec:deblurring}), finding the solution of this inverse problem
in the Maximum a Posteriori (MAP) (\citet{2005opas.conf..397T,1998MNRAS.301..419P})
sense involves minimizing a two-term cost function:\begin{equation}
\Qmap(\V a)=\Qdata(\V a)+\mu\,\Qprior(\V a)\,,\label{eq:objectif}\end{equation}
with respect to the parameters $\V a$. In the right hand side of
equation~(\ref{eq:objectif}), the term $\Qdata(\V a)$ enforces
agreement of the model with the data, whereas $\Qprior(\V a)$ is
a regularization term used to enforce our prior knowledge about the
sought fields, and $\mu\ge0$ is a Lagrange multiplier used to tune
the relative importance of the prior with respect to the data.

For errors with a centered Gaussian distribution, the likelihood term
writes:\[
\Qdata(\V a)=\sum_{j,k}W_{j_{1},k_{1},j_{2},k_{2}}\, e_{k_{1}}(\hat{\V n}_{j_{1}})\, e_{k_{2}}(\hat{\V n}_{j_{2}})\,,\]
with $e_{k}(\hat{\V n}_{j})=g_{k}(\hat{\V n}_{j})-\gamma_{k}(\hat{\V n}_{j})/[1-\kappa(\hat{\V n})]$
and $\M W=\M C^{-1}$ with $C_{j_{1},k_{1},j_{2},k_{2}}=\avg{e_{k_{1}}(\hat{\V n}_{j_{1}})\, e_{k_{2}}(\hat{\V n}_{j_{2}})}$.
If the errors are further uncorrelated, the likelihood simplifies
to: \begin{equation}
\Qdata(\V a)=\sum_{j,k}w_{j,k}\,\left[g_{k}(\hat{\V n}_{j})-\frac{\gamma_{k}(\hat{\V n}_{j})}{1-\kappa(\hat{\V n}_{j})}\right]^{2}\,,\label{eq:data-penalty}\end{equation}
where the sum is carried over the index $j$ of the sampled sky directions
$\hat{\V n}_{j}$ (so called sky \emph{pixels}) and index $k$ of
the two components of, say, the Q and U polarization fields respectively
(see Appendix~\ref{sec:Detailled-model} for an explicit formulation
with all the relevant indices) and the weights are related to the
variance of the noise:\begin{equation}
w_{j,k}=\mathrm{Var}\left(e_{k}(\hat{\V n}_{j})\right)^{-1}\,.\label{eq:weights}\end{equation}
This allow us to account for non uniform noise on the sky and also
cuts (the galaxy, bright stars, \emph{etc}.) for which the variance
can be considered as infinite and thus the corresponding weights set
to zero. Note that setting the weights in this statistically
consistent way yields no such biases as those which would result from
interpolation or inpainting methods used to replace missing data \citep{pires},
{see also \cite{starckforever}
for such implementation in the context of CMB experiments).
}
For this recovery problem, our prior is that the field $\V{\kappa}$
must be as smooth as possible in the limit that the model remains
compatible with observables within the error bars, that is equation~(\ref{eq:model})
must be valid. To that end, the regularization is written as a penalty
based on the second order spatial derivatives (Laplacian) $\nabla^{2}\kappa$
of the field $\kappa$: \begin{equation}
\Qprior(\V a)=\norm{\nabla^{2}\V{\kappa}}\,.\label{eq:regul-penalty}\end{equation}
Equation~(\ref{eq:laplacian_of_kappa}) in Appendix~\ref{sec:Detailled-model}
gives the expression of $\nabla^{2}\kappa$ as a function of the unknown
$\V a$. In order to enforces smoothness while preserving some sharp
features in the $\kappa$ map, quadratic and non quadratic norms of
the Laplacian have been considered for the regularization, see Appendix
\ref{sec:Detailled-model}. 
\begin{figure}
\includegraphics[width=0.9\columnwidth]{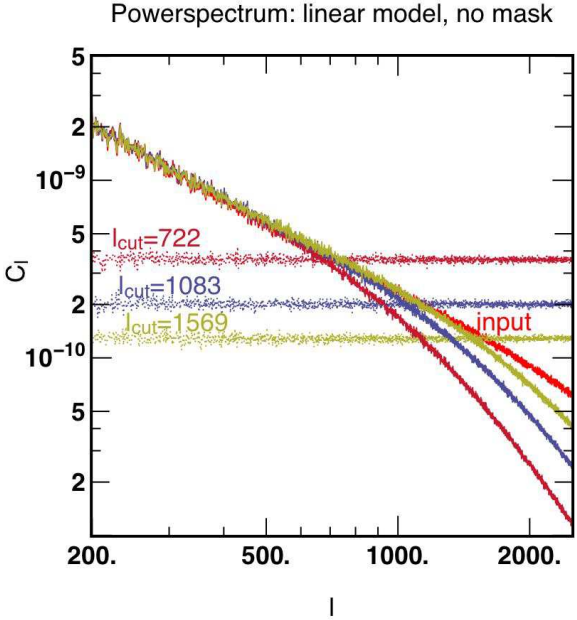}
\caption{a zoom on the power spectra of the three reconstructions of $_{2048}S_{{\rm FS}}^{\ell_{{\rm cut}}}$
for $\ell_{{\rm cut}}=722,1083,$ and $1569$  (resp. 24, 44 and 78 ngal/$\square$ arcmin), together with the power
spectra of the noise. Note that the level of smoothing decreases with
increasing signal to noise, in parallel to the bias in the corresponding
power spectrum.\label{fig:a-zoom-on}}
\end{figure}
\subsection{Generating the virtual data set}
Let us first describe in turn  the simulation  used to model the full
sky $\kappa$ map, and  the generation of the corresponding map.
\subsubsection{The simulation}
The \noun{Horizon} 4$\Pi$ (\citet{teyssier2008}, \citet{prunet2008})
simulation was used, a $\Lambda$CDM dark matter simulation using
the WMAP 3 cosmogony with a box size of $2h^{-1}$Gpc on a grid of
$4096^{3}$ cells. The 70 billion particles were evolved using the
Particle Mesh scheme of the \emph{\noun{RAMSES}} code (\citet{2002A&A...385..337T})
on an adaptively refined grid (AMR) with about 140 billions cells,
reaching a formal resolution of 262144 cells in each direction (roughly
7 kpc/h comoving). The simulation covers a sufficiently large volume
to compute a full-sky convergence map, while resolving Milky-Way size
halos with more than 100 particles, and exploring small scales deeply
into the non-linear regime. The dark matter distribution in the simulation
was integrated in a light cone out to redshift 1, around an observer
located at the center of the simulation box. 
\subsubsection{Mock data}
This light cone was then used to calculate the corresponding full
sky lensing convergence field, which is mapped using the \noun{Healpix}
pixelisation scheme with a pixel resolution of $\Delta\theta\simeq0.74\,\mathrm{arcmin^{2}}$
($n_{{\rm {\rm side}}}=4096$). Specifically, the convergence $\kappa(\hat{\V n})$
at the sky coordinate $\hat{\V n}$ is computed from the density contrast,
$\delta(\V x,z)$ in the Born approximation using:
\begin{equation}
\kappa(\hat{\V n})=\frac{3}{2}\Omega_{m}\!\!\int_{0}^{z_{s}}\!\!\frac{\mathrm{d}z}{E(z)}\frac{\mathcal{D}(z)\mathcal{D}(z,z_{s})}{\mathcal{D}(z_{s})}\frac{1}{a(z)}\delta(\frac{c}{H_{0}}\mathcal{D}(z)\hat{\V n},z)\,,\label{eq:kappa-delta}\end{equation}
which is valid for sources at a single redshift $z_{s}=1$, and $\mathcal{D}(z)=H_{0}\,\chi(z)/c$
is the adimensional comoving radial coordinate, hence $\mathrm{d}\mathcal{D}=\mathrm{d}z/E(z)$.
The detailed procedure to construct such maps from the simulation
using equation (\ref{eq:kappa-delta}) is described in Appendix~\ref{sec:lightcone} (chosing the 
sampling strategy) and \ref{sec:from-kappa-to}
and in \citet{teyssier2008}. In practice, a set of degraded maps
of $\kappa$ was generated from the full resolution, $n_{{\rm side}}=4096$
down to $n_{{\rm side}}=128$ in powers of 2, together with the corresponding
masks (see Figure \ref{fig:mask}). Different levels of noise (corresponding
to $700\le\ell_{{\rm cut}}<2500$) and maps with/without Galactic
masks are considered. The corresponding simulations are labeled as
$_{n_{{\rm nside}}}S_{{\rm FS/GC}}^{\ell_{{\rm cut}}}$ . Cartesian
maps are also used, labeled as $_{n_{{\rm pixel}}}C_{{\rm NL/lin}}^{{\rm SNR}}$
corresponding to Cartesian sections of the full-sky maps, where for
commodity, the experiments involving high resolution where calibrated.
Here the flag ${\rm NL/lin}$ refers to whether or not the non-linear
model is accounted for.
\subsubsection{Penalty weight}
In this paper, the weight of the penalty, $\mu$, in equation (\ref{eq:objectif})
is chosen so that the $\ell_{2}$ cutoff corresponds to the scale,
$\ell_{{\rm crit}}$ at the intersection of the signal and the noise
power spectra, see e.g. Figure \ref{fig:a-zoom-on}. Specifically 
\[
\mu\propto1/\ell_{{\rm crit}}^{2}\,.
\]
In a more realistic situation, when the power spectrum of the signal
is unknown, generalized cross validation could be used to find this
scale. When $\ell_{1}-\ell_{2}$ penalty is implemented (see Section
\ref{sec:L2-L1-penalty}), the $\ell_{1}$ parameter $\epsilon$ entering
equation (\ref{eq:l2-l1-norm}) is chosen so that it cuts off the
tail of the PDF of the Laplacian of the recovered field at the 3-$\sigma$
level.
\subsection{Optimization \& Performance}
Let us now turn to 
the optimization procedure and the
performance of the algorithm.
\subsubsection{Optimization}
Recall that the procedure assumes here a sampling strategy, since
the noisy $\mathbf{g}$ field is given on a pixelisation of the sphere.
To solve the optimization problem, we used the algorithm \noun{vmlm}
from \noun{OptimPack} \noun{\citep{Thiebaut:spie2002:bdec}} which
only involves computing the objective function $\Qmap(\V a)$ and
its partial derivative with respect to the parameters $\V a$. \noun{Vmlm}
is an unconstrained version of \noun{vmlmb} which has been used for
the deblurring problem and which is described in some details in section
\ref{sec:inverse-deblurring-problem}. The optimization of equation~(\ref{eq:objectif})
is carried by computing in turn equation (\ref{eq:defk}) and Equations
(\ref{eq:kappa-delta}) and (\ref{eq:grad-chi2}) using \noun{Healpix
(\citet{1999elss.conf...37G})} in \noun{OpenMP} or \noun{MPI}. 
\subsubsection{Overall Performance}
Each back and forth transform takes respectively $0.1$, $0.5$, $2$,
$8$, $32$ and $128$ seconds on an octo \noun{opteron} for $n_{{\rm side}}$
equal to $128$, $256$, $512$, $1024$, $2048$ and $4096$, see
Table~\ref{tab:the-performance-of}. The linearized problem without
mask converges typically in a dozen iterations (which typically only
involve a back and forth transform, unless the convergence is poor).
The linearized mask problem takes a few hundred iterations, see Table
\ref{tab:perfGC}, and so does the non-linear problem (or the linearized
problem with a non-linear $\ell_{1}-\ell_{2}$ penalty function). 
\begin{figure}
\includegraphics[width=1\columnwidth]{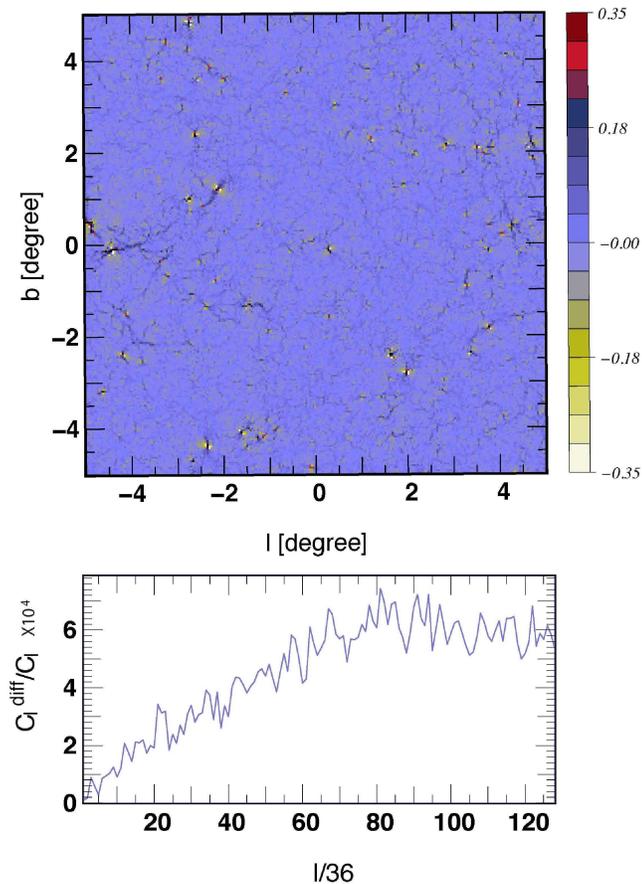}\caption{\emph{Top panel}: a map of the $100$ times difference between the
recovered map with the non-linear model and $\ell_{1}-\ell_{2}$ penalty,
and the recovered map without accounting for the non-linearity. As
expected the difference is largest at high frequencies near the cluster
and along the filaments.\emph{ Bottom panel:} the power spectrum of
the relative difference as a function of $\ell$. \label{Non linear}}
\end{figure}
%%%%%%%%%%%%%%%%%%%%%%%%%%%%%%%%%%%
\section{Validation and post analysis}\label{sec:Validation}
%%%%%%%%%%%%%%%%%%%%%%%%%%%%%%%%%%%
%%%%%%%%%%%%%%%%%%%%%%%%%%%%%%%%%%

Let us illustrate on a sequence of statistical tests several crucial features of the ASKI map making algorithm:
its ability to fill gaps, its ability to preserve the geometry and sharpness of clusters and maintain the gravitational nature of the signal in the presence  of masks, and the freedom to choose strong/weak prior on the two-points correlation. 
These properties are important in various contexts of the weak lensing studies, such as the  estimation of cosmological parameters, the physics of clusters,  the interpretation of tomographic data from upcoming surveys,  
constraining the dark energy equation of state through the redshift evolution of statistical and topological tracers.
We chose a selection of statistical tests that are sensitive to different aspects of map-making.

\subsection{One point statistics}
%%%%%%%%%%%%%%%%%%%%%%%%%
\subsubsection{Cluster counts}
\begin{figure}
\includegraphics[width=1.05\columnwidth]{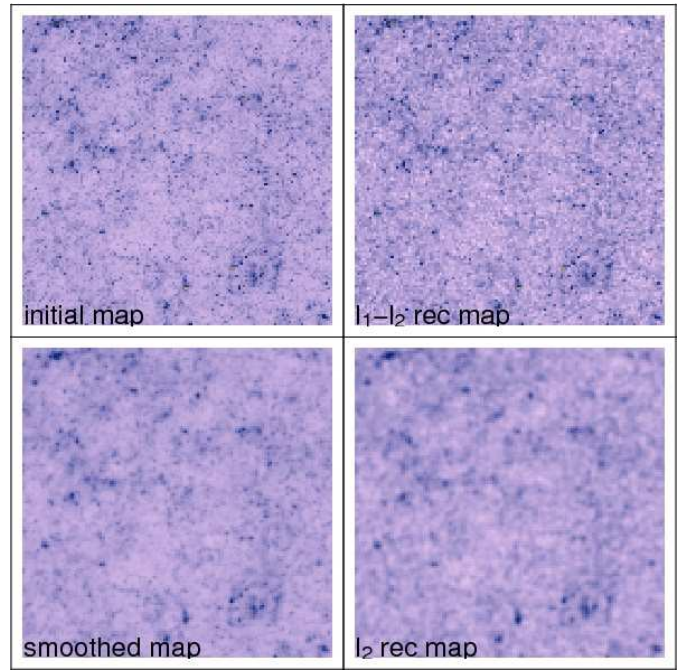}
\caption{\emph{top left panel:} a zoom of the original map at coordinates $(l,b)=(0^{\circ},0^{\circ})$;\emph{
top right panel:} reconstruction with $\ell_{1}-\ell_{2}$ penalty
using the non-linear model. \emph{bottom left panel:} input map smoothed
at a FHWM of 1.5 pixels.\emph{ bottom right panel:} reconstruction
with $\ell_{2}$ penalty using the non-linear model. The color table
is linear. The edge-preserving penalty appears qualitatively to preserve
much better the amplitude and the number of high peaks in the $\kappa$
map, as shown quantitatively in Figure~\ref{fig:L1L2-penalty2.}. \label{fig:L1L2-penalty.}}
\end{figure}
\begin{figure}
\includegraphics[width=1\columnwidth]{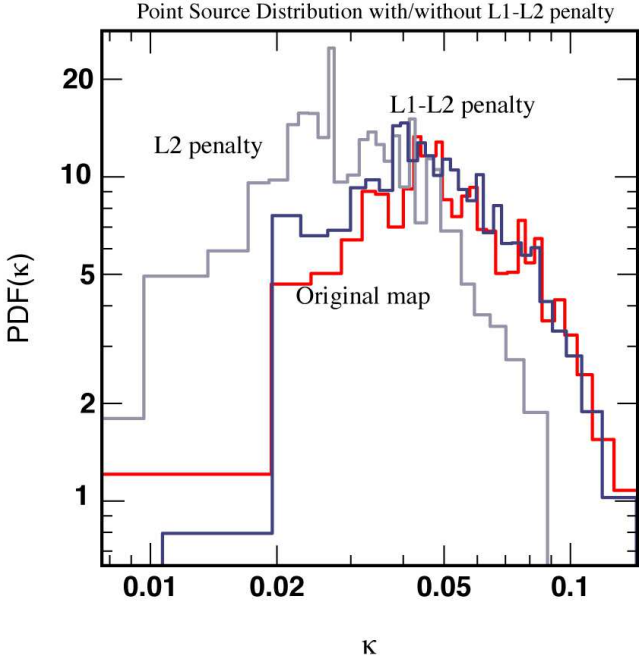}
\caption{The PDF of point source as defined in Section \ref{sub:Point-source-extraction}
corresponding to the maps of Figure \ref{fig:L1L2-penalty.} recovered
with $\ell_{1}-\ell_{2}$ penalty, and $\ell_{2}$ penalty respectively.
The improvement with an edge-preserving penalty is significant. \label{fig:L1L2-penalty2.}}
\end{figure}

One of the main assets of high resolution full-sky lensing maps is
to probe multiple scales: it then becomes possible to sample the non
linear transition scale and, e.g. study the shape of clusters. Figure~\ref{fig:L1L2-penalty.}
illustrates this feature while displaying the result of the inversion
with $\ell_{2}$ and $\ell_{1}-\ell_{2}$ penalties. For this experiment,
a Cartesian subset at galactic coordinates $(l,b)=(0^{\circ},0^{\circ})$
was extracted. The corresponding non-linear shear field $\mathbf{g}$
was generated via Fourier transform, and noised with a white additive
noise of SNR of 1. This set was then inverted while assuming $\ell_{2}$
(bottom right) and $\ell_{1}-\ell_{2}$ (top right) penalties. The
choice for the two penalty weights, $\mu$ and $\epsilon$ was made
on the basis of least square residual in the inverted $\kappa$ maps.
The improvement of $\ell_{1}-\ell_{2}$ over $\ell_{2}$ penalty is
significant. This statement is made more quantitative in Figure~\ref{fig:L1L2-penalty2.}
which displays the PDF of the peaks within that image for the initial
map (top left panel of Figure \ref{fig:L1L2-penalty.}) computed following
the peak patch prescription described in Section \ref{sub:Point-source-extraction}.
The agreement between the input and the recovered distribution is
significantly enhanced by the optimal $\ell_{1}-\ell_{2}$ (top right)
penalty.

\subsubsection{Skewness and Kurtosis}
\begin{figure*}
\includegraphics[width=0.9\columnwidth]{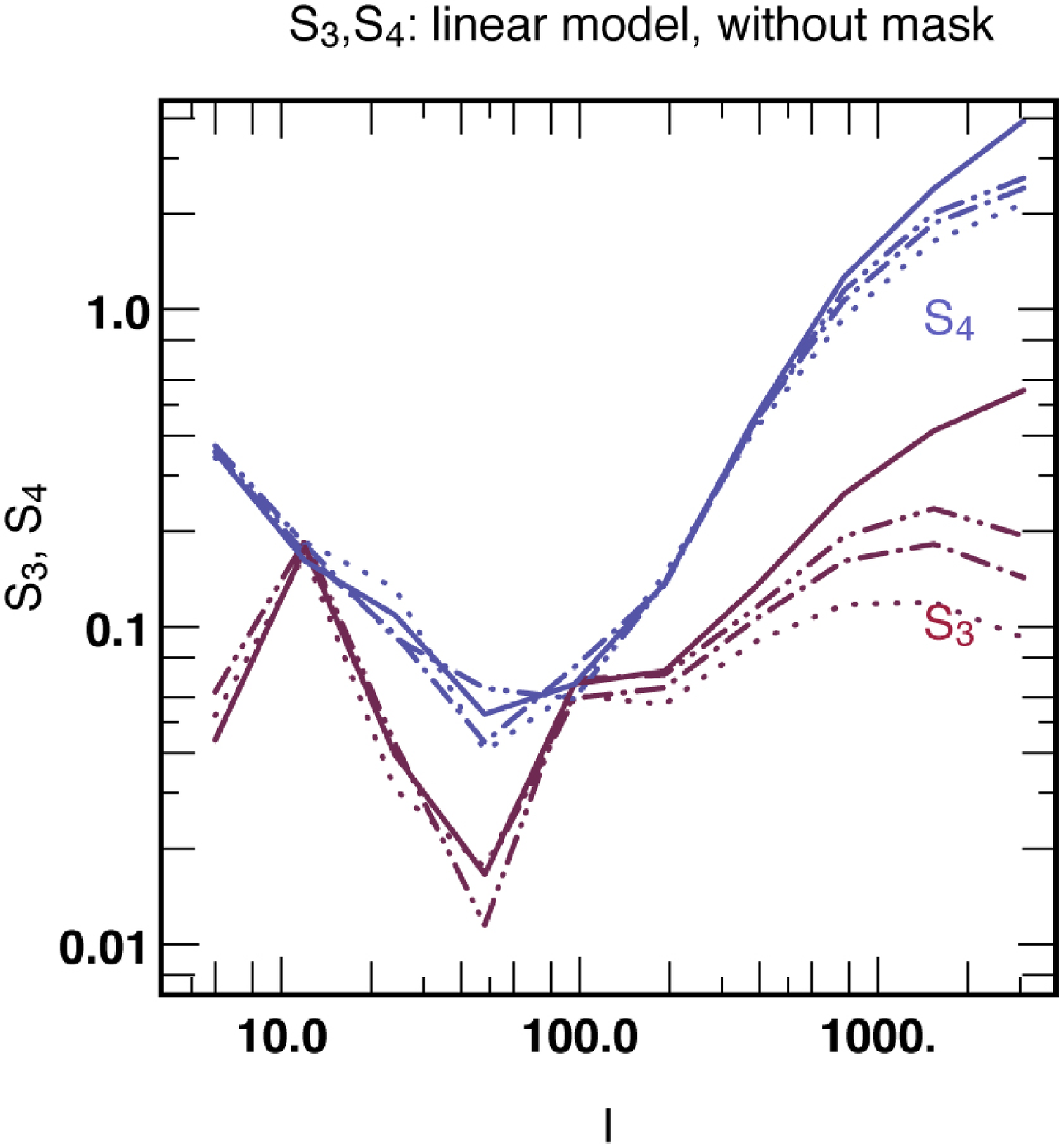}\hskip 0.5cm\includegraphics[width=0.9\columnwidth]{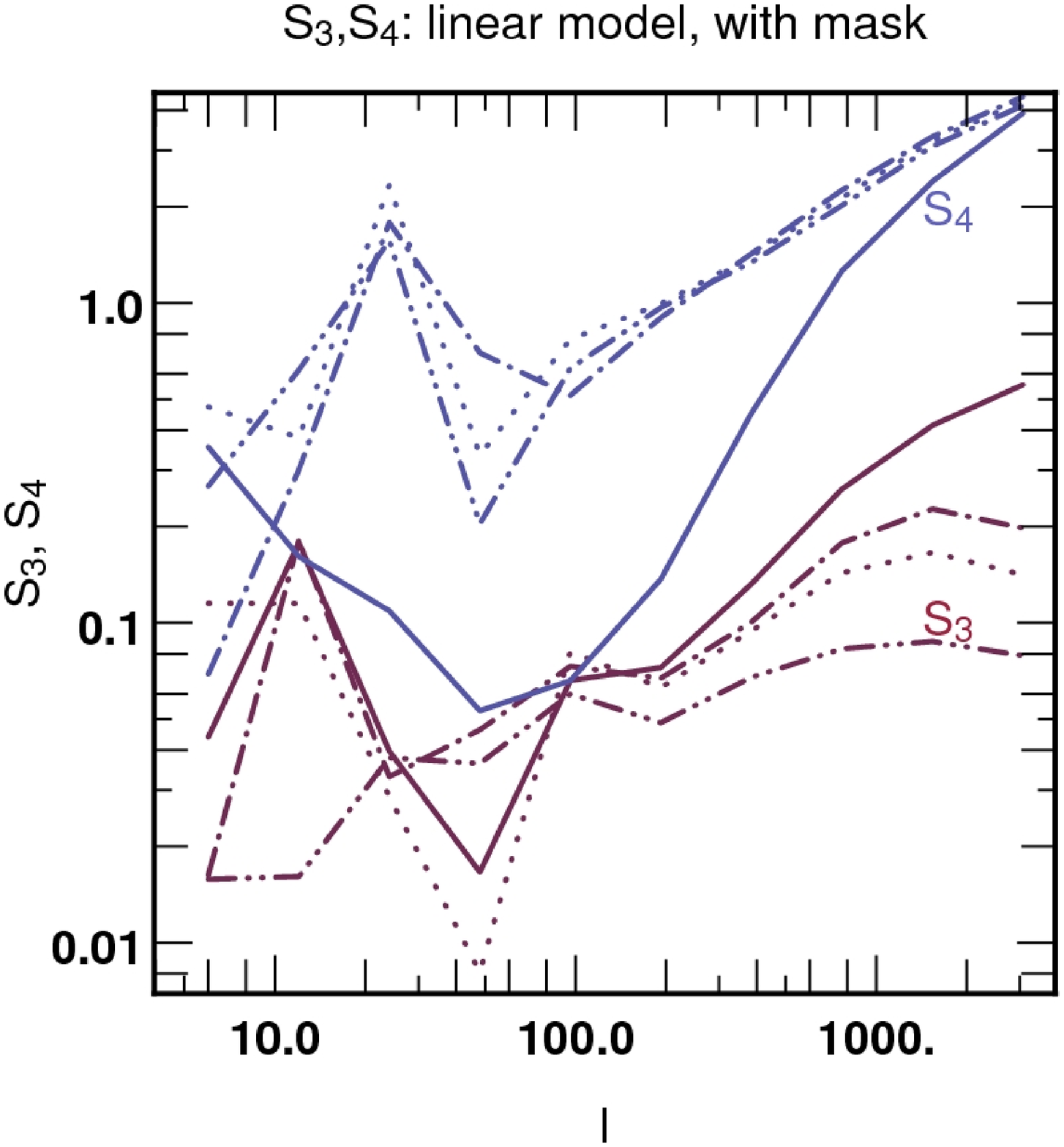}
\caption{\emph{left panel}: skewness, $S_{3}$ and kurtosis, $S_{4}$ as a
function of scale (using sharp top hat filtering) for the model (\emph{plane
line}) and the recovered $\kappa$ maps of a simulation $_{2048}S_{{\rm FS}}^{\ell_{{\rm cut}}}$
(\emph{dotted, dot-dashed, dot-dot-dashed line} for $\ell_{{\rm cut}}=722,1083,$
and $1569$,  resp. 24, 44 and 78 ngal/$\square$ arcmin); \emph{right panel}: same as top panel, but for the $_{2048}S_{{\rm GC}}^{\ell_{{\rm cut}}}$set.
Note that the kurtosis of the cut is significantly different at small
$\ell$. \label{fig:Skewness-and-kurtosis}}
\end{figure*}
The simplest statistics to explore the non linear transition is the skewness,
$S_{3}$ and the kurtosis, $S_{4}$ of the PDF of the recovered maps.
Furthermore, it has been shown that these parameters provide a powerful
tool to measure the underlying cosmological parameters (\citet{Bernardeau:1997p5919,Takada:2002p5791,Takada:2004p5780}).
Figure \ref{fig:Skewness-and-kurtosis} displays the evolution of
these numbers as a function of scale in the initial and recovered
maps, with and without galactic masking. The top hat filter used
here is of width $[2^{i},2^{i+1}]$, while the harmonic number of each
band is the mean of its boundary: $\bar{i}=(2^{i}+2^{i+1})/2$.
The recovery of skewness and kurtosis is good in the case of unmasked
data. Of course it degrades with the scale as we reach $\ell_{\rm cut}$.
Using the reconstructed map is not the optimal way of measuring the
3 and 4 point functions at small scale. However, an optimal dedicated
estimator can be built upon the same regularization technique. The
masked case is not as good. There, a dedicated estimator, acting only
on small, clean, pieces of the sky will probably yield better results. 

 \subsubsection{Accounting for a non-linear model}
Figure \ref{Non linear} shows the effect of accounting for the non
linearity in equation (\ref{eq:model}). Here a set of Cartesian simulations
is used $_{{\rm 256}}C_{{\rm NL/lin}}^{1}$. This map represents (a
100 times) the difference between the recovered map while accounting
for $1-\kappa$ in equation (\ref{eq:model}) in the inversion, and
the recovered map while neglecting this factor. The difference is
small in amplitude, but shows as expected the strongest bias near
the clusters and the filaments, where $\kappa$ is largest. The bottom
panel represents the corresponding relative power spectrum, $C_{\ell}[{\rm NL-{\rm lin}}]/C_{\ell}[{\rm input}]$
as a function of $\ell$. Again the larger discrepancy occurs at higher
$\ell$, corresponding to the sharp peaks at the positions of the
clusters. Hence the non-linearity should be accounted for in the model
if the shape of the cluster is an issue (see also \cite{White,Dodelson,2009ApJ...696..775S}). 
For all practical purposes, we have therefore demonstrated that at scales below $\ell_{{\rm max}}<4096$
solving the linearized problem is de facto equivalent to the general
non-linear problem when $\kappa$ is neglected at the denominator
in equation (\ref{eq:model}).

\subsection{Two points statistics}
%%%%%%%%%%%%%%%%%%%%%%%%%%%

Since \noun{ASKI} was constructed to provide the optimal {\sl map} given the measured shear,
 we do not expect that it will yield {\sl the} optimal estimator for non-linear functions of these maps, such as the powerspectrum,  bispectrum etc..
Nevertheless it is of interest to compare the two point statistics of input and recovered maps, to see how \noun{Aski} deals with masks and how
it affects the occurrence of spurious B modes. 

\subsubsection{Optimal Wiener filtering}\label{sub:Optimal-Wiener-filterning}
\begin{figure}
\includegraphics[width=1\columnwidth]{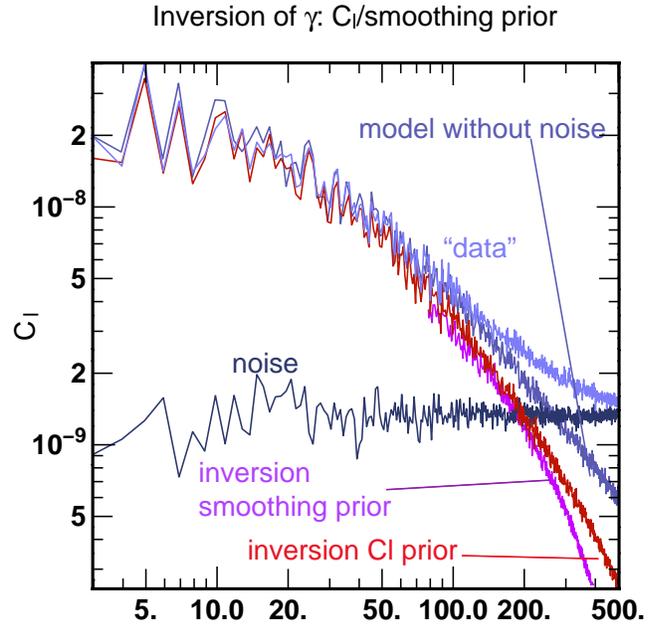}
\caption{The power spectrum of the input and recovered $\kappa$, (with smoothing
and $C_{\ell}$ prior, see equation~(\ref{eq:stationnary-prior})) as a function of $\ell$, together with the
power spectrum of the noise and the noisy equivalent $\kappa$ using
a simulation, $_{512}S_{{\rm FS}}^{224}$ (i.e ngal$/\square {\rm arcmin}= 5$). Note that the recovered
power spectrum departs from the power spectrum of the input field
roughly at the cutoff frequency when a quadratic smoothing penalty
is applied.
\label{Wiener}}
\end{figure}

Throughout this paper the prior $C_{\ell}\equiv\ell^{-1}(\ell+1)^{-1}$
(``Laplacian prior'') is used in equation (\ref{eq:laplacian-prior}).
Let us briefly investigate how a customized (Wiener) prior for $C_{\ell}$
changes the reconstruction at small scales. Figure \ref{Wiener} shows
that the corresponding power
spectra of the reconstructed kappa maps, as expected, differ mostly for scales where the signal to
noise is smaller than one. However, when the smoothing (Laplacian) prior amplitude
(see equation~\ref{eq:regul-penalty}) is tuned to minimize the
reconstruction error as in the figure, the power spectra of the
reconstructed maps for the two different priors (Laplacian and Wiener)
are quite similar. \\
It is interesting to note that the power of the reconstructed map with
the Wiener prior (light brown line in Figure~\ref{Wiener})
is systematically biased low as compared to the input power
spectrum. This reflects the fact that an optimal (minimum variance) estimation of the
power spectrum \emph{is not equivalent to} a power spectrum estimation
on an optimal (minimum variance) reconstructed map. However, in the
simple case where we have noisy data without masks, the bias of the
power spectrum of the minimum variance map reconstruction is known, it
is simply given by $C_{\ell} / (C_{\ell}+N_{\ell})$ where $C_\ell$ is
the power spectrum of the underlying kappa map (without noise), and
$N_\ell$ is the noise power spectrum in ``kappa'' space, which is
given approximately in our case by $\sigma^2\Omega_{pix}$, where
$\sigma^2$ is the noise variance per pixel in the shear field, and
$\Omega_{pix}$ is the solid angle of a pixel. \\
In Figure~\ref{Wiener},
the noise level is shown by the horizontal dark blue line. One can see
in particular in the figure, that when the model power spectrum
(without noise) crosses the noise power spectrum, the power spectrum
of the minimum variance map (golden line) is lower than that of the
model by a factor of $~2$, as expected from the considerations above,
even in the presence of masks.
Thus an approximate, but simple way to get an unbiased estimate of the
kappa power spectrum is to correct the minimum variance map power spectrum
by the ratio $C_\ell/(C_\ell+N_\ell)$. Note however that a true
minimum variance power spectrum estimation of the kappa field is not the aim of the
present method (see e.g. \cite{Pen2003} for the flat-sky case).

Nevertheless elsewhere in this paper, a smoothing prior which is not customized
to the specific problem is preferred. 

\subsubsection{Filling gaps within masks}
\begin{table}
\begin{tabular}{|r|r|r|r|r|r|}
\hline 
$n_{{\rm side}}$ & 128 & 256 & 512 & 1024 & 2048 \tabularnewline
\hline 
time for one step (s)  & 0.121 & 0.121 & 0.502 & 1.88 & 8.53 \tabularnewline
\hline 
number of steps (s) & 252 & 313 & 315 & 377 & 325 \tabularnewline
\hline 
total time (s)  & 40.1 & 50.3 & 200 & 989 & 3340 \tabularnewline
\hline
\end{tabular}
\caption{same as Table \ref{tab:the-performance-of} with Galactic masks\noun{.
\label{tab:perfGC}}}
\end{table}
\begin{figure*}
\includegraphics[clip,width=0.65\columnwidth]{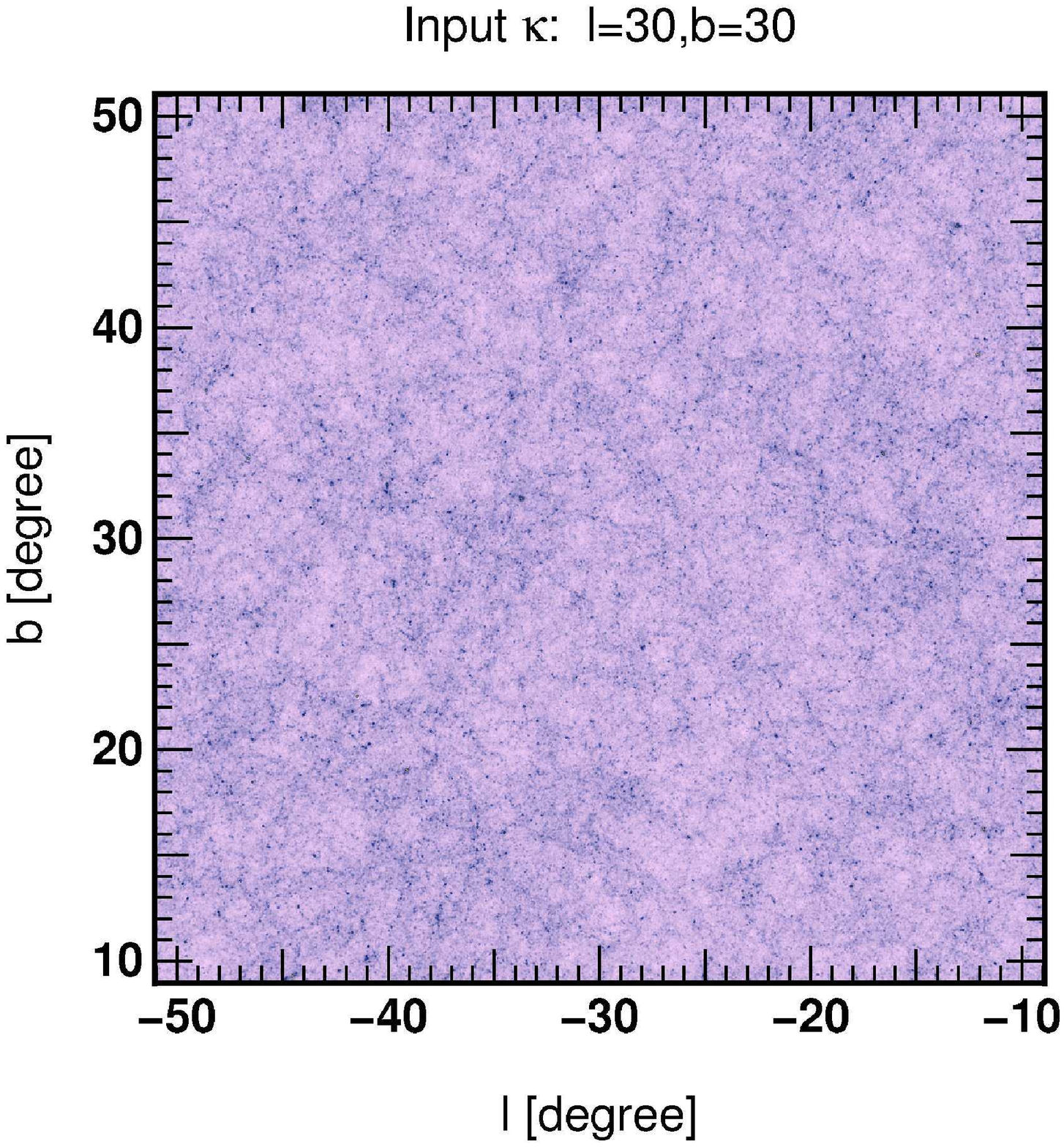}
\includegraphics[clip,width=0.67\columnwidth]{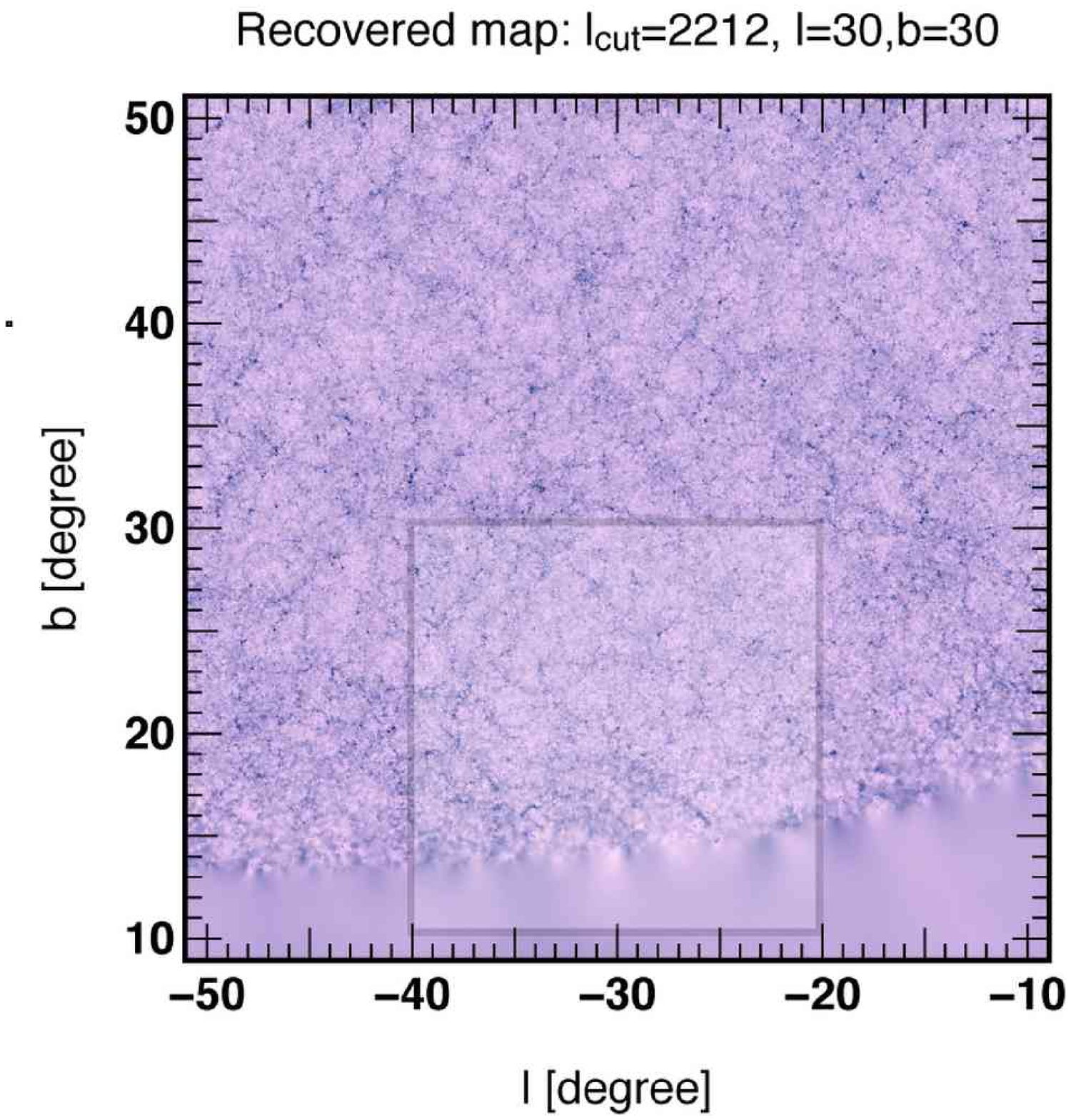}
\includegraphics[clip,width=0.67\columnwidth]{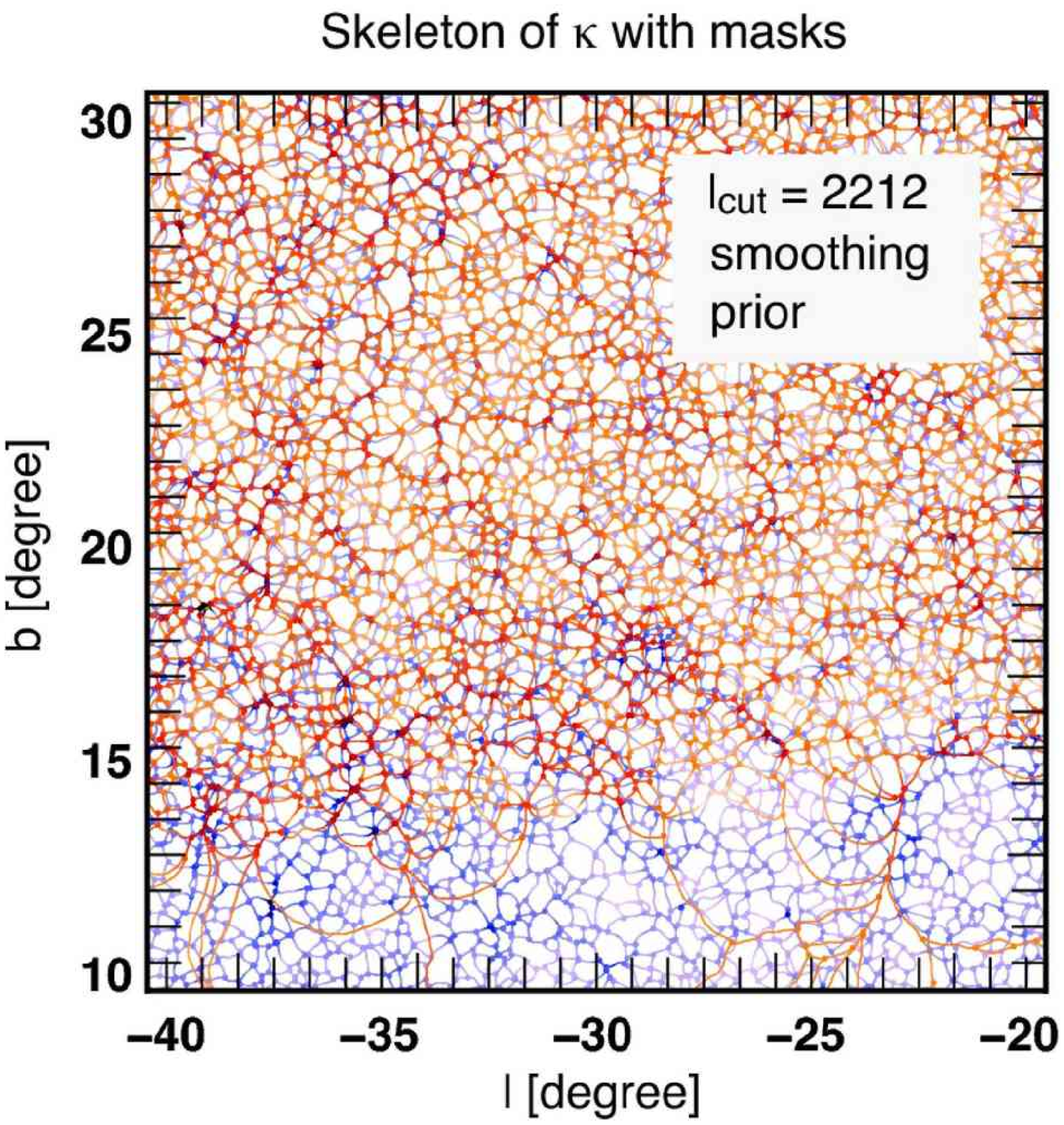}
\caption{\emph{left panel:} the initial $\kappa$ map in the region with bright
stars masking shown in Figure~\ref{fig:mask} at coordinates $(l,b)=(30^{\circ},30^{\circ})$;
\emph{middle panel: }the corresponding recovered $\kappa$ map\emph{
}of a simulation $_{2048}S_{{\rm GC}}^{2212}$ Note that the gaps
have been nicely filled up to the very edge of the mask;\emph{ right
panel:} the corresponding two skeletons (\emph{color coded by $\kappa$
in purple: }input skeleton;\emph{ in orange:} recovered skeleton)
for the inner region (marked as a square on the middle panel), when
masking is present. Note the clear gradient away from the mask in
the quality of the match between the two skeletons; Recall that most
of this field is partially shielded by stars, as seen in Figure \ref{fig:mask}.
\label{fig:gaps}}
\end{figure*}
Let us first compare visually  the recovered map to the input map. Figure \ref{fig:gaps} illustrates a
feature of the penalized reconstruction: it interpolates quite well 
 and provides means to fill the gaps corresponding
to the galactic cuts. 
For a more quantitative comparison we also plot on this figure the ridges of both fields (using the skeleton, see below), which match very well
up to the very edge of the mask.
The smoothing penalty also induces a level of
extrapolation, best seen in the residuals, see Figure \ref{fig:Leak}.
The masking (or more generally, non uniform weights, $w_{i}$) nevertheless
biases the reconstructed map, as seen on Figure \ref{fig:Skewness-and-kurtosis}
and \ref{fig:gaps2}. 
\begin{figure}
\includegraphics[width=1.\columnwidth]{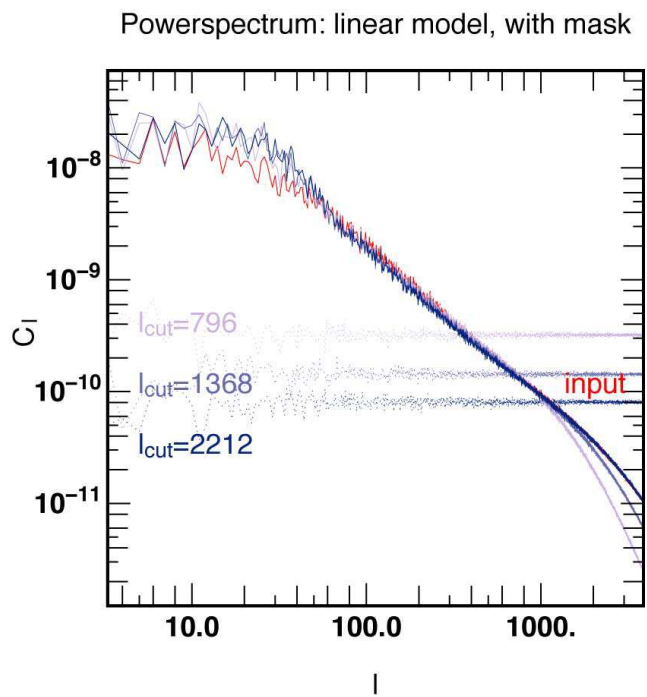}
\caption{the power spectra of three high resolution reconstructions corresponding to Figure~\ref{fig:gaps}
for $\ell_{{\rm cut}}=796,1368,$ and $2212$ (resp. 28, 63 and 130 ngal/$\square$ arcmin corresponding to a low,  intermediate and high end values) together with the power
spectra of the noise. Note that the the recovered power spectrum has
extra power at large scales and less power at intermediate scales,
an artifact of the mask which can be corrected for by accounting for
the prior knowledge of the auto-correlation of the mask.\label{fig:gaps2}}
\end{figure}
Note finally that when masks
are accounted for, it is straightforward to correct for them when
computing the powerspectrum as the harmonic transform of the autocorrelation,
which in turn is derived by correcting for the autocorrelation of
the masks:
 (see \cite{spice,Hivon,Chon} for details).
 When  seeking the  three-point  correlations one could also
 proceed  accordingly, and  divide by  the three-point correlation  of the
 mask.   Indeed a three-point reduced  correlation is  simply one  plus the
 excess  probability of  finding  triplets,  which in  turn  is computed  by
 counting the number  of found triplets and dividing  by the expected number
 of such triplets given the shape of  the mask (\cite{Szapudi2}). This also applies if the mask
 is grey.

\subsubsection{Residual B modes}\label{sub:Residual-B-modes} 
Let us investigate the effect of leaking of B modes with the following
experiment: the noise in the transform of the B channel is boosted
by some fixed amount over a map which has Galactic cuts. This corresponds
to the case where the $B$ is significantly larger than the noise,
yet uncorrelated with the $E$ mode, corresponding to e.g. a systematic
bias in the ellipticity extraction for example. It is expected that,
due to masks, this $B$ mode will leak in $E$. An example of such
leak, for $B$ modes as large or up to $32$ times larger than the
noise, is shown in Figure \ref{fig:Leak}. The power spectrum of the
residuals in the corresponding $\kappa$ map is computed while masking
in the residual the exact regions corresponding to the cuts. When
this boost is zero, (bottom curve in Figure \ref{fig:Here-the-residual})
the power spectrum of these residuals is flat and corresponds to the
noise powerspectrum. In contrast, the stronger the boost the larger
the scale below which this power spectrum is colored. Note that it
was checked that, as expected, these coherent residuals disappear
completely if the galactic cuts are ignored. It would also be interesting
to compare the distribution of the shape of dark matter in input/recovered
clusters.

Finally, note that Appendix \ref{sec:noise} discusses briefly the effect of noise 
in powerspectrum estimation.

%%%%%%%%%%%%%%%%%
\subsection{Alternative statistics: critical sets}
%%%%%%%%%%%%%%%%%
Let us close this section with a quantitative  comparison  of the input and the recovered map
using more exotic probes  to  estimate the quality of the reconstruction,  and the prospect it offers for dark energy measurements.
Indeed, the predictions of the perturbative hierarchical clustering model are often given
through the hierarchy of the differences between the
moments to their Gaussian limit. Yet higher order moments 
are generally difficult to test directly in real-life observations,
due to their sensitivity to very rare events. As argued in \cite{PGP} the geometrical
analysis of the critical sets in the field (extrema counts, Genus, critical lines etc..)  may provide more robust measures
of non-Gaussianity, and is becoming elsewhere an active 
field of investigation \citep{Gott,Park}.
 
\subsubsection{Peak patch counts and area}\label{sub:Point-source-extraction}
\begin{figure*}
\includegraphics[width=0.9\columnwidth]{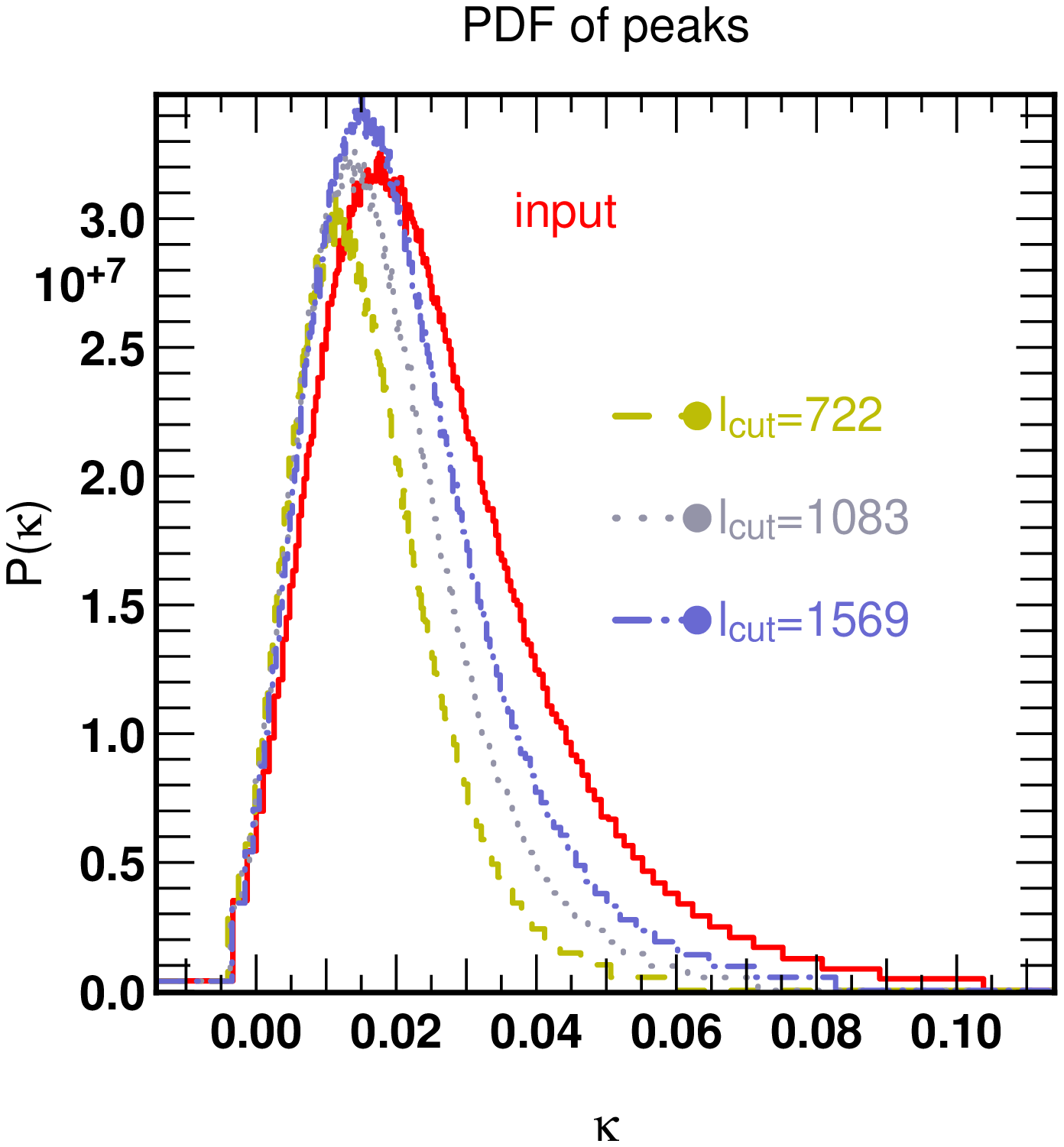}\hskip 1cm
\includegraphics[width=0.9\columnwidth]{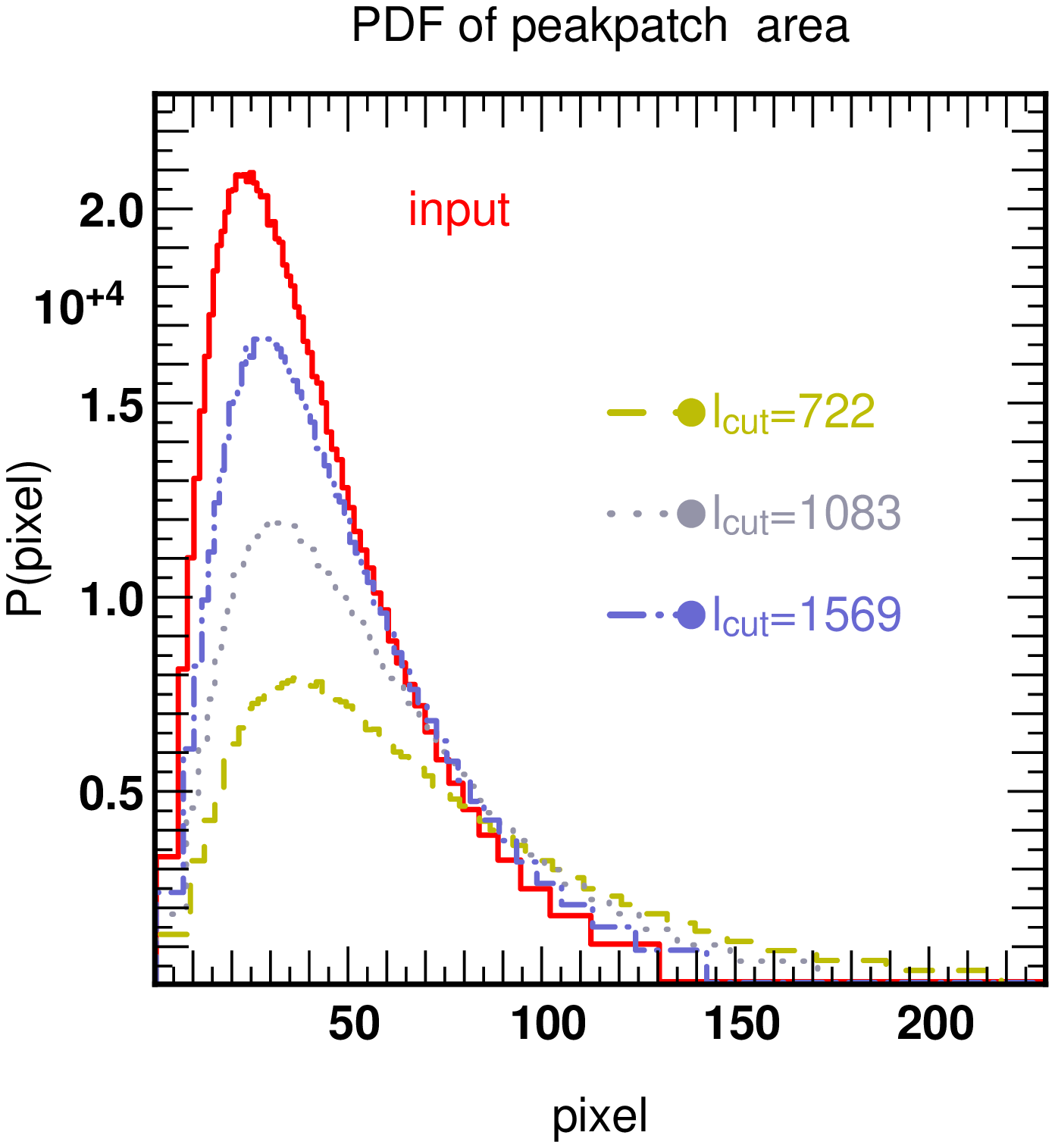}
\caption{\label{fig:Point source}\emph{left panel}: the PDFs of $\kappa_{{\rm max}}$
at point sources before and after reconstruction of set of simulations
$_{2048}S_{{\rm FS}}^{\ell_{{\rm cut}}}$ (\emph{dashed, dotted, dot-dashed
line} for $\ell_{{\rm cut}}=722,1083,$ and $1569$  (resp. 24, 44 and 78 ngal/$\square$ arcmin);\emph{ right
panel}: the PDFs of the area of peak patches (see Figure \ref{fig:Peak})
before and after reconstruction for the same set of simulations. Note
that, as expected, the recovered distribution of peaks is less skewed
than the original, whereas conversely, the PDF of the area of the peak
patches for the low SNR reconstruction is more skewed towards larger
patches.}
\end{figure*}
Even though many tools are available to identify peaks within the reconstructed map, 
let us validate here our reconstruction using  a segmentation of both the initial and the recovered maps
using peak patches on the sphere, which 
 are a segmentation of the map based on the attraction patches of the $\kappa$ map when following its 
 gradient (see \citet{rsex}).
Within each peak patch (see Figure \ref{fig:Peak}), the brightest
pixel is assigned a mass corresponding to the enclosed mass within
the peak patch. This quantity is gravitationally motivated (as the patch corresponds to the attraction region of the cluster) and 
is both robust (as the geometry of the patch
 only depends on the imposed smoothing length, which in turn is fixed by the resolution of the 
survey) and sensitive  to small features in the map; it is therefore a good indicator of the quality of the reconstruction. 
 Figure \ref{fig:Point source} (\emph{left panel})
displays the corresponding PDFs %for resolutions corresponding to $n_{{\rm side}}=256,512,1024$
before and after reconstruction. As expected, the recovered point
source PDF has a shifted mode and is less skewed than the original
distribution. This trend decreases with increasing SNR. For realistic galaxy counts of  40 {ngal}$/\square$ arcmin,
the agreement between the input and the recovered PDF is fairly good, and the corresponding residual bias can be modeled 
(as the reconstruction is essentially a smoothing of the underlying map).
This could lead to interesting constraints on $\Omega_{\rm m}$ and $D(z)$  when used in conjunction with weak lensing tomography in order to probe its redshift evolution.
The right panel of Figure \ref{fig:Point source} focuses on a different quantity, the area of the patches, which when compared
to the area of the corresponding void patches, could also be used as a measure of the  gravitationally induced non gaussianities, together with their shape (higher moments of $\kappa$ within a patch).
Again, the reconstruction seems to recover this distribution well enough to suggest that such a tool could be used in the future
to study the cosmic evolution of the projected web.
\subsubsection{Topology \& geometry: critical lines}\label{sub:Topology}

\begin{figure}
\includegraphics[width=0.9\columnwidth]{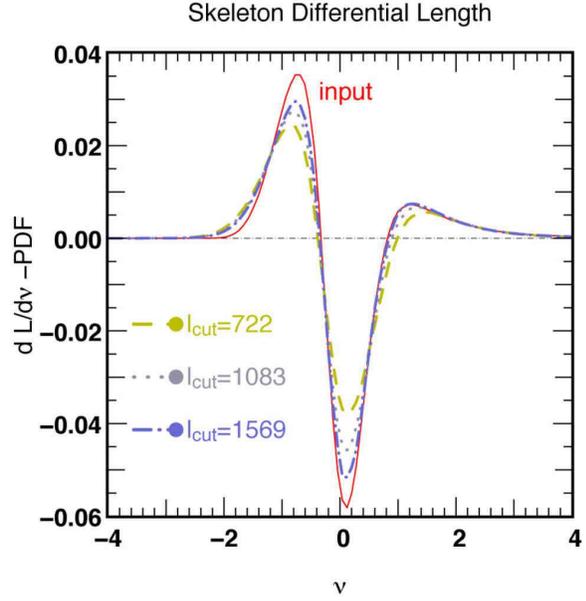}\caption{The input skeleton differential length (a tracer of $\Omega_{m}$) with its recovered counterparts
as a function of the normalized $\kappa$ contrast, $\nu\equiv(\kappa-\bar{\kappa})/\sigma_{\kappa}$
for the set $_{2048}S_{{\rm FS}}^{\ell_{{\rm cut}}}$ with $\ell_{{\rm cut}}=722,1083,$
and $1569$  (resp. 24, 44 and 78  ngal/$\square$ arcmin) . Here the PDF of the normalized $\kappa$ contrast was
subtracted to the differential length for clarity. As expected, the
agreement is best at large convergence. This figure is complementary
to Figure \ref{fig:gaps} which shows that the {\sl geometry} of the field
is well preserved on average. \label{fig:peakpatch-diff-length}}
\end{figure}

\begin{figure}
\includegraphics[width=1\columnwidth]{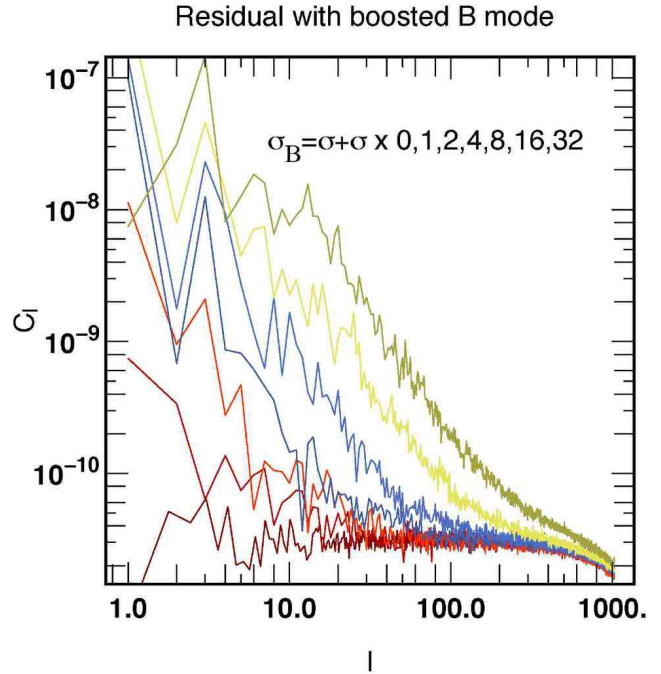}
\caption{Power spectrum of the mask weighted residual error on $\kappa$ as
a function of the harmonic number, $\ell$. The different curves correspond
to boost of the $B$ modes of increasing relative strength. The low
order modes are polluted by leaks from the masks (see also Figure~\ref{fig:Leak});
here $l_{{\rm cut}}=752$  (60 ngal/$\square$ arcmin). \label{fig:Here-the-residual}}
\end{figure}

\begin{figure}
\includegraphics[width=1\columnwidth]{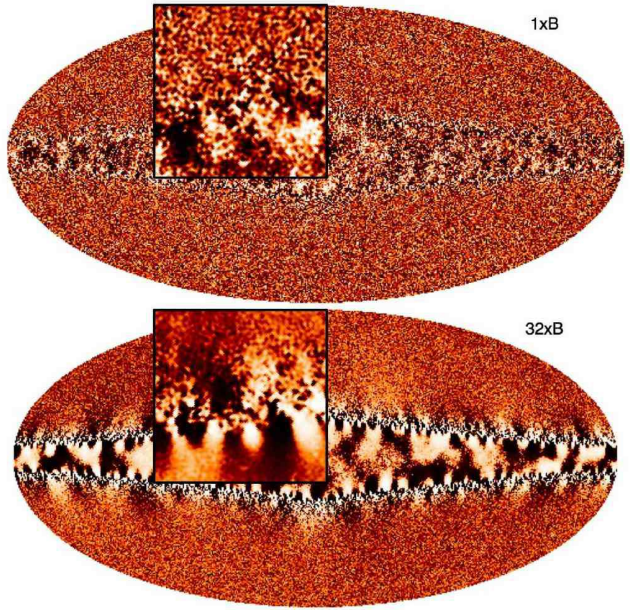}\caption{shows an example of full-sky leak of the B modes when masks are accounted
for; \emph{top panel}: the residuals corresponding to $\sigma_{B}=\sigma+\sigma$;
the inner box corresponds to a zoom near the edge of the galactic
cut at $(b,l)=(30,20)$; \emph{bottom panel}: same residual and box
for $\sigma_{B}=\sigma+32\sigma$. Note that for the latter case,
the extend of the leakage is much larger and coherent. \label{fig:Leak}}
 \end{figure}

Let us now compare the shape of the recovered map to the initial map
from the point of view of its critical lines.  
For this
purpose, let us use here the skeleton as a geometric probe  (\citet{2006MNRAS.366.1201N,2007arXiv0707.3123S}).
It is defined in 2D as the boundary of the void patches, 
which in turn are a segmentation of the map based on the valleys of the $\kappa$ map (corresponding to the peak patches (defined above) of {\sl  minus} the field). The
skeleton of the initial field and the recovered fields for simulation
$_{2048}S_{{\rm GC}}^{\ell_{{\rm cut}}}$ is computed, and represented
in Figure \ref{fig:gaps}. The recovered skeletons are qualitatively
fairly close to the original skeleton, which demonstrates that the
local topology and geometry of the field is well recovered. Let us
make this comparison more quantitative. The differential length per
unit area of the recovered field (\emph{the set }$_{2048}S_{{\rm FS}}^{\ell_{{\rm cut}}}$
with $\ell_{{\rm cut}}=722,1083,$ and $1569$\emph{ as labeled})\footnote{ Note that 
ngal$/\square {\rm arcmin}=40(l_{\rm cut}/1000)^{1.5}$.
}
over the initial $\kappa$ map (\emph{thin line}) as a function of
density threshold is also shown in Figure \ref{fig:peakpatch-diff-length}, while Figure \ref{fig:L2-full-sky} shows
the corresponding maps for similar runs, together with a map of the orientation of the $\mathbf g$ field.
The agreement increases at larger density thresholds, which suggests
that the topology of dense regions is well recovered\footnote{In fact the relative distance between the recovered 
and the input skeleton could also be used as  an alternative to the differential length see \citet{caucci}.}.  
The total length
was shown (\citet{rsex}) to trace well the underlying
shape parameter of the powerspectrum and has been used in 3D to constaint the
dark matter content of the universe (\citet{skel1}).
%  The differential length can also be used as a non gaussianity test \citep{skel2D}.
As shown in \citet{skllpogo} this would work  for two dimensional maps and could therefore be used 
with $\kappa$ maps such as those reconstructed via the present method.  
The redshift evolution of this differential count, when tomographic data is available, could complement e.g. Genus measurements as means of constraining the dark energy equation of state in a manner which 
could be more robust than direct cumulant estimation.
Eventually, the skeleton could also be used to characterize the connectivity of
clusters (i.e. the number of connected projected filaments), as it will also depend
on the cosmic dark energy content of the universe (\citet{connex}). 

This rapid review has shown that, depending on the final objective (cosmological parameters, cross correlation with other maps etc.), a variety of estimators can be extracted from the recovered maps. \noun{Aski} was
 shown to perform rather well  with respect to these 
 estimators.
Defining the best combination of these estimators, -- and the optimal penalty associated --
 will be one of the 
key topic lensing research for the coming years.

 %%%%%%%%%%%%%%%%%%%%%%%%%%%%%
\section{Conclusion \& Discussion}\label{sec:Discussion}
%%%%%%%%%%%%%%%%%%%%%%%%%%%%%
%%%%%%%%%%%%%%%%%%%%%%%%%%%%%
This paper sketched possible solutions to issues that a full-sky weak-lensing
pipeline will have to address, and presented an inverse method
implementing  the debluring of the image  and the map making step.

Weak lensing surveys require measuring statistical distributions of
the morphological parameters (ellipticity, orientation, ...) of a
very large number of galaxies. This paper demonstrated that these
parameters can be measured with a better accuracy and strongly reduced
bias if the deep sky images are properly deblurred prior to the shape
measurements. Using a {\sl relative} figure of merit (the recovered \noun{SExtractor}
ellipticity) we have shown that this deblurring could in crowded fields
improve more than tenfold the accuracy of the recovered ellipticities.
This deblurring is critical in crowded regions, where the overlapping
of stars and galaxies otherwise prevents accurate morphological estimation.
Henceforth dealing with such regions is important for a full-sky survey.
Since such surveys will require the processing of a great number of
large images, the calibration of these techniques is automated on
the images themselves via cross validation {\sl after} identification and removal of the 
stars within the field (see Figure~1). In particular the level of regularization,
$\mu$ and the $\ell_{1}-\ell_{2}$ threshold are {\sl automatically} tuned
in order to deal with the noise level and the dynamics of the raw images.
The gap-filling interpolation feature of the inversion would apply
even more efficiently in this regime than in the map reconstruction
regime described in Section \ref{sec:A-Model-for}. The algorithm
described here scales well since it only relies on DFTs: hence it
could be applied to very large images  such as
those produced by modern surveys.  \noun{Aski} uses
the efficient variable metric limited memory algorithm \noun{OptimPack}, which
allows both optimizations to scale to high resolutions. The deblurring is implemented on
Cartesian maps as large as  $16\,384^{2}$ pixels.
% while the inversion is
%carried on the sphere for HEALPix resolutions as large as $n_{{\rm side}}=4096$. 
 Generalized cross validation was
shown to yield a quantitative threshold in order to remove accurately
the point sources within the field, hence imposing the optimal level
of smoothing for the galaxies only. In this paper, the focus was put
on blurred 8-meter ground-based observations, but the implementation
for EUCLID-like space missions should be straightforward.
The above described improvements could clearly be reproduced if alternative 
  state  of the  art shear estimators were to be used (as compared by the SHear Testing Program).

This paper also demonstrated that optimization in the context of Maximum
A Posteriori provided a consistent framework for the optimal reconstruction
of $\kappa$ maps on the sphere. 
{The main asset of the 
\noun{Aski} algorithm
 is that the penalty can be applied in model
space, while
the optimization iterates back and forth between data space  and model space. This freedom  allows it to deal simultaneously with masks (in data space) and edge preserving penalties.}
 Providing $\kappa$ maps is critical both
in its own right, as it maps the dark matter distribution of our universe
and gives access to the underlying powerspectrum at large scale. Such
maps are also interesting when cross-correlated with other surveys
(optical surveys, CMB maps, lensing reconstruction and distribution
of SZ clusters from the Planck mission, redshift evolution of 
X-ray sources counts etc..) in order
to explore the evolution of the large-scale structure, and in the
case of the surveys mapping the baryonic matter, to better understand
biasing as a function of scale. Finally, though not optimally, it
can be used to compute second and higher order statistics, and noticeably 
the three-point statistics, the Genus, cluster counts or the skeleton,  which may constrain more
efficiently the dark energy equation of state, as they are less sensitive to rare events. It should be stressed
once more that while the reconstructed $\kappa$ maps yield biased
estimates of the power-spectrum and higher order statistics, the technique
described in this paper can be adapted to build dedicated optimal
estimators for each of those observables.

 Section \ref{sec:Validation} demonstrated
the quality and limitations of the reconstruction using various statistical
tools on a full-sky simulation of $\mathbf{g}$ with resolutions of
up to $12\times4096^{2}=201\,326\,592$ pixels thanks again to the efficiency of \noun{Optimpack}. In particular, it
identified point sources of the fields, analyzed their PDF and
showed that $\ell_{1}-\ell_{2}$ penalty was critical at small scales.
It also investigated the effect of leakage of $B$ modes when Galactic
cuts are present. It presented a method to
probe the topology and geometry of a field on the sphere, the peak
patches and the skeleton, and applied it to compare the recovered
field to the initial field. Such tools allow us to quantify the differences between the two maps and
act as an efficient source segmentation algorithm. Indeed, the degeneracy
between the cosmological parameters $(\Omega_{{\rm M}},\sigma_{8})$
is for instance best lifted with cluster counts.
They may  also turn out to be of importance when probing the  dark energy equation of state
as  they are less sensitive to rare events.
 The Cartesian dual formulation of \noun{aski}
was also implemented and may prove useful for surveys where sky coverage
is sufficiently small. 

In short, \noun{Aski} accounts for the possible building blocks that a full-scale
pipeline aiming at sampling the dark matter distribution over the
whole sky should provide. Specifically it allows for (i) automatically  deblurring
very large images using non-parametric self-calibrated
edge-preserving $\ell_{1}-\ell_{2}$ deconvolution with positivity;
(ii) carrying the large \emph{non-linear }inverse problem of reconstructing
the convergence $\kappa$ from the shear $\mathbf{g}$ using equation
(\ref{eq:model}): the back and forth iterations between model and
data are  consistent with constraints in both spaces, and allow for
an accurate recovery of cluster profiles and shapes; (iii) \emph{non-uniform} weighting and masking: consistent with realistic Galactic
cuts (and bright stars masking) and non-uniform sampling of the different
regions of the sky, dealing transparently with the issue of the boundary;
(iv) edge-preserving $\ell_{1}-\ell_{2}$ penalty yielding quasi point-like 
cluster reconstruction. Finally (v) it introduced peak patches
and the skeleton on the sphere, together with its statistics.

Possible improvements/investigation beyond the scope of this paper
involve:
(i) comparing the absolute gain in shear estimation using alternative  tools to \noun{Sextractor} (such as \cite{Massey:2007p176}) with more realistics galactic shapes; 
 (ii) deblurring the images with a variable PSF within the
field; (iii) building optimal estimators for the power spectrum $C_{\ell}^{\kappa}$,
or the asymmetry $S_{3}$ (a possible option would be to rely on perturbation
theory, and invert the non-linear problem for both $C_{\ell}^{\kappa}$
and $S_{3}$); (iv) inverting for $\gamma$ \emph{and} $\kappa$
\emph{simultaneously} and checking a posteriori the amplitude of the
$B$ modes (an alternative to the model described in equation (\ref{eq:defF1});
the issue of unicity of the solution will be a challenge); (v) carrying
the deprojection while assuming prior knowledge of a complete distribution
of source planes in equation (\ref{eq:kappa-delta}) (the corresponding
inverse problem remains linear, with an effective kernel which depends
on the optical configuration and the distribution of galaxies as a
function of redshift); (vi) moving away from the Born approximation,
which involves solving Poisson's equation for each slice, and ray-tracing 
back to the source while solving for the lens equation though
all the slices; (vii) implementing a more realistic noise modeling
(which amounts to changing the cost function, equation (\ref{eq:objectif}));
(viii) studying the shape of dark matter distribution in clusters and
groups: typically this would also involve cross-correlating the corresponding
distribution with the light at various wavelengths,
(ix) defining the post analysis which most sensitive to dark energy,  given the feature of the surveys to come
 and finally (x)
propagating the analysis up to the cosmic figure of merit for the
dark energy parameters.
\subsection*{Acknowledgments}
\emph{We thank Dmitry Pogosyan, Dominique Aubert, Eric Hivon, Martin Kilbinger and Yannick Mellier for
comments and suggestions, the} \noun{Horizon} 4$\Pi$ \emph{team and
the staff at the CCRT} \emph{for their help in producing the simulation,
and D.~Munro for freely distributing his }\emph{\noun{Yorick}}\emph{
programming language and }\noun{opengl} \emph{interface (available
at }\texttt{http://yorick.sourceforge.net/}) \emph{The galactic mask
was provided to us by Adam Amara. This work was carried within the
framework of the} \noun{horizon} \emph{project}: \texttt{www.projet-horizon.fr}. 
\bibliographystyle{mn2e}
\bibliography{aski}

\begin{thebibliography}{}

\bibitem[\protect\citeauthoryear{{Abrial}, {Moudden}, {Starck}, {Fadili},
  {Delabrouille} \& {Nguyen}}{{Abrial} et~al.}{2008}]{starckforever}
{Abrial} P.,  {Moudden} Y.,  {Starck} J.-L.,  {Fadili} J.,  {Delabrouille} J.,
    {Nguyen} M.~K.,  2008, Statistical Methodology, 5, 289

\bibitem[\protect\citeauthoryear{{Aubert}, {Pichon} \& {Colombi}}{{Aubert}
  et~al.}{2004}]{aubert04}
{Aubert} D.,  {Pichon} C.,    {Colombi} S.,  2004, \mnras, 352, 376

\bibitem[\protect\citeauthoryear{Aubert \& Kornprobst}{Aubert \&
  Kornprobst}{2008}]{image}
Aubert G.,  Kornprobst P.,  2008, Mathematical Problems in Image Processing:
  Partial Differential Equations and the Calculus of Variations (Applied
  Mathematical Sciences), first edition edn.
Springer Verlag

\bibitem[\protect\citeauthoryear{{Bartelmann}, {Narayan}, {Seitz} \&
  {Schneider}}{{Bartelmann} et~al.}{1996}]{k14}
{Bartelmann} M.,  {Narayan} R.,  {Seitz} S.,    {Schneider} P.,  1996, \apjl,
  464, L115+

\bibitem[\protect\citeauthoryear{Bartelmann \& Schneider}{Bartelmann \&
  Schneider}{2001}]{2001PhR...340..291B}
Bartelmann M.,  Schneider P.,  2001, \physrep, 340, 291

\bibitem[\protect\citeauthoryear{Benabed \& Scoccimarro}{Benabed \&
  Scoccimarro}{2005}]{Benabed:2005p2447}
Benabed K.,  Scoccimarro R.,  2005, Arxiv preprint astro-ph

\bibitem[\protect\citeauthoryear{Bernardeau, Mellier \& van
  Waerbeke}{Bernardeau et~al.}{2002}]{Bernardeau:2002p2223}
Bernardeau F.,  Mellier Y.,    van Waerbeke L.,  2002, Arxiv preprint astro-ph

\bibitem[\protect\citeauthoryear{Bernardeau, van Waerbeke \&
  Mellier}{Bernardeau et~al.}{1997}]{Bernardeau:1997p5919}
Bernardeau F.,  van Waerbeke L.,    Mellier Y.,  1997, Astronomy and
  Astrophysics, 322, 1

\bibitem[\protect\citeauthoryear{{Bertin} \& {Arnouts}}{{Bertin} \&
  {Arnouts}}{1996}]{sex}
{Bertin} E.,  {Arnouts} S.,  1996, aaps, 117, 393

\bibitem[\protect\citeauthoryear{Bradac, Schneider, Lombardi \& Erben}{Bradac
  et~al.}{2005}]{Bradac:2005p3138}
Bradac M.,  Schneider P.,  Lombardi M.,    Erben T.,  2005, arXiv, astro-ph

\bibitem[\protect\citeauthoryear{Bridle \& Abdalla}{Bridle \&
  Abdalla}{2007}]{Bridle:2007p1602}
Bridle S.,  Abdalla F.,  2007, The Astrophysical Journal

\bibitem[\protect\citeauthoryear{Bridle, Hobson, Lasenby \& Saunders}{Bridle
  et~al.}{1998}]{Bridle:1998p2820}
Bridle S.,  Hobson M.,  Lasenby A.,    Saunders R.,  1998, Mon. Not. R. Astron.
  Soc.

\bibitem[\protect\citeauthoryear{Cacciato, Bartelmann, Meneghetti \&
  Moscardini}{Cacciato et~al.}{2006}]{Cacciato:2006p3147}
Cacciato M.,  Bartelmann M.,  Meneghetti M.,    Moscardini L.,  2006, arXiv,
  astro-ph

\bibitem[\protect\citeauthoryear{{Caucci}, {Colombi}, {Pichon}, {Rollinde},
  {Petitjean} \& {Sousbie}}{{Caucci} et~al.}{2008}]{caucci}
{Caucci} S.,  {Colombi} S.,  {Pichon} C.,  {Rollinde} E.,  {Petitjean} P.,
  {Sousbie} T.,  2008, \mnras \, in press, pp 000--000

\bibitem[\protect\citeauthoryear{{Chen} \& {Szapudi}}{{Chen} \&
  {Szapudi}}{2005}]{Szapudi2}
{Chen} G.,  {Szapudi} I.,  2005, \apj, 635, 743

\bibitem[\protect\citeauthoryear{{Chon}, {Challinor}, {Prunet}, {Hivon} \&
  {Szapudi}}{{Chon} et~al.}{2004}]{Chon}
{Chon} G.,  {Challinor} A.,  {Prunet} S.,  {Hivon} E.,    {Szapudi} I.,  2004,
  \mnras, 350, 914

\bibitem[\protect\citeauthoryear{Crittenden, Natarajan, Pen \&
  Theuns}{Crittenden et~al.}{2002}]{Crittenden:2002p2070}
Crittenden R.,  Natarajan P.,  Pen U.,    Theuns T.,  2002, The Astrophysical
  Journal

\bibitem[\protect\citeauthoryear{Dodelson \& Zhang}{Dodelson \&
  Zhang}{2005}]{Dodelson}
Dodelson S.,  Zhang P.,  2005, Phys. Rev. D, 72, 083001

\bibitem[\protect\citeauthoryear{{Erben}, {Van Waerbeke}, {Bertin}, {Mellier}
  \& {Schneider}}{{Erben} et~al.}{2001}]{skymaker}
{Erben} T.,  {Van Waerbeke} L.,  {Bertin} E.,  {Mellier} Y.,    {Schneider} P.,
   2001, \aap, 366, 717

\bibitem[\protect\citeauthoryear{Fu \& et al.}{Fu \& et~al.}{2008}]{Fu:2007}
Fu et al. 2008, Astronomy \& Astrophysics

\bibitem[\protect\citeauthoryear{Girard}{Girard}{1989}]{Girard-1989-fast_monte%
_carlo_cross_validation}
Girard D.~A.,  1989, Numr. Math., 56, 1

\bibitem[\protect\citeauthoryear{Golub, Heath \& Wahba}{Golub
  et~al.}{1979}]{Golub:Heath:Wahba:1979}
Golub G.~H.,  Heath M.,    Wahba G.,  1979, Technometrics, 21, 215

\bibitem[\protect\citeauthoryear{{G{\'o}rski} \& {et al.}}{{G{\'o}rski} \& {et
  al.}}{1999}]{1999elss.conf...37G}
{G{\'o}rski} K.~M.,  {et al.} 1999, in {Banday} A.~J.,  {Sheth} R.~K.,   {da
  Costa} L.~N.,  eds, Evolution of Large Scale Structure : From Recombination
  to Garching {Analysis issues for large CMB data sets}.
pp 37--+

\bibitem[\protect\citeauthoryear{{Gott}, {Choi}, {Park} \& {Kim}}{{Gott}
  et~al.}{2009}]{Gott}
{Gott} J.~R.,  {Choi} Y.-Y.,  {Park} C.,    {Kim} J.,  2009, \apjl, 695, L45

\bibitem[\protect\citeauthoryear{Halkola, Seitz \& Pannella}{Halkola
  et~al.}{2006}]{Halkola:2006p3208}
Halkola A.,  Seitz S.,    Pannella M.,  2006, arXiv, astro-ph

\bibitem[\protect\citeauthoryear{{Heymans}}{{Heymans}}{2006}]{2006MNRAS.368.13%
23H}
{Heymans} e.~a.,  2006, \mnras, 368, 1323

\bibitem[\protect\citeauthoryear{Hirata \& Seljak}{Hirata \&
  Seljak}{2004}]{Hirata:2004p1549}
Hirata C.,  Seljak U.,  2004, Physical Review D

\bibitem[\protect\citeauthoryear{{Hivon}, {G{\'o}rski}, {Netterfield}, {Crill},
  {Prunet} \& {Hansen}}{{Hivon} et~al.}{2002}]{Hivon}
{Hivon} E.,  {G{\'o}rski} K.~M.,  {Netterfield} C.~B.,  {Crill} B.~P.,
  {Prunet} S.,    {Hansen} F.,  2002, \apj, 567, 2

\bibitem[\protect\citeauthoryear{H\"ogbom}{H\"ogbom}{1974}]{Hogbom-1974-CLEAN}
H\"ogbom J.~A.,  1974, \aaps, 15, 417

\bibitem[\protect\citeauthoryear{{Hu}}{{Hu}}{2000}]{2000PhRvD..62d3007H}
{Hu} W.,  2000, \prd, 62, 043007

\bibitem[\protect\citeauthoryear{Jee, Ford, Illingworh, White \& et al.}{Jee
  et~al.}{2007}]{Jee:2007p996}
Jee M.,  Ford H.,  Illingworh G.,  White R.,    et al. 2007, The Astrophysical
  Journal

\bibitem[\protect\citeauthoryear{Kilbinger \& Schneider}{Kilbinger \&
  Schneider}{2005}]{Kilbinger:2005p2383}
Kilbinger M.,  Schneider P.,  2005, Arxiv preprint astro-ph

\bibitem[\protect\citeauthoryear{Kitching, Heavens, Taylor, Brown \& et
  al.}{Kitching et~al.}{2006}]{Kitching:2006p542}
Kitching T.,  Heavens A.,  Taylor A.,  Brown M.,    et al. 2006, Arxiv preprint
  astro-ph

\bibitem[\protect\citeauthoryear{Marshall, Hobson, Gull \& Bridle}{Marshall
  et~al.}{2002}]{Marshall:2002p2821}
Marshall P.,  Hobson M.,  Gull S.,    Bridle S.,  2002, Monthly Notices of the
  Royal Astronomical Society

\bibitem[\protect\citeauthoryear{{Massey}}{{Massey}}{2007}]{2007MNRAS.376}
{Massey} e.~a.,  2007, \mnras, 376, 13

\bibitem[\protect\citeauthoryear{Massey, Rhodes, Leauthaud, Capak, Ellis \& et
  al.}{Massey et~al.}{2007}]{Massey:2007p176}
Massey R.,  Rhodes J.,  Leauthaud A.,  Capak P.,  Ellis R.,    et al. 2007,
  Arxiv preprint astro-ph

\bibitem[\protect\citeauthoryear{Mugnier, Fusco \& Conan}{Mugnier
  et~al.}{2004}]{Mugnier_Fusco_etal-2004-JOSAA-Mistral}
Mugnier L.~M.,  Fusco T.,    Conan J.-M.,  2004, 21, 1841

\bibitem[\protect\citeauthoryear{Nocedal \& Wright}{Nocedal \&
  Wright}{2006}]{Nocedal_Wright-2006-numerical_optimization}
Nocedal J.,  Wright S.~J.,  2006, Numerical Optimization, second edition edn.
Springer Verlag

\bibitem[\protect\citeauthoryear{{Novikov}, {Colombi} \& {Dor{\'e}}}{{Novikov}
  et~al.}{2006}]{2006MNRAS.366.1201N}
{Novikov} D.,  {Colombi} S.,    {Dor{\'e}} O.,  2006, \mnras, 366, 1201

\bibitem[\protect\citeauthoryear{{Park}, {Choi}, {Vogeley}, {Gott}, {Kim},
  {Hikage}, {Matsubara}, {Park}, {Suto} \& {Weinberg}}{{Park}
  et~al.}{2005}]{Park}
{Park} C.,  {Choi} Y.-Y.,  {Vogeley} M.~S.,  {Gott} J.~R.~I.,  {Kim} J.,
  {Hikage} C.,  {Matsubara} T.,  {Park} M.-G.,  {Suto} Y.,    {Weinberg} D.~H.,
   2005, \apj, 633, 11

\bibitem[\protect\citeauthoryear{{Pen}}{{Pen}}{2003}]{Pen2003}
{Pen} U.-L.,  2003, \mnras, 346, 619

\bibitem[\protect\citeauthoryear{{Pichon} \& {al.}}{{Pichon} \&
  {al.}}{2009}]{connex}
{Pichon} C.,  {al.} 2009, \mnras \, in prep., pp 000--000

\bibitem[\protect\citeauthoryear{{Pichon} \& {Bernardeau}}{{Pichon} \&
  {Bernardeau}}{1999}]{PB}
{Pichon} C.,  {Bernardeau} F.,  1999, Astronomy and Astrophysics, 343, 663

\bibitem[\protect\citeauthoryear{{Pichon} \& {Thi\'ebaut}}{{Pichon} \&
  {Thi\'ebaut}}{1998}]{1998MNRAS.301..419P}
{Pichon} C.,  {Thi\'ebaut} E.,  1998, \mnras, 301, 419

\bibitem[\protect\citeauthoryear{{Pichon}, {Vergely}, {Rollinde}, {Colombi} \&
  {Petitjean}}{{Pichon} et~al.}{2001}]{2001MNRAS.326..597P}
{Pichon} C.,  {Vergely} J.~L.,  {Rollinde} E.,  {Colombi} S.,    {Petitjean}
  P.,  2001, \mnras, 326, 597

\bibitem[\protect\citeauthoryear{{Pires}, {Starck}, {Amara}, {Teyssier},
  {Refregier} \& {Fadili}}{{Pires} et~al.}{2008}]{pires}
{Pires} S.,  {Starck} J.~.,  {Amara} A.,  {Teyssier} R.,  {Refregier} A.,
  {Fadili} J.,  2008, ArXiv e-prints

\bibitem[\protect\citeauthoryear{{Pogosyan}, {Gay} \& {Pichon}}{{Pogosyan}
  et~al.}{2009}]{PGP}
{Pogosyan} D.,  {Gay} C.,    {Pichon} C.,  2009, ArXiv e-prints

\bibitem[\protect\citeauthoryear{{Pogosyan}, {Pichon}, {Gay}, {Prunet},
  {Cardoso}, {Sousbie} \& {Colombi}}{{Pogosyan} et~al.}{2009}]{skllpogo}
{Pogosyan} D.,  {Pichon} C.,  {Gay} C.,  {Prunet} S.,  {Cardoso} J.~F.,
  {Sousbie} T.,    {Colombi} S.,  2009, \mnras, 396, 635

\bibitem[\protect\citeauthoryear{{Prunet}, {Pichon}, {Aubert}, {Pogosyan},
  {Teyssier} \& {Gottloeber}}{{Prunet} et~al.}{2008}]{prunet2008}
{Prunet} S.,  {Pichon} C.,  {Aubert} D.,  {Pogosyan} D.,  {Teyssier} R.,
  {Gottloeber} S.,  2008, \apjs, 178, 179

\bibitem[\protect\citeauthoryear{{Richardson}}{{Richardson}}{1972}]{1972JOSA..%
.62...55R}
{Richardson} W.~H.,  1972, Journal of the Optical Society of America
  (1917-1983), 62, 55

\bibitem[\protect\citeauthoryear{Schirmer, Erben, Hetterscheidt \&
  Schneider}{Schirmer et~al.}{2007}]{schirmer:2007p920}
Schirmer M.,  Erben T.,  Hetterscheidt M.,    Schneider P.,  2007, Astronomy \&
  Astrophysics, 462

\bibitem[\protect\citeauthoryear{Schneider, van Waerbeke, Kilbinger \&
  Mellier}{Schneider et~al.}{2002}]{2002A&A...396....1S}
Schneider P.,  van Waerbeke L.,  Kilbinger M.,    Mellier Y.,  2002, \aap, 396,
  1

\bibitem[\protect\citeauthoryear{Schwarz}{Schwarz}{1978}]{Schwarz:1978:CLEAN}
Schwarz U.~J.,  1978, \aap, 65, 345

\bibitem[\protect\citeauthoryear{{Seitz}, {Schneider} \& {Bartelmann}}{{Seitz}
  et~al.}{1998}]{ka12}
{Seitz} S.,  {Schneider} P.,    {Bartelmann} M.,  1998, \aap, 337, 325

\bibitem[\protect\citeauthoryear{Seitz, Schneider \& Bartelmann}{Seitz
  et~al.}{1998}]{Seitz:1998p2768}
Seitz S.,  Schneider P.,    Bartelmann M.,  1998, Arxiv preprint astro-ph

\bibitem[\protect\citeauthoryear{{Shapiro}}{{Shapiro}}{2009}]{2009ApJ...696..7%
75S}
{Shapiro} C.,  2009, \apj, 696, 775

\bibitem[\protect\citeauthoryear{{Sheth} \& {Tormen}}{{Sheth} \&
  {Tormen}}{1999}]{ShT}
{Sheth} R.~K.,  {Tormen} G.,  1999, \mnras, 308, 119

\bibitem[\protect\citeauthoryear{{Skilling}, {Strong} \& {Bennett}}{{Skilling}
  et~al.}{1979}]{1979MNRAS.187..145S}
{Skilling} J.,  {Strong} A.~W.,    {Bennett} K.,  1979, \mnras, 187, 145

\bibitem[\protect\citeauthoryear{Soulez, Denis, Thi\'ebaut, Fournier \&
  Goepfert}{Soulez et~al.}{2007}]{Soulez_et_al-2007-out_of_field_detection}
Soulez F.,  Denis L.,  Thi\'ebaut E.,  Fournier C.,    Goepfert C.,  2007, J.
  Opt. Soc. Am. A, 24, 3708

\bibitem[\protect\citeauthoryear{{Sousbie}, {Pichon} \& {Colombi}}{{Sousbie}
  et~al.}{2008}]{rsex}
{Sousbie} T.,  {Pichon} C.,    {Colombi} 2008, \mnras \,, pp 000--000

\bibitem[\protect\citeauthoryear{{Sousbie}, {Pichon}, {Colombi}, {Novikov} \&
  {Pogosyan}}{{Sousbie} et~al.}{2007}]{2007arXiv0707.3123S}
{Sousbie} T.,  {Pichon} C.,  {Colombi} S.,  {Novikov} D.,    {Pogosyan} D.,
  2007, ArXiv e-prints, 707

\bibitem[\protect\citeauthoryear{{Sousbie}, {Pichon}, {Courtois}, {Colombi} \&
  {Novikov}}{{Sousbie} et~al.}{2006}]{skel1}
{Sousbie} T.,  {Pichon} C.,  {Courtois} H.,  {Colombi} S.,    {Novikov} D.,
  2006, ArXiv Astrophysics e-prints

\bibitem[\protect\citeauthoryear{Starck, Pires \& Refregier}{Starck
  et~al.}{2005}]{Starck:2005p2878}
Starck J.,  Pires S.,    Refregier A.,  2005, Arxiv preprint astro-ph

\bibitem[\protect\citeauthoryear{{Szapudi}, {Prunet} \& {Colombi}}{{Szapudi}
  et~al.}{2001}]{spice}
{Szapudi} I.,  {Prunet} S.,    {Colombi} S.,  2001, \apjl, 561, L11

\bibitem[\protect\citeauthoryear{Takada \& Jain}{Takada \&
  Jain}{2002}]{Takada:2002p5791}
Takada M.,  Jain B.,  2002, Monthly Notice of the Royal Astronomical Society,
  337, 875

\bibitem[\protect\citeauthoryear{Takada \& Jain}{Takada \&
  Jain}{2003}]{Takada:2003p2287}
Takada M.,  Jain B.,  2003, Monthly Notices of the Royal Astronomical Society

\bibitem[\protect\citeauthoryear{Takada \& Jain}{Takada \&
  Jain}{2004}]{Takada:2004p5780}
Takada M.,  Jain B.,  2004, Monthly Notices of the Royal Astronomical Society,
  348, 897

\bibitem[\protect\citeauthoryear{{Tarantola} \& {Valette}}{{Tarantola} \&
  {Valette}}{1982}]{1982RvGSP..20..219T}
{Tarantola} A.,  {Valette} B.,  1982, Reviews of Geophysics and Space Physics,
  20, 219

\bibitem[\protect\citeauthoryear{{Teyssier}}{{Teyssier}}{2002}]{2002A&A...385.%
.337T}
{Teyssier} R.,  2002, \aap, 385, 337

\bibitem[\protect\citeauthoryear{{Teyssier}, {Pires}, {Prunet}, {Aubert},
  {Pichon}, {Amara}, {Benabed}, {Colombi}, {Refregier} \& {Starck}}{{Teyssier}
  et~al.}{2008}]{teyssier2008}
{Teyssier} R.,  {Pires} S.,  {Prunet} S.,  {Aubert} D.,  {Pichon} C.,  {Amara}
  A.,  {Benabed} K.,  {Colombi} S.,  {Refregier} A.,    {Starck} J.-L.,  2008,
  ArXiv e-prints

\bibitem[\protect\citeauthoryear{Thi\'ebaut}{Thi\'ebaut}{2002}]{Thiebaut:spie2%
002:bdec}
Thi\'ebaut E.,  2002, in Starck J.-L.,  Murtagh F.~D.,  eds, Astronomical Data
  Analysis {II} Vol.~4847, Optimization issues in blind deconvolution
  algorithms.
pp 174--183

\bibitem[\protect\citeauthoryear{{Thi{\'e}baut}}{{Thi{\'e}baut}}{2005}]{2005op%
as.conf..397T}
{Thi{\'e}baut} E.,  2005, in {Foy} R.,  {Foy} F.~C.,  eds, NATO ASIB Proc. 198:
  Optics in astrophysics {Introduction to Image Reconstruction and Inverse
  Problems}.
pp 397--+

\bibitem[\protect\citeauthoryear{{van Waerbeke}, {Bernardeau} \&
  {Mellier}}{{van Waerbeke} et~al.}{1999}]{ka11}
{van Waerbeke} L.,  {Bernardeau} F.,    {Mellier} Y.,  1999, \aap, 342, 15

\bibitem[\protect\citeauthoryear{{Wahba}}{{Wahba}}{1990}]{1990smod.conf.....W}
{Wahba} G.,  ed. 1990, {Spline models for observational data}

\bibitem[\protect\citeauthoryear{{White}}{{White}}{2005}]{White}
{White} M.,  2005, Astroparticle Physics, 23, 349

\end{thebibliography}

\appendix
\section{Efficient Star Removal}\label{sec:Efficient-Star-Removal}
We have observed that for realistic deep field images, generalized
cross validation (GCV) yields an hyper-parameter value which is relevant
to regularize the higher part of the dynamic (mainly due to stars,
i.e. point-like objects which concentrate their luminous energy in
a very small area) but which is much too low to regularize the lower
parts ot the dynamic where galaxies remain. \textcolor{black}{Indeed,
when dealing with images with a large dynamical range, GCV yields
a value of the regularization level $\mu$ which is necessarily a
compromise between not smoothing too much the sharp features and sufficient
smoothing of low contrasted structures to avoid noise amplification.
}The solution to the problem of underestimating the regularization
weight can be solved by applying the GCV method onto the image with
no stars. We want to find structures of known shape $s(x)$ but unknown
position and intensity in the image $\V y$. In our case, $s(x)$
is the PSF since we want to detect stars. This reasoning could however
be generalized to other kind of objects. If a single object of this
shape is present in the image, this could be achieved by considering
the objective function:\[
\phi_{\mathrm{full}}(\alpha,t)=\sum_{k}w_{k}\,[\alpha\, s(x_{k}-t)-y_{k}]^{2}\]
to be minimized w.r.t. the weight $\alpha$ and the offset $t$, here
a 2-D vector. In fact, since $\V y$ may be crowded with similar structures
(or with other fainter structures), a better strategy is to limit
the local fit to a small region of interest (ROI) around the structure.
This is achieved by minimizing:\[
\phi(\alpha,t)=\sum_{k}w_{k}\, r(x_{k}-t)\,[\alpha\, s(x_{k}-t)-y_{k}]^{2}\,,\]
where $r(\delta\V x)$ is equal to 1 within the region of interest
(ROI) and equal to 0 outside the ROI. Minimization of $\phi(\alpha,t)$
w.r.t. $\alpha$ yields the best intensity for a local fit around
$t$:\begin{equation}
\frac{\partial\phi}{\partial\alpha}=0\quad\Longleftrightarrow\quad\alpha^{\star}=\frac{\sum_{k}w_{k}\, r(x_{k}-t)\, s(x_{k}-t)\, y_{k}}{\sum_{k}w_{k}\, r(x_{k}-t)\, s(x_{k}-t)^{2}}\,.\label{e:best-intensity}
\nonumber \end{equation}
Inserting $\alpha^{\star}$ in the objective function yields:
\begin{eqnarray*}
\phi^{\star}(t) \!\!\! & \triangleq & \left.\phi(\alpha,t)\right|_{\alpha=\alpha^{\star}}\,,\\
\!\!\! & = & \hskip -0.25cm \sum_{k}w_{k}\, r(x_{k}-t)\, y_{k}^{2}
% \\&& 
  -\frac{\left(\sum_{k}w_{k}\, r(x_{k}-t)\, s(x_{k}-t)\, y_{k}\right)^{2}}{\sum_{k}w_{k}\, r(x_{k}-t)\, s(x_{k}-t)^{2}}\,. \nonumber \end{eqnarray*}
Since $r(\delta\V x)^{2}=r(\delta\V x)$, by defining $s_{\RM{ROI}}(\delta\V x)\equiv r(\delta\V x)\, s(\delta\V x)$,
the local criterion and local best intensity can be rewritten as:\begin{eqnarray*}
\phi^{\star}(t) & = & \sum_{k}r(x_{k}-t)\, w_{k}\, y_{k}^{2}-\frac{\left(\sum_{k}s_{\RM{ROI}}(x_{k}-t)\, w_{k}\, y_{k}\right)^{2}}{\sum_{k}s_{\RM{ROI}}(x_{k}-t)^{2}\, w_{k}}\,,\\
\alpha^{\star}(t) & = & \frac{\sum_{k}s_{\RM{ROI}}(x_{k}-t)\, w_{k}\, y_{k}}{\sum_{k}s_{\RM{ROI}}(x_{k}-t)^{2}\, w_{k}}\,.\end{eqnarray*}
These parameters can be computed for all shifts by an integer number
of pixels by means of FFT's (cross-correlation product). Unfortunately,
the overall minimum of $\phi^{\star}(t)$ is not the best choice for
removing the brightest structures since there is no warranty that
this minimum corresponds to a bright object. It is better to select
the the location which yields the brightest structure, i.e. the maximum
of $\alpha^{\star}(t)$.
\textcolor{black}{After removal of the contribution $\alpha^{\star}(t^{\star})\, s(x-t^{\star})$
from the data, this technique can be repeated to detect the second
brightest source, and so on. The corresponding algorithm is very similar
to the }\textcolor{black}{\noun{clean}\,}\textcolor{black}{ method \citep{Hogbom-1974-CLEAN,Schwarz:1978:CLEAN}
with the further refinement of accounting for non-stationary noise
and missing data. It has been shown that it achieves sub-pixel precision
\citep{Soulez_et_al-2007-out_of_field_detection} and that it could
be used to detect (and remove) out of field sources \citep{Soulez_et_al-2007-out_of_field_detection}.}
\section{Model on the sphere}\label{sec:Detailled-model} 
Let us describe in more details the model used for the inversion of
Section \ref{sec:Inverse-problem}.
\begin{figure*}
\includegraphics[width=0.8\textwidth]{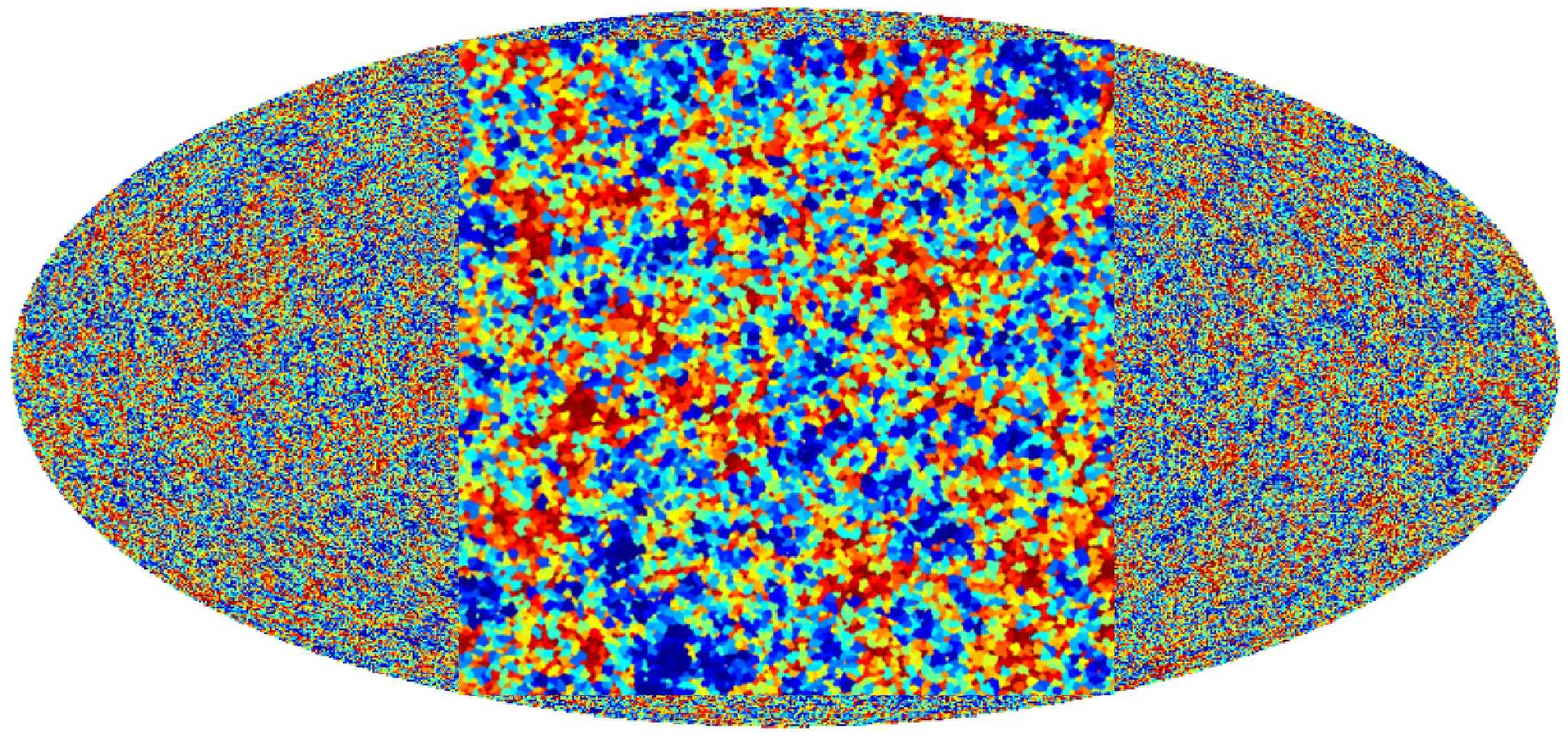}
\caption{Peak patch of the recovered $\kappa$ map. The inner box zooms the
central region. The color coding corresponds loosely to the density
of the different peak patches. The PDF of the area of these patches
is described in Figure \ref{fig:Point source} while the maxima mentioned
in this figure are found within each patch. \label{fig:Peak}}
\end{figure*}
\subsection{Discretization and Sampling}
After discretization and using explicit indices, the model in equation~(\ref{eq:model})
writes:\[
g_{j,k}=\frac{\gamma_{j,k}}{1-\kappa_{j}}+e_{j,k}\,,\]
where the index $j$ runs over the sky coordinates $\hat{\V n}_{j}=(x_{j},y_{j})$,
index $k$ corresponds to the two components $U$ and $\Qmap$ of
the polarization, whereas $\ell$ and $m$ are the harmonic indices
and $p$ refers to the two components of the spinned 2-harmonic. In
words, the discretization yields:
\[
g_{j,k}\equiv g_{k}(\hat{\V n}_{j}),\,\,\,\gamma_{j,k}\equiv\gamma_{k}(\hat{\V n}_{j}),\,\,\,\kappa\equiv\kappa(\hat{\V n}_{j}),\,\,\, e_{j,k}\equiv e_{k}(\hat{\V n}_{j})\,.\]
Here the fields $\V{\kappa}$ and $\V{\gamma}$ are linear functions
of the complex field $\V a$ of the spherical harmonic coefficients
of $\V{\kappa}$. Using the matrix notation of the paper, $\V{\kappa}$
and $\V{\gamma}$ write:\[
\V{\kappa}=\M K\cdot\V a\,,\quad\V{\gamma}=\M G\cdot\V a\,,\]
where $\M K=\mathbf{Y}$ and $\M G=\Spin{\mathbf{Y}}\cdot\M J$;
with explicit index notations:
\[
\kappa_{j}=\sum_{\ell,m}K_{j,\ell,m}\, a_{\ell,m}=\sum_{\ell,m}\mathbf{Y}{}_{j,\ell,m}\, a_{\ell,m}\,,\]
and \begin{eqnarray*}
\gamma_{j,k} & = & \sum_{\ell,m}G_{j,k,\ell,m}\, a_{\ell,m}
%\\& = & 
= \sum_{\ell,m,p}\Spin{\mathbf{Y}}^{}{}_{j,k,\ell,m,p}\,\left(\M
  J\cdot\V a\right)_{\ell,m,p}\,,\end{eqnarray*}

To get the detailed expression of the operator $\M J$ we start from
the relationship between the lensing potential, the convergence and
the shear fields on the sphere. To do this, we need first to define
the null diad, based on the polar coordinates unit vectors:
\begin{equation}
\M m_{\pm}=\frac{(\hat{\M e}_\theta \mp i \hat{\M e}_\phi)}{\sqrt{2}}\,.
\end{equation}
Given this diad, the lensing potential, convergence and shear are related through:
\begin{eqnarray*}
\nabla_i\nabla_j\Phi &=& \kappa g_{ij}
 %\\&+&
 +
 (\gamma_1+i\gamma_2)(\M m_+\otimes\M m_+)_{ij} \\
&+& (\gamma_1-i\gamma_2)(\M m_-\otimes\M m_-)_{ij}\,,
\end{eqnarray*}
where $g_{ij}$ is the spherical metric tensor, and $\nabla$ the spherical covariant derivative. 
Now, using the following expression of the second covariant derivative of a scalar spherical
harmonic:
\begin{eqnarray*}
\nabla_i\nabla_j\M Y_{\ell,m} &=& {1\over 2}\sqrt{(l+2)!\over(l-2)!}\left[\frac{}{} {}_2\M Y_{\ell,m}(\M m_+\otimes\M m_+) \right.\\
&&\!\!\!\!\!\! +\left. {}_{-2}\M Y_{\ell,m}(\M m_-\otimes\M m_-)\frac{}{}\right]_{ij} -{\ell(\ell+1)\over 2}\M Y_{\ell,m}g_{ij}\,, 
\end{eqnarray*}
we can relate the convergence and shear fields to the spherical harmonic coefficients $\Phi_{\ell m}$ of the lensing potential:
\begin{eqnarray}
\kappa(\hat{n}) &=& -\sum\limits_{\ell m}{1\over
  2}\ell(\ell+1)\Phi_{\ell,m}\M Y_{\ell,m}(\hat{n})\,, \\
(\gamma_1\pm i\gamma_2)(\hat{n}) &=& \sum\limits_{\ell m} {1\over 2}\sqrt{(l+2)!\over(l-2)!}
\Phi_{\ell,m} {}_{\pm 2}\M Y_{\ell,m}(\hat{n})\,, \\
&=&\!\!\!\! -\sum\limits_{\ell m}(a_{\ell,m,E}\pm ia_{\ell,m,B}){}_{\pm 2}\M Y_{\ell,m}(\hat{n})\,,
\end{eqnarray}
where the last equality defines the $E$ and $B$ modes coefficients.
Relating the latter coefficients to the spherical harmonics decomposition of $\kappa$ written above,
we get the following expression for the $\M J$ operator coefficients:
\begin{eqnarray}
\left(\M J\cdot\V a\right)_{\ell,m,E} & = & \sqrt{\frac{(\ell+2)(\ell-1)}{(\ell+1)\ell}}\, a_{\ell,m}\,,\label{eq:defF1a}\\
\left(\M J\cdot\V a\right)_{\ell,m,B} & = & 0\,,\label{eq:defF2a}
\end{eqnarray}
where $a_{\ell,m}$ are the spherical harmonics coefficients of the convergence field.

\subsection{Likelihood}
The data related term in the cost function is\[
\Qdata=\sum_{j,k}w_{j,k}\,\left(\frac{\gamma_{j,k}}{1-\kappa_{j}}-g_{j,k}\right)^{2}\,.\]
The gradient of this term is needed to find the solution of the inverse
problem: \begin{eqnarray}
%\frac{\partial\Qdata(\V a)}{\partial a_{\ell,m}} & = & 2\,\sum_{j,k}H_{\ell,m,j,k}\,\frac{r_{j,k}}{1-\kappa_{j}}+
%\,\nonumber \\ &  &
% 2\,\sum_{j}\,\M Y_{\ell,m,j}^{*}\frac{\sum_{k}r_{j,k}}{\left(1-\kappa_{j}\right)^{2}}\,,\label{eq:grad-chi2}
\frac{\partial\Qdata(\V a)}{\partial a_{\ell,m}} \!\!\! & = & \!\!\!\! 2\,\sum_{j,k}H_{\ell,m,j,k}\,\frac{r_{j,k}}{1-\kappa_{j}}+
%\,\nonumber \\ &  &
 2\,\sum_{j}\,\M Y_{\ell,m,j}^{*}\frac{\sum_{k}\gamma^*_{j,k}r_{j,k}}{\left(1-\kappa_{j}\right)^{2}}\,,\label{eq:grad-chi2} \nonumber
 \end{eqnarray}
where\[
r_{j,k}=w_{j,k}\,\left(\frac{\gamma_{j,k}}{1-\kappa_{j}}-g_{j,k}\right)\]
are the weighted residuals, and where\begin{equation}
%H_{\ell,m,j,k}=\Spin{\M Y}_{\ell,m,1,j,k}^{*}+\sqrt{\frac{(\ell+2)(\ell-1)}{(\ell+1)\ell}}\,\Spin{\M Y}_{\ell,m,2,j,k}^{*}\,.\label{eq:H-operator}\end{equation}
H_{\ell,m,j,k}=\sqrt{\frac{(\ell+2)(\ell-1)}{(\ell+1)\ell}}\,\Spin{\M Y}_{\ell,m,1,j,k}^{*}\,.\label{eq:H-operator}\end{equation}
%The \emph{exact} harmonic transform has the properties that:\[
%\left(\M Y^{-1}\right)^{\mathrm{H}}=\frac{n_{\mathrm{pix}}}{2\,\pi}{\M Y}\,,\quad\quad\left(\Spin{\M Y}^{-1}\right)^{\mathrm{H}}=\frac{n_{\mathrm{pix}}}{2\,\pi}\,\Spin{\M Y}\,,\]
%where the H exponent denotes the conjugate transpose of the linear
%operator. Remarking that $\left(\M Y^{-1}\right)^{\mathrm{H}}$ and
%$\left(\Spin{\M Y}^{-1}\right)^{\mathrm{H}}$ are the operators that
%appear in equation~(\ref{eq:grad-chi2}) and equation~(\ref{eq:H-operator}).
%These properties can be used to compute the gradient of $\Qdata$.
%
\subsection{Regularization}
The aim of the regularization is to avoid ill-conditioning and noise
amplification in the inversion. Following a  Bayesian prescription, this
can be achieved by requiring the field $\V{\kappa}$ to obey some
known \emph{a priori} statistics, or while assuming a roughness penalty for $\Qprior$.
\subsubsection{Wiener filter and $\ell_{2}$ penalty}
%
%\Xtophe{Eric: est ce qu'on peut rendre cette section un tout petit  peu moins verbieuse ?}
Assuming the field $\V{\kappa}$ has Gaussian distribution with mean
$\bar{\V{\kappa}}=\avg{\V{\kappa}}$ and covariance $\M C_{\V{\kappa}}=\avg{\left(\V{\kappa}-\bar{\V{\kappa}}\right)\cdot\left(\V{\kappa}-\bar{\V{\kappa}}\right)\TR}$,
the prior penalty should write:
\[\mu\equiv 1 \quad {\rm and}\quad
\Qprior=\left(\V{\kappa}-\bar{\V{\kappa}}\right)\TR\cdot\M C_{\V{\kappa}}^{-1}\cdot\left(\V{\kappa}-\bar{\V{\kappa}}\right)\,.\]
For a field with zero mean ($\bar{\V{\kappa}}=\V 0$) and stationary
isotropic statistics, the regularization can be expressed in terms
of the harmonic coefficients:
\begin{equation}
\Qprior(\V a)  =  \Norm{\M C^{-1/2}\cdot\V a}^{2}\label{eq:stationnary-prior}
  =  \sum_{\ell}\frac{\sum_{m}\Abs{a_{\ell,m}}^{2}}{C_{\ell}}\,,\end{equation}
with\begin{equation}
C_{\ell}=\Avg{\Abs{a_{\ell,m}}}^{2}\,,\label{eq:stationnary-covariance}\end{equation}
where the angular brackets denote here the expected value taken over
the index $m$ of the harmonic coefficients. The gradient of the stationnary
isotropic Gaussian regularization in equation~(\ref{eq:stationnary-prior})
is:\[
\frac{\partial\Qprior(\V a)}{\partial a_{\ell,m}}=2\,\frac{a_{\ell,m}}{C_{\ell}}\,.\]
Note that the regularization in equation~(\ref{eq:stationnary-prior})
with a known power spectrum $C_{\ell}$ for the field $\V{\kappa}$
yields the so-called Wiener filter. 

When the power spectrum of $\V{\kappa}$
is not exactly known, a quadratic prior can alternatively be used. For instance:\begin{equation}
\Qprior(\V a)= %\lambda\,
\Norm{\M R^{-1/2}\cdot\V a}^{2}=
%\lambda\,
\sum_{\ell}\frac{\sum_{m}\Abs{a_{\ell,m}}^{2}}{R_{\ell}}\,.\label{eq:quadratic-prior}\end{equation}
%where $\lambda\ge0$ is a regularization parameter that must be properly
% tuned. 
In our framework, effective regularization is achieved by requiring
the field $\V{\kappa}$ to be somewhat smooth. In practice, this is
obtained by requiring $R_{\ell}$ to be a positive non-decreasing
function of the index $\ell$. Note that, from a Bayesian viewpoint,
the regularization in equation~(\ref{eq:quadratic-prior}) corresponds
to the prior that $\V{\kappa}$ is a stationary isotropic centered
Gaussian field with mean power spectrum  %$C_{\ell}=R_{\ell}/\lambda$,
$C_{\ell}=R_{\ell}$,
which is similar to the Wiener filter except that the exact statistics
is not known in advance (because some parameters of the regularization
have to be tuned; for instance, $\mu$ need not be equal to one). The gradient of $\Qprior$ in equation~(\ref{eq:quadratic-prior})
reads:\[
\frac{\partial\Qprior(\V a)}{\partial a_{\ell,m}}=2\,
%\lambda\,
\frac{a_{\ell,m}}{R_{\ell}}\,.\]
The quadratic prior in equation~(\ref{eq:quadratic-prior}) can be
expressed in terms of $\V{\kappa}$:\[
\Qprior= %\lambda\,
\Norm{\M R^{-1/2}\cdot\V a}^{2}=
%\lambda\,
\Norm{\M D\cdot\V{\kappa}}^{2}\,,\]
where $\M D=\M R^{-1/2}\cdot\M Y^{\#}$ is some finite difference
operator which gives an estimate of the local fluctuation of the
field, and $\mathbf{Y}^{\#}$ is the (pseudo-)inverse of the scalar
spherical harmonics matrix. In our framework, we choose to measure
the amplitude of the local fluctuations of the field $\V{\kappa}$
by its Laplacian $\nabla^{2}\V{\kappa}$ and to express the regularization
penalty as:\begin{equation}
\Qprior(\V a)= %\lambda\,
\sum_{j}\phi\left(\left(\nabla^{2}\V{\kappa}\right)_{j}\right)\label{eq:laplacian-prior}\,,
\end{equation}
where the cost function $\phi(r)$ is an increasing function of $\abs r$.
When $\phi(r)=r^{2}$, our regulariztion is a quadratic penalty similar
to equation~(\ref{eq:quadratic-prior}). Using matrix notation, the
Laplacian of the field $\V{\kappa}$ write :\begin{equation}
\nabla^{2}\V{\kappa}=\mathbf{Y}\cdot\M L^{-1/2}\cdot\V a\,,\label{eq:laplacian_of_kappa}
\quad
{\rm with} \quad
\left(\M L^{-1/2}\cdot\V a\right)_{\ell,m}=\frac{a_{\ell,m}}{\sqrt{L_{\ell}}}\,, \nonumber\end{equation}
where $L_{\ell}\equiv\ell^{-2}(\ell+1)^{-2}$. In order to perform
the minimization, the gradient of the regularization must be computed.
By the chain rule:\begin{eqnarray}
\frac{\partial\Qprior(\V a)}{\partial a_{\ell,m}} & = & %\lambda\,
\sum_{j}\phi'\left(\left(\nabla^{2}\V{\kappa}\right)_{j}\right)\,\frac{\partial\left(\nabla^{2}\V{\kappa}\right)_{j}}{\partial a_{\ell,m}}\,,\nonumber \\
 & = & % \lambda
 \sum_{j}\frac{\M Y_{\ell,m,j}^{*}}{\sqrt{L_{\ell}}}\,\phi'\left(\left(\nabla^{2}\V{\kappa}\right)_{j}\right),\label{eq:regul-grad}\end{eqnarray}
where $\phi'(r)$ is the derivative of $\phi(r)$.
\subsubsection{$\ell_{2}-\ell_{1}$ penalty}\label{sec:L1-L2-kappa}
As for the image restoration, quadratic regularization yields spuriours
ripples in the regularized $\V{\kappa}$ map. To avoid them, we propose
to use a $\ell_{2}-\ell_{1}$ cost function $\phi$ applied to the
Laplacian of $\V{\kappa}$. The details of the $\ell_{2}-\ell_{1}$
cost function are discussed in section \ref{sec:L2-L1-penalty}.Taking
$\Qprior(\V a)=\sum_{j}\phi\left(\left(\nabla^{2}\V{\kappa}\right)_{j}\right)$,
with $\phi$ given in equation (\ref{eq:l2-l1-norm}), yields:\begin{eqnarray*}
\frac{\partial\Qprior(\V a)}{\partial a_{\ell,m}} & = & \sum_{j}\frac{2\,\varepsilon\,\left(\nabla^{2}\V{\kappa}\right)_{j}}{\varepsilon+\Abs{\left(\nabla^{2}\V{\kappa}\right)_{j}}}\,\frac{\partial\left(\nabla^{2}\V{\kappa}\right)_{j}}{\partial a_{\ell,m}}\,,\\
 & = & 2\,\varepsilon\,\sum_{j}\frac{\M Y_{\ell,m,j}^{*}}{\sqrt{C_{\ell}}}\,\frac{\left(\nabla^{2}\V{\kappa}\right)_{j}}{\varepsilon+\Abs{\left(\nabla^{2}\V{\kappa}\right)_{j}}}\,.\end{eqnarray*}
{In practice, we use GCV to set the level of the
regularization, possibly after cluster removal (as explained in Appendix~\ref{sec:Efficient-Star-Removal})
and the $\ell_{1}-\ell_{2}$ threshold is set to be $\varepsilon=\alpha\,\sigma$
where $\alpha\sim2-3$ and $\sigma$ is the standard deviation of
the histogram of spatial finite differences.}

\section{From the sphere to the plane}\label{sec:From-the-sphere}
In section~\ref{sec:flat-sky} we sketched the correspondence between the fullsky and the 
flat sky approximation of the lens equation. Let us derive it here precisely and use it to investigate the 
effect of shot noise in the estimation of $\kappa$.
\subsection{Derivation}\label{sec:sphere-plane}
Following closely \citet{2000PhRvD..62d3007H}, let us start with
a scalar field on the sphere, and its decomposition on the usual spherical
harmonics:\begin{equation}
X(\hat{n})=\sum_{\ell m}X_{\ell,m}\M Y_{\ell,m}\,,\label{eq:defXY}\end{equation}
and let us define\begin{equation}
X({\bf l})=\sqrt{\frac{4\pi}{2\ell+1}}\sum_{m}i^{-m}X_{\ell,m}e^{im\phi_{\ell}}\,,\label{eq:Fourier}\end{equation}
together with the inverse relation\[
X_{\ell,m}=\sqrt{\frac{2\ell+1}{4\pi}}i^{m}\int\frac{d\phi_{\ell}}{2\pi}X({\bf l})e^{-im\phi_{l}}\,,\]
where $\phi_{\ell}$ is the polar angle of the ${\bf l}$ vector in Fourier
space. Let us show that $X({\bf l})$ corresponds to the Fourier
decomposition of the field in the flat-sky limit (small angles near
the pole). Indeed, taking the asymptotic behavior of the spherical
harmonics\[
\M Y_{\ell,m}\approx J_{m}(\ell\theta)\sqrt{\frac{\ell}{2\pi}}e^{im\phi}\,,\]
together with the plane-wave expansion in terms of Bessel functions\begin{equation}
e^{i{\bf
    l}.\hat{n}}=\sum_{m}i^{m}J_{m}(\ell\theta)e^{im(\phi-\phi_{\ell})}\approx\sqrt{\frac{2\pi}{\ell}}\sum_{m}i^{m}\M
Y_{\ell,m}e^{im\phi_{l}}\,. \nonumber
\label{eq:planewave}
\end{equation}
We get from equation (\ref{eq:defXY})\begin{eqnarray*}
X(\hat{n}) & \approx & \sum_{\ell}\frac{\ell}{2\pi}\int\frac{d\phi_{\ell}}{2\pi}X({\bf l})\sum_{m}J_{m}(\ell\theta)i^{m}e^{im(\phi-\phi_{\ell})}\,,\\
 & \approx & \int\frac{d^{2}\ell}{(2\pi)^{2}}X({\bf l})e^{i{\bf l}.\hat{n}}\,.\end{eqnarray*}
For a spin-2 field, let us proceed likewise. We start from the all-sky
definition of a spin-2 tensor field, and its decomposition in spin-2
spherical harmonics:\begin{equation}
_{\pm}X(\hat{n})=\sum_{\ell m}{}_{\pm}X_{\ell,m}{}_{\pm2}\M Y_{\ell,m}\,,\label{eq:defXYspin}\end{equation}
where $_{\pm}X(\hat{n})$ is defined in the spherical tangent
coordinates $e_{\theta},e_{\phi}$. We define, as in equation (\ref{eq:Fourier}),
the Fourier modes of the components of the spin-2 field as $_{\pm}X({\bf l})$.
We have in the flat-sky limit the following asymptotic form for the
spin-2 spherical harmonics:\begin{equation}
_{\pm2}\M
Y_{\ell,m}\approx\frac{1}{\ell^{2}}e^{\mp2i\phi}(\partial_{x}\pm
i\partial_{y})^2\M Y_{\ell,m}\,.\label{eq:Ylm2}\end{equation}
Plugging equation (\ref{eq:defXYspin}) into equation (\ref{eq:Ylm2})
yields:\begin{eqnarray*}
_{\pm}X(\hat{n}) & \approx &
\sum_{\ell}\frac{\ell}{2\pi}\int\frac{d\phi_{\ell}}{2\pi}X({\bf l})e^{\mp2i\phi}\frac{1}{\ell^{2}}(\partial_{x}\pm
i\partial_{y})^{2}e^{i{\bf l}\cdot\hat{n}}\,,\\
 & \approx & -\int\frac{d^{2}\ell}{(2\pi)^{2}}{}_{\pm}X({\bf l})e^{\pm2i(\phi_{\ell}-\phi)}e^{i{\bf l}\cdot\hat{n}}\,.\end{eqnarray*}
Redefining the spin-2 field in the fixed coordinate system such that
the first axis ($e_{x}$) is aligned with $\phi=0$, we obtain:\begin{equation}
_{\pm}X'(\hat{n})\approx-\int\frac{d^{2}\ell}{(2\pi)^{2}}{}_{\pm}X({\bf l})e^{\pm2i\phi_{\ell}}e^{i{\bf l}\cdot\hat{n}}\,,\label{eq:EB-def}\end{equation}
where $\ell_{x}+i\ell_{y}=\ell e^{i\phi_{\ell}}$. Expanding
$_{\pm}X({\bf l})=E({\bf l})\pm iB({\bf l})$,
we can relate these rotationally invariant quantities to the Fourier
transforms of the spin-2 field individual components. In the case of
weak lensing, we get the following flat sky limits:
\begin{eqnarray}
\kappa(\hat{n}) &\approx& -{1\over 2}\int{{\rm d}^2\ell\over (2\pi)^2} 
\ell^2\Phi({\bf l})e^{i{\bf l} . \hat{n}} \\
(\gamma_1\pm i\gamma_2)'(\hat{n}) &\approx& -{1\over 2}
\int{{\rm d}^2\ell\over (2\pi)^2} \ell^2\Phi({\bf l})
e^{\pm 2i\phi_\ell} e^{i{\bf l} . \hat{n}}
\end{eqnarray}
After identification, we thus get the limits for the operator $\M J$:
\begin{equation}
\M J = (\M 1,\M 0)
\end{equation}
independently of the Fourier mode modulus.

\subsection{ SNR investigation in the plane}\label{sec:noise}

\begin{figure}
\includegraphics[width= \columnwidth]{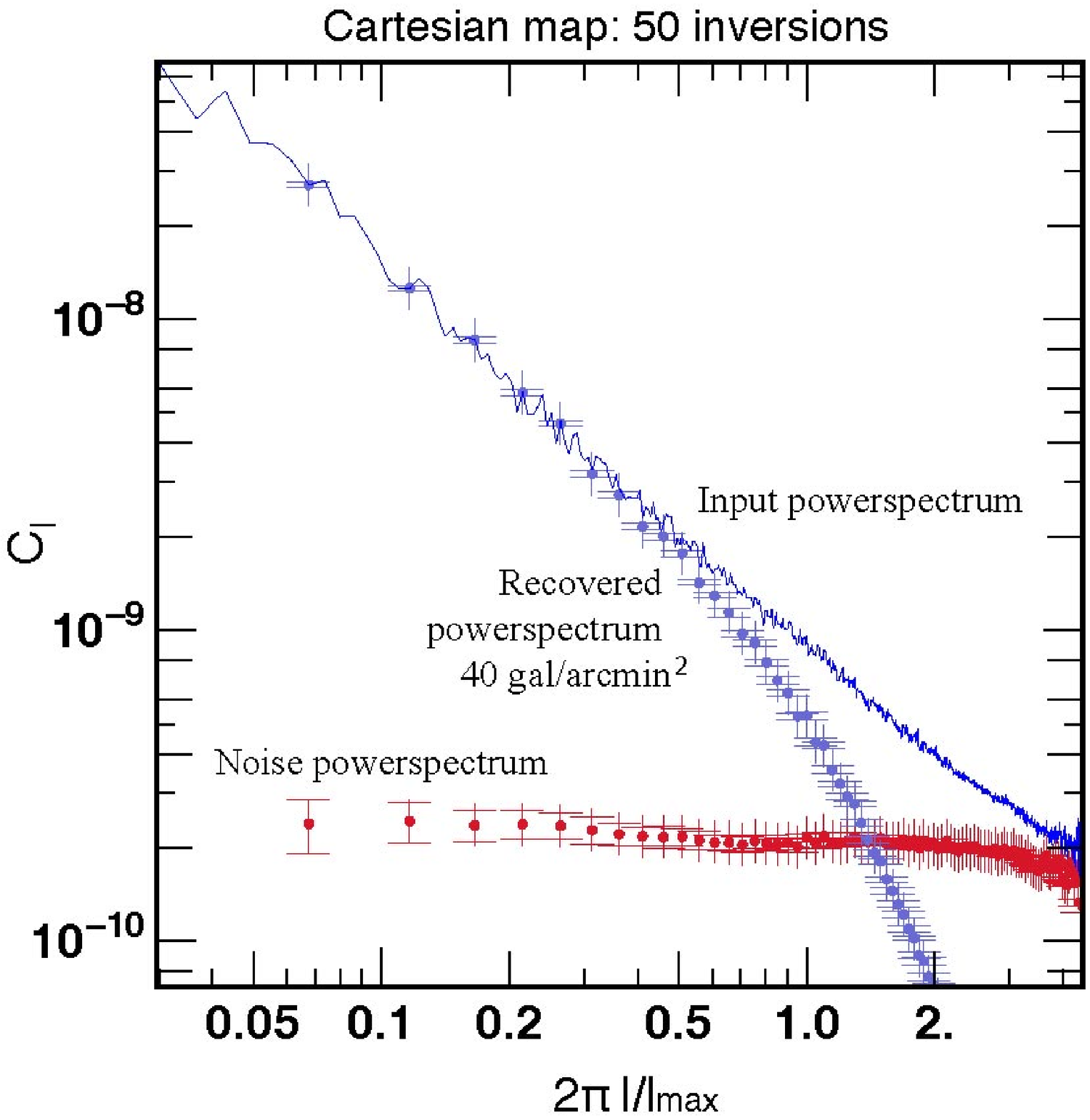}
\caption{The effect of noise on the reconstruction of the powerspectrum for a set of  50 realizations of the noise for the map  
$_{\rm 1024}C_{{\rm lin}}^{\ell}$ ($\ell_{\rm max}=1200$). Note that at these scales, the spread in the recovered powerspectra for the
 different realizations is only visible above the cutoff frequency.
}
\label{fig:stark-noise}
\end{figure}
Let us briefly investigate the effect of noise on the recovery of the $\kappa$ powerspectrum arising from the finite number of 
sources per unit area. For this purpose, let us consider the simplest setting corresponding to a cartesian map without mask
which can therefore be inverted linearly following equations~(\ref{eq:yeblocal})-(\ref{eq:local-shear}). 
In this regime, the regularized solution is simply given in Fourier space  by
\be
\hat{\kappa}=\frac{1}{1+\mu(\ell_{x}^{2}+\ell_{y}^{2})}\left[\hat{g}_{x}\frac{(\ell_{
x}^{2}-\ell_{y}^{2})}{(\ell_{x}^{2}+\ell_{y}^{2})}+\frac{2\ell_{x}\ell_{y}}{(\ell_{x}^{2}+\ell_{y}^{2
})}\hat{g}_{y}\right]\,,
\ee
where $\hat g_x$,  $\hat g_y$ and $\hat \kappa$ are the Fourier transform of the observed shear and convergence, and 
$\mu$ the penalty hyperparameter.
In 
Figure~\ref{fig:stark-noise} we make use of the simulation $_{\rm 1024}C_{{\rm lin}}^{\ell}$, whose residues  (after non linear inversion) are 
shown in Figure~(\ref{Non linear}). Here
 50  Monte Carlo realizations of the noise corresponding 40 galaxies/$\square$arcmin are averaged to produce an estimate of 
 the corresponding errors. Clearly the shot noise remains small at all considered frequencies.cpp

\section{ Convergence maps}\label{sec:simu-2-map}

The inversion technique described in the main text was validated using the mocks extracted from the 
\noun{horizon}-4$\pi$ simulation \citep{prunet2008}. Let us briefly describe here how this simulation was used to generate mock slices and 
$\kappa$ maps.
\subsection{Light cone generation}\label{sec:lightcone}
\begin{figure}
\includegraphics[width= \columnwidth]{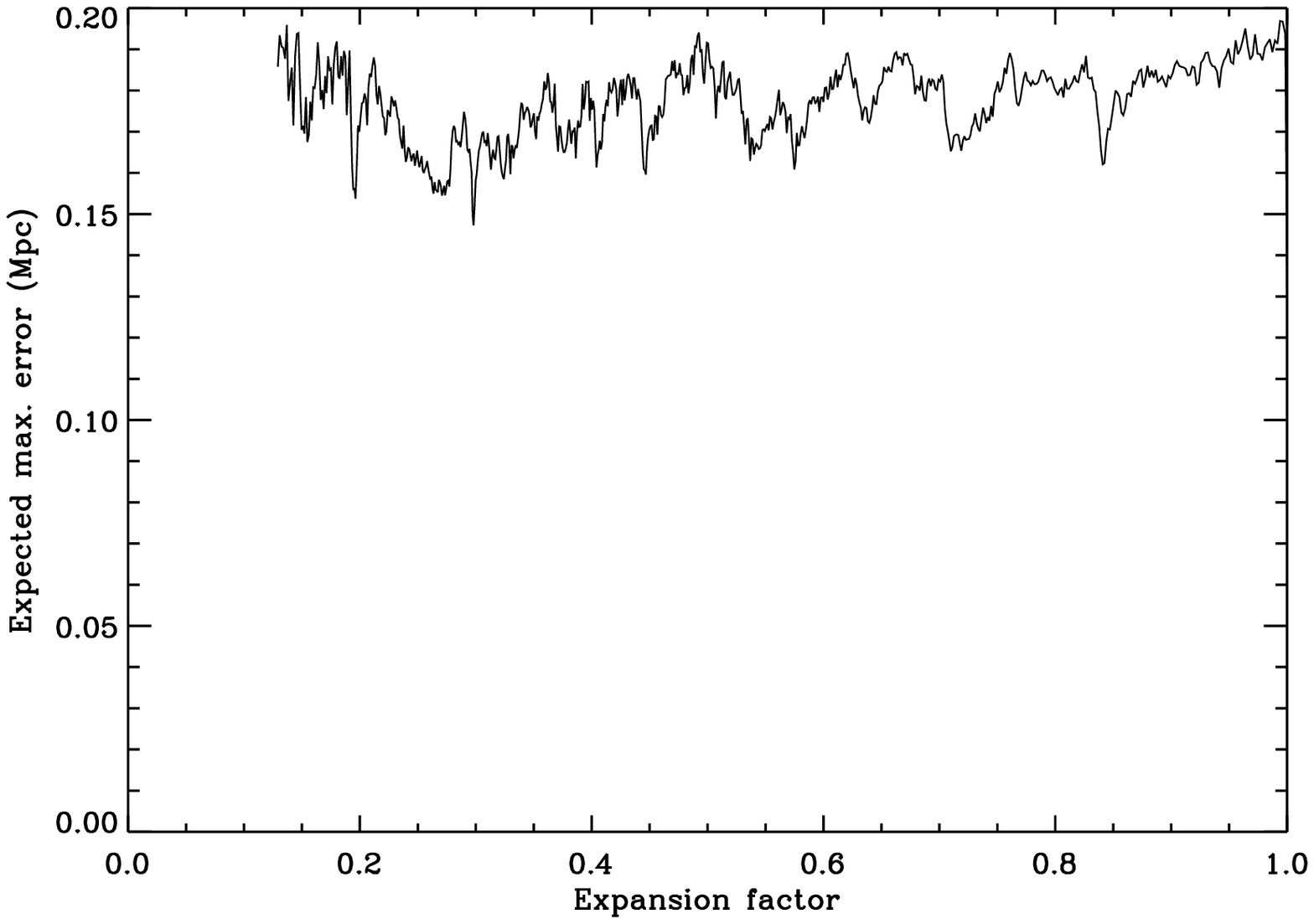}
\caption{the expected maximum uncertainty on particle positions due to
the method used to create the light cone as a function of the expansion factor.
It is computed according to equation (\ref{eq:errorexp}) with a velocity $v$
estimated to be 3 times the Virial velocity of the largest cluster in
the simulation.}
\label{fig:errordiscz}
\end{figure}
The generation of a light cone during run time can be performed easily 
at each coarse time step of the simulation. Given a choice of the observer position in the simulation
box, that we  suppose here for simplicity to be at the origin of coordinates,
it is easy to select the particles that belong to the slice in between
redshifts $z_2 < z_1$ corresponding to two successive coarse time steps: if $(x,y,z)$
are the comoving coordinates of a particle, and $d=\sqrt{x^2+y^2+z^2}$ its comoving
distance from the observer, we must have $d_{\rm dist}(z_2) < d \leq  d_{\rm dist}(z_1)$
for the particle to be selected, where $d_{\rm dist}(z)$ is the comoving distance that a photon
covers between redshift $z$ and present time in the simulation box: 
$d_{\rm dist}= \int c \rm{d}t/a(t)$, where $c$ is the speed of light and $a$ the expansion factor. 
The problem is that structures evolve during a coarse time step, so there are necessarily
some discontinuities at the border between two successive light cone slices.
These discontinuities are due to large scale motions of particles plus their thermal
velocity within dark matter halos. Given the large size of the simulation considered here,
thermal motion within the largest cluster are expect to bring the most significant effects of
discontinuity. For a particle with peculiar velocity $v$, the largest discontinuity
to be expected, i.e. the largest possible difference between expected and actual position of the
particle is given by
\begin{equation}
\Delta= (v/c) [d_{\rm dist}(z_1)-d_{\rm dist}(z_2)].
\label{eq:errorexp}
\end{equation}
In equation~(\ref{eq:errorexp}), we performed a linear Lagrangian approximation, i.e. we neglected
variations of the velocity of the particle during the coarse time step. 
Using Press \& Schechter formalism, or the improved formula of Sheth \& Tormen (1999),
the mass of the largest cluster in the Horizon simulation solves approximately
the implicit equation
\begin{equation}
\Omega_0 \rho_{\rm c} L^3 F[M_{\rm max}(z),z]/M_{\rm max}(z)=1,
\label{eq:mymmax}
\end{equation}
where $\rho_{\rm c}$ is the critical density of the Universe and $F$ is the
fraction of mass in the Universe in objects of mass larger than $M$.
Basically, this equation states that the mass in objects of mass larger
than $M$ is equal to $M$, which means that we are left with only one cluster
of mass $M$, the largest detectable cluster in our cube of size $L$. 
We can compute $F(M,z)$ with the usual formula, e.g.
\begin{equation}
F(M,z)=\int_{\normalsize\mu > \nu(M,z)} f(\mu) \rm{d}\mu,
\end{equation}
with $ \nu=1.686/\sigma(M,z)$ where $\sigma(M,z)$ is the linear variance at redshift $z$ corresponding to 
mass scale $M$, and $f(\mu)$ is given by equation (10) of \cite{ShT}.
Performing these calculations, we find that the largest cluster at present time
in a cube of size $L=2000 h^{-1}$ Mpc should have a typical mass of
$M_{\rm max}(z=0) \simeq 1.47\times 10^{15} M_{\odot}$. With a standard Friend-of-friend
algorithm using a linking parameter $b=0.2$, we find that the most massive halo
detected in the simulation presents a somewhat larger mass, $M=5.4\times 10^{15} M_{\odot}$.
Yet, in that rare events regime,
we cannot expect our theoretical estimate to be more accurate.
What matters, though, is the thermal velocity rather than the mass. Applying the Virial
theorem, we have (e.g., Peacock, 1999)
$
v^2 \simeq{G M_{\rm max}}/{R_{\rm vir}}, 
$
with
\begin{equation}
\frac{4}{3} \pi R_{\rm vir}^3 \rho_{\rm vir} = M_{\rm max}, \quad \rho_{\rm vir}\simeq 178 \Omega_0 \rho_{\rm c} (1+z)^3/\Omega(z)^{0.7}, \nonumber
\end{equation}
where $\Omega(z)$ is the density parameter as a function of redshift ($\Omega(0) \equiv \Omega_0$).
These expressions are given in physical coordinates hence the factor $(1+z)^3$ in the expression of $\rho_{\rm vir}$. 
This reads, at $z=0$, $v \simeq 1570$ km/s for $M_{\rm max}(z=0) \simeq 1.47\times 10^{15} M_{\odot}$.
In the largest cluster of the simulation, the overall velocity dispersion is of the order of $2100$ km/s, a slightly
larger value that reflects the actual value of the mass. To be conservative, we estimate the expected
errors in equation (\ref{eq:errorexp}) with the Virial velocity rescaled by a factor $2100/1570$, and with
a further multiplication by a factor 3 to be in the 3$\sigma$ regime. The corresponding maximal expected discontinuity
displacement is shown in Mpc as a function of the expansion factor on Figure~\ref{fig:errordiscz}. As expected
from the dynamically self-consistent calculation of the coarse time step (which is basically determined by
a Courant condition using the velocity field), the comoving error does not change significantly with redshift and remains below
the very conservative limit of 200 kpc. Obviously, we expect in practice the errors brought by discontinuities
to be in general much smaller than that, as for $z=0$ the present errors 
corresponds to unrealistic velocities as large as about 6000 km/s!

\subsection{From slices to $\kappa$ maps}\label{sec:from-kappa-to}
In the main text, the expression for $\kappa$ as a function of the
density contrast in the simulation is given in equation (\ref{eq:kappa-delta})
in the geometric optic approximation. Let us rearrange this formula
in a form that is more suited to integration over redshift slices
in a simulation.\[
\kappa(\hat{\mathbf{n}}_{{\rm pix}})\approx\frac{3}{2}\Omega_{m}\sum_{b}W_{b}\frac{H_{0}}{c}\int_{\Delta z_{b}}\frac{cdz}{H_{0}E(z)}\delta(\frac{c}{H_{0}}\mathcal{D}(z)\hat{\mathbf{n}}_{{\rm pix}},z)\,,\]
where \[
W_{b}=\left(\int_{\Delta z_{b}}\frac{dz}{E(z)}\frac{\mathcal{D}(z)\mathcal{D}(z,z_{s})}{\mathcal{D}(z_{s})}\frac{1}{a(z)}\right)/\left(\int_{\Delta z_{b}}\frac{dz}{E(z)}\right) \]
 is a slice-related weight, and the integral over the density contrast,
$\delta$, reads\begin{eqnarray*}
I \!\!& = & \!\! \int_{\Delta z_{b}}\frac{cdz}{H_{0}E(z)}\delta(\frac{c}{H_{0}}\mathcal{D}(z)\hat{\mathbf{n}}_{{\rm pix}},z)\,,\\
\!\! & = & \!\!\!\int_{\Delta\chi_{b}}d\chi\delta(\chi\hat{n}_{{\rm pix}},\chi)
  \approx  \frac{V({\rm simu})}{N_{{\rm part}}({\rm simu})}\left(\frac{N_{{\rm part}}(\theta_{{\rm pix}},z_{b})}{S_{{\rm pix}}(z_{b})}-1\right)\,,\end{eqnarray*}
where \[
S_{{\rm pix}}(z_{b})=\frac{4\pi}{N_{{\rm pix}}}\frac{c^{2}}{H_{0}^{2}}\mathcal{D}^{2}(z_{b})\]
is the comoving surface of the spherical pixel. Putting all together,
we get the following formula for the convergence map:\begin{equation}
\kappa(\theta_{{\rm pix}}) \!\! = \!\! \frac{3}{2}\Omega_{m}\frac{N_{{\rm pix}}}{4\pi}\left(\frac{H_{0}}{c}\right)^{3}\!\frac{V({\rm simu})}{N_{{\rm part}}({\rm simu})} \sum_{b}W_{b}\frac{N_{{\rm part}}(\theta_{{\rm pix}},z_{b})}{\mathcal{D}^{2}(z_{b})}\,.
\nonumber
\end{equation}
  Once the $\kappa$ map is available it is straightforward to build the corresponding $\mathbf g$ using equation~(\ref{eq:model}).

\end{document}